\documentclass[a4paper,12pt]{JHEP3}
\usepackage{epsf}
\usepackage{graphicx}

\def\bea{\begin{eqnarray}}
\def\eea{\end{eqnarray}}
\def\beq{\begin{equation}}
\def\eeq{\end{equation}}

\def\lsim{\mathrel{\rlap{\lower4pt\hbox{\hskip1pt$\sim$}}
    \raise1pt\hbox{$<$}}}                

\preprint{
LPT-ORSAY/03-82 \\
ECT*-03-06 \\
IPNO DR 03-09
}

\title{\vspace{-0.3cm}
Resumming QCD vacuum fluctuations in
three-flavour Chiral Perturbation Theory}

\author{S.~Descotes-Genon\\
Laboratoire de Physique Th\'eorique,
91405 Orsay Cedex, France\\
E-mail: \email{descotes@th.u-psud.fr}}
\author{N.~H.~Fuchs\\
Department of Physics, Purdue University, West Lafayette IN 47907, USA\\
E-mail: \email{nhf@physics.purdue.edu}}
\author{L.~Girlanda\\
European Center for Theoretical Studies in Nuclear Physics and
related areas, Strada delle Tarabelle 286, 38050 Trento, Italy\\
E-mail: \email{girlanda@ect.it}}
\author{J.~Stern\\
Groupe de Physique Th\'eorique, IPN, 91405 Orsay Cedex, France\\
E-mail: \email{stern@ipno.in2p3.fr}}

\abstract{
Due to its light mass of order $\Lambda_{\rm QCD}$, the strange
quark can play a special role in Chiral Symmetry Breaking ($\chi$SB):
differences in the pattern of $\chi$SB in the limits $N_f=2$ ($m_u,m_d\to 0$, 
$m_s$ physical) and $N_f=3$ ($m_u,m_d,m_s\to 0$) may arise due to
vacuum fluctuations of $s\bar{s}$ pairs, related
to the violation of the Zweig rule in the scalar sector
and encoded in particular in the $O(p^4)$ low-energy constants $L_4$ and 
$L_6$. In case of large fluctuations, 
we show that the customary treatment of $SU(3)\times SU(3)$ chiral
expansions generate instabilities upsetting their convergence.
We develop a systematic program to cure these instabilities 
by resumming nonperturbatively vacuum fluctuations of $s\bar{s}$ pairs,
in order to extract information about $\chi$SB from experimental
observations even in the presence of large fluctuations. 
We advocate a Bayesian framework for treating the
uncertainties due to the higher orders. As an application,
we present a three-flavour analysis of
the low-energy $\pi\pi$ scattering and show that
the recent experimental data imply a  lower bound on the quark mass
ratio $2m_s/(m_u+m_d)\geq 14$ at 95\% confidence level. We outline how
additional information may be incorporated to further constrain 
the pattern of $\chi$SB in the $N_f=3$ chiral limit.

\vspace{-0.4cm}
}


\begin{document}

\section{Introduction} 

Light quarks have their own hierarchy of masses. On one hand, $m_u$
and $m_d$ are much smaller than any intrinsic QCD scale, and their
non-zero values induce only small corrections to the $SU(2) \times
SU(2)$ chiral limit, in which $m_u = m_d = 0$.  A systematic expansion
in $m_u$ and $m_d$, keeping all remaining quark masses at their
physical values, defines the two-flavour Chiral Perturbation
Theory ($\chi$PT)~\cite{GL1}. On the other hand,
the mass $m_s$ of the strange quark is considerably higher (see
e.g., Ref.~\cite{ms}
and references therein for recent determinations); indeed, it
is nearly of the order of $\Lambda_{\rm QCD}$, the characteristic
scale describing the running of the QCD effective
coupling. Nevertheless, a slowly convergent $SU(3) \times SU(3)$
chiral expansion is conceivable~\cite{GL2}.  This is suggested from a
comparison of the kaon mass $M_K$ with the mass scale $\Lambda_H \sim$
1 GeV of (strange) QCD bound states not protected by chiral
symmetry: $M^2_K/\Lambda^2_H \sim 0.25 - 0.30$.  Due to the rather
specific value of $m_s$, the strange quark plays a special role
among all six quarks:
\begin{enumerate}
\item[i)] $m_s$ is small enough to be used as an expansion parameter
(at least in some restricted sense) and to relate properties of QCD
vacuum in the $SU(3) \times SU(3)$ chiral symmetry limit $m_u = m_d =
m_s = 0$ to observable quantities.

\item[ii)] Unlike $m_u , m_d$, the strange quark mass is
sufficiently large, $m_s \sim \Lambda_{\rm QCD}$, to influence the
magnitude of order parameters characteristic of the $SU(2) \times
SU(2)$ chiral limit $m_u, m_d = 0$ with $m_s$ fixed at its physical
value.

\item[iii)] At the same time, $m_s$ is not large enough to suppress
loop effects of massive $\bar ss$ vacuum pairs. This is to be
contrasted with heavy quarks $Q = c, b, t$ for which $ m_Q \gg
\Lambda_{\rm QCD}$ and the effect of $\bar QQ$ pairs on the vacuum
structure is expected to be tiny.
\end{enumerate}

The above remarks single out the role of massive strange sea quarks,
and suggest a possibly different behaviour for $N_f=2$ and $N_f=3$
chiral dynamics. The origin of this difference clearly appears in
connection with the possibility that in the vacuum channel
($J^{PC}=0^{++}$) the Zweig rule and the $1/N_c$ expansion break
down. This is strongly suggested by scalar meson spectroscopy~\cite{scalar},
sum rule studies~\cite{Bachir1,Bachir2,DS,D}, 
as well as by instanton-inspired models~\cite{instant}\footnote{It
should also be visible in fully unquenched lattice simulations.}.
Furthermore, the enhancement of Zweig-rule violating effects of $\bar
ss$ pairs on chiral order parameters has a natural theoretical
interpretation as a consequence of fluctuations of the lowest
eigenvalues of Euclidean QCD Dirac operator, in particular of their
density~\cite{dirac}. 
These fluctuations would only affect quantities dominated by
the infrared end of the Dirac spectrum~\cite{DGS}; 
however, this is precisely the
case of chiral order parameters such as the quark condensate and the
pion decay constant. (Most of observables not protected by
chiral symmetry are not especially sensitive to small Dirac
eigenvalues and they have no particular reason to break the Zweig rule
or the $1/N_c$ expansion.)  Fluctuations of small Dirac eigenvalues
lead to a large long-range correlation between $0^+$ massive $\bar ss$
and massless $ \bar{u}u + \bar{d}d$ pairs. This correlation enhances
the $SU(2)\times SU(2)$ order parameters
\begin{eqnarray}
\Sigma(2)&=&  - \lim_{m_u,m_d\to 0}\langle\bar uu\rangle 
  |_{m_s={\rm physical}}\\
F(2)^2  &=& \lim_{m_u,m_d\to 0} F^2_{\pi}|_{m_s={\rm physical}}
\end{eqnarray}
by a contribution which is induced from vacuum $\bar ss$ pairs.  The
``induced condensate'' and ``induced decay constant''~\cite{DGS,GST}
are proportional
to $m_s$ and vanish in the $SU(3) \times SU(3)$ chiral limit $m_u =
m_d = m_s = 0$.  As a result the $N_f=3$ condensate $\Sigma(3)$ and
the decay constant $F(3)^2$ can be substantially suppressed compared
to the corresponding two-flavour order parameters,
\begin{eqnarray} \label{parasig}
\Sigma(2)>& \Sigma(3)=& \Sigma(2)|_{m_s=0} \\
F(2)^2   >& F(3)^2   =& F(2)^2|_{m_s=0} \ . \label{paraf}
\end{eqnarray}
The existence of this paramagnetic effect and its sign can be expected
on general theoretical grounds~\cite{DGS}, but its magnitude depends on the
size of fluctuations of small Dirac eigenvalues, which is hard to
infer from first principles.  A general discussion of the interplay
between chiral order and fluctuations in the QCD vacuum can be found
in Ref.~\cite{ordfluc}.

The main question to be asked is how can the effect of vacuum
fluctuations on chiral symmetry breaking be detected experimentally.
Recall that two-flavour order parameters are most easily accessible
via low-energy $\pi\pi$ scattering.  Using accurate recent data~\cite{E865},
we have inferred values for the $N_f=2$ condensate and decay constant;
expressed in suitable physical units, we found~\cite{DFGS}
\begin{eqnarray}
X(2) =& \displaystyle 
   \frac{(m_u+m_d)\Sigma(2)}{F^2_\pi M^2_\pi}&= 0.81 \pm 0.07 \\
Z(2) =& \displaystyle
   \frac{F(2)^2}{F^2_{\pi}} &= 0.89 \pm 0.03 \ .
\end{eqnarray}
The fact that both $X(2)$ and $Z(2)$ are rather close to one indicates
that, as long as $m_s$ is kept at its physical value, the effect of
nonzero $m_u,m_d$ is indeed small.  This in turn suggests that the
standard two-flavour $\chi$PT is a well-behaved expansion~\cite{GL1}; 
its leading
order, described by the decay constant $F^2 \equiv F(2)^2$ and by the
$N_f=2$ quark condensate $\Sigma(2)\equiv F^2 B$, is dominant.  On the
other hand, the three-flavour order parameters $\Sigma(3)$ and
$F(3)^2$ are more difficult to pin down, since they require an
extrapolation to $m_s=0$.  The latter necessitates the use of
three-flavor $\chi$PT, including more observables such as the masses
and decay constants of the whole octet of Goldstone bosons, the
$K-\pi$ form factors, $K-\pi$ scattering amplitude, etc.  The $N_f=3$
$\chi$PT involves more low-energy constants starting in order
$O(p^4)$, and higher orders are likely to be more important than in
the two-flavour case.  Most existing analyses~\cite{GL2,BEG,ABT,twoloop} 
are based on an
explicit assumption that the effect of vacuum fluctuations of $\bar
ss$ pairs on order parameters is small: it is usually assumed that the
two parameters of the $O(p^2)$ Lagrangian $ F_0 \equiv F(3)$ and $\Sigma(3)
\equiv F_0^2B_0$ are such that $F_0 \approx F_{\pi}$ and
$(m_u+m_d)\Sigma(3) \approx F_\pi^2 M^2_{\pi}$, i.e., that the
corrections due to nonvanishing $m_s$ can be treated as a small
perturbation.

A closely related assumption concerns the smallness of the two
$O(p^4)$ LEC's $L^r_6(\mu)$ and $L^r_4(\mu)$ which describe the
large-$N_c$ suppressed and Zweig-rule violating effects of
fluctuations in the vacuum channel.  Experimental information on the
actual size of these two constants has been rather scarce; for long
time it was customary to posit $L^r_6(M_{\rho})=(-0.2 \pm 0.3)\times
10^{-3}$ and $L^r_4(M_{\rho})= (-0.3 \pm 0.5)\times 10^{-3}$ as an
input to both one-loop~\cite{GL2,BEG} and two-loop
calculations~\cite{ABT,twoloop}. More
recently, attempts of indirect estimates of $L_4$ and $L_6$ have
appeared, all pointing towards a small positive values compared to the
old Zweig-rule based estimates mentioned above.  Rapidly convergent
sum rules for the correlator
$\langle(\bar uu)(\bar ss)\rangle$~\cite{Bachir1,Bachir2,DS,D}  yield a rough
estimate $L^r_6(M_{\rho}) =(0.6 \pm 0.3)\times 10^{-3}$, while from
the analysis of $\pi-K$ sum-rules~\cite{Kpi} it has been 
concluded that
$L^r_4(M_{\rho})=(0.2 \pm 0.3)\times 10^{-3}$. The last conclusion has
been confirmed in a recent two-loop fit to the scalar
form-factors~\cite{bijscal}.  The point is that the effect of such
small shifts 
on order parameters is amplified by large coefficients: with the above
estimates, $\Sigma(3)$ and $F(3)^2$ can be suppressed compared to
$\Sigma(2)$ and $F(2)^2$ respectively by as much as a factor of 2. In
this way, vacuum fluctuations of $\bar ss$ pairs could lead to a
particular type of instability in three-flavour $\chi$PT.
                 
The main purpose of the present work is to investigate instabilities
in $N_f=3$ $\chi$PT that would specifically arise from the (partial)
suppression of order parameters $\Sigma(3)$ and $F^2_0$, and to
propose a systematic nonperturbative modification (resummation) of the
standard $\chi$PT recipe that could solve the problem.  We assume that
the whole expansion of relevant observables in powers of $m_s$ is
globally -- though slowly and at most asymptotically -- convergent.
The problem may occur with the enhancement of particular terms of the
type $m_s L_6$ or $m_s L_4$ that appear with large coefficients and
can be identified as arising from fluctuation of vacuum $\bar ss$
pairs.  These terms are responsible for the important ``induced
contributions'' to $\Sigma(2)$ and $F(2)^2$, explaining why
$\Sigma(2)$ and $F(2)^2$ at physical $m_s$ could be substantially
larger than their $m_s=0$ limit $\Sigma(3)$ and $F(3)^2$ respectively.
We show that in order to solve this particular problem it is not
necessary to modify the standard chiral counting rules as in 
generalized $\chi$PT~\cite{GChPT}.  
The modification we propose is more modest: within the
standard expansion scheme in powers of quark masses and external
momenta, it appears sufficient to resum the fluctuation terms driven
by $m_s L_4$ and $m_s L_6$ in the usual perturbative reexpression of
order parameters $m_s
\Sigma(3)$ and $F^2_0$ in terms of observables such as $M_{\pi},
M_K, \ldots$ and physical decay constants. This resummation is of
importance for the purpose of extracting the value of $N_f=3$ order
parameters from experiment.

The possible effects of vacuum fluctuations in three-flavour $\chi$PT
and their resummation are discussed in Sections 2 and 3. These two
sections are focused on Goldstone boson masses and decay constants,
which are the observables directly entering the reexpression of order
parameters $m \Sigma(3)$ and $F^2_0$. In our approach, the influence
of higher $\chi$PT orders ($O(p^6)$ and higher) is encoded into a few
parameters referred to as ``NNLO remainders'', which are kept through
the whole analysis whatever their values. The latter depend on the
model one takes for the higher order counterterms and one hopes they
remain reasonably small independently of the model used. The result
of this part of our article is an exact expression of $L_4, L_5, L_6$
and $L_8$ in terms of three fundamental parameters
\begin{equation}
X(3)= \frac{(m_u+m_d)\Sigma(3)}{F^2_{\pi}M^2_{\pi}}\,, \qquad
Z(3)= \frac {F_0^2}{F^2_{\pi}}\,, \qquad
r   = \frac {2m_s}{m_u+m_d}\,,
\end{equation}
and four NNLO remainders. Using these expressions inside the $\chi$PT
formulae for various additional observables, one can hope to pin down
the values of $X(3), Z(3)$ and $r$ for a given set of assumptions
about higher orders (NNLO remainders). The logical structure of the
problem naturally calls for a Bayesian statistical type approach~\cite{bayes}.

As a first application we consider the three-flavour analysis of
$\pi\pi$ scattering, since today rather accurate data exist in this
case and we know from past studies~\cite{DGS,ordfluc} that a strong correlation
exists between the value of $r$ and the characteristics of the
two-flavour chiral limit as revealed in low-energy $\pi\pi$
scattering.  In Section 4 a quantitative analysis of this correlation
is presented for the first time. On the other hand, the $N_f=3$ order
parameters $X(3)$ and $Z(3)$ cannot be extracted from the $\pi\pi$
data alone. In Section 5 we survey some possibilities of learning
about these fundamental order parameters from $\pi-K$ scattering,
$\eta \to 3 \pi$ decays, OPE condensates and sum rules and, last but
not least, from lattice simulations with three fully dynamical
fermions: we present the corresponding extrapolation formulae using
our resummed $\chi$PT formulation.

\section{Convergence and instabilities of $N_f = 3$ chiral expansion}
 \label{secconvinst}

We first recall the general structure of three-flavour $\chi$PT~\cite{GL2}, 
emphasising where and how the instabilities due to vacuum
fluctuations of $\bar ss$ pairs~\cite{DGS} could possibly show up. Unless
stated otherwise, a typical quantity subject to the expansion in powers of
running quark masses  $m_u , m_d , m_s$ will be thought of as a connected QCD
correlation function of quark currents ($ V , A , S , P$) with external
momenta fixed somewhere in a low-energy region away from the singularities
generated by Goldstone bosons. We will take as a working hypothesis that
the usual low-energy observables, e.g., Goldstone boson masses, decay constants,
form factors and scattering amplitudes (at particular kinematical
points), when linearly expressed through such QCD correlation functions
exhibit optimal convergence properties. While a similar assumption 
is implicitly made in the standard off-shell formulation of 
$\chi$PT~\cite{GL1,GL2}, we will shortly argue 
that in the presence of important vacuum fluctuations
this assumption should be understood as a restriction:               
observables that are not \emph{linearly} expressible
in terms of QCD correlators, e.g., \emph{ratios} of Goldstone boson masses,
need not admit a well convergent perturbative treatment and they should be
treated with a particular care. This selects for instance $F_\pi^2
M_\pi^2$, $F_\pi^2$ and $F_\pi^2 F_K^2 A_{\pi K}$ (where $A_{\pi K}$
denotes the $\pi K$-scattering amplitude), but rules out $M_\pi^2$.

\subsection{The bare $\chi$PT series}

       The chiral expansion of symmetry-breaking observables in terms of
the three lightest quark masses $ m_u , m_d , m_s $ is actually not a
genuine power series expansion, due to the presence of chiral
logarithms, which reflect infrared singularities characteristic of the
chiral limit. One can nevertheless give an unambiguous
scale-independent meaning to the renormalized coefficients of each
power of individual quark masses. An observable $A$ can be
represented as a formal series
\begin{equation}\label{expone}
A = \sum\limits_{j,k,l} m_u^j m_d^k m_s^l  A_{jkl}
[ m_u,m_d,m_s ; B_0,F_0 ; L_1^r(\mu)\ldots L_{10}^r(\mu) ; C_1^r(\mu)\ldots
C_{90}^r(\mu) ;\ldots ],
\end{equation}
where the coefficients $A_{jkl}$ are defined  in terms of the constants
contained in the effective Lagrangian: \emph{i)} The basic order
parameters $B_0$ and $F_0$ which are related to the three-flavour chiral 
limit of the quark condensate and  of the pion decay constant 
respectively,           
\begin{equation}\label{OP}
\Sigma(3)=-\lim_{m_u,m_d,m_s\to 0}\langle\bar{u}u\rangle\,, \qquad
F_0\equiv F(3)=\lim_{m_u,m_d,m_s\to 0} F_\pi\,, \qquad
B_0 = \frac{\Sigma(3)}{F(3)^2}\,;
\end{equation}
\emph{ii)} the 10  $O(p^4)$  LEC's
$L_i^r(\mu)$, \emph{iii)} the  90  $O(p^6)$  LEC's  $C_i^r(\mu)$~\cite{twoloopgen}, 
and eventually higher-order counterterms. All LEC's are renormalized 
at the scale $\mu$. In addition, the 
$A_{jkl}$ depend logarithmically on the quark masses through the Goldstone
boson masses in the loops, and this dependence      
is such that for each $jkl$
the coefficient $A_{jkl}$ is independent of the scale $\mu$. 
The representation (\ref{expone}) has been explicitly worked out 
for some observables
to one~\cite{GL2} and two loops~\cite{ABT,twoloop,bijscal}
and there is no doubt that it extends to all orders of the chiral
expansion. We shall
refer to the expansion expressed in the form (\ref{expone}) as a
\emph{bare expansion}, to emphasize that it is entirely written in terms of
the parameters of the effective Lagrangian -- no reexpression of the latter
in terms of observable quantities has been performed. It is crucial that
even before one starts rewriting and reordering  the series (\ref{expone})
in powers of Goldstone boson masses, the full renormalisation of the bare
expansion (\ref{expone}) can be performed order by order in quark masses.
Consequently, the coefficients $A_{jkl}$ are finite as well as cut-off and
renormalisation scale-independent for all values of quark masses and of
(renormalized) LEC's in the effective Lagrangian.

                In view of possible applications, we are concerned with
practical questions related to the convergence properties of the
bare $\chi$PT expansion (\ref{expone}) in QCD. The latter will 
depend on the values of running quark masses and on the values of the LEC's
at the typical hadronic scale $\Lambda_H \sim M_{\rho}$ set 
by the masses of non-Goldstone hadrons. In particular, one
should question the convergence of the  bare chiral expansion for the actual  
values of quark masses
and not just in the infinitesimal vicinity of the chiral limit. In the real
world, all three quarks $u d s$ are sufficiently light,
\begin{equation}\label{qm}
m_u(\Lambda_H),m_d(\Lambda_H) \ll m_s(\Lambda_H) \ll \Lambda_H\,,
\end{equation}
to expect a priori some (at least asymptotic) convergence of the
three-flavour bare $\chi$PT series.
As pointed out in Refs.~\cite{DS,DGS}, 
instabilities of the latter can nevertheless
occur due to fluctuations of massive $\bar ss$ pairs in the vacuum. 
The importance of such pairs is measured by 
the strength of the effective QCD coupling;         
i.e., comparing $m_s$ with $\Lambda_{QCD}$, rather 
than with the hadronic scale $\Lambda_H$.
Furthermore, the impact of these fluctuations is proportional to $m_s$. Hence,
instabilities due to fluctuations of vacuum quark-antiquark pairs
turn out to be particularly relevant for strange quarks and could
manifest themselves when two- and three-flavour chiral
expansions are compared. 

It has been argued~\cite{DS,DGS} that fluctuations of $\bar ss$ pairs
 lead to a partial suppression of the three-flavour
condensate $\Sigma(3)$, reducing the relative importance 
of the first term in the
bare expansion of the Goldstone boson masses. We can consider for instance
the Ward identity related to the mass of the pion (from now on
we neglect isospin breaking and take $m_u = m_d = m$):
\begin{eqnarray}\label{expgold}
F_{\pi}^2 M^2_{\pi}&=& 2 m \Sigma(3)+ 2 m m_s Z^s\\ \nonumber
&& + 4 m^2 \left[ A + Z^s +
    \frac{B^2_0}{32\pi^2}
   \left(3\log\frac{M_K^2}{M_\pi^2}+\log\frac{M_\eta^2}{M_K^2}\right)\right]
   + F_{\pi}^2 M_{\pi}^2 d_{\pi}\,.
\end{eqnarray}
The parameters $Z^s$ and $A$ are defined in terms of the LEC's
$L_6(\mu) , L_8(\mu)$ and  logarithms of Goldstone boson masses (their
expression is recalled in App.~\ref{appident}).
Vacuum fluctuations of $\bar ss$-pairs  show up in the term $m_s Z^s$.
For the physical value of $m_s \sim \Lambda_{QCD}$, the corresponding $O(p^4)$
term $ 2m m_s Z^s$ can be as important~\cite{Bachir1,DS,D} 
as the leading-order
condensate term $2m \Sigma(3)$. Even then, the remainder
$d_{\pi}$, which collects all $O(p^6)$ and higher contributions, can still be
small: $d_{\pi} \ll 1$. In other words, vacuum fluctuations need not affect
the \emph{overall} convergence of the bare chiral expansions  such as
(\ref{expone}) or (\ref{expgold}) at least for some well-defined selected
class of observables.

\subsection{The role of NNLO remainders} \label{secNNLO}
              
	      Let us write a generic bare expansion (\ref{expone}) in a
concise form
\begin{equation}\label{LNLR}
A = A_{\rm LO} + A_{\rm NLO} + A \,\delta\! A .
\end{equation}
Eq.~(\ref{LNLR}) is an identity: $A_{\rm LO}$ collects leading powers in quark
masses in the bare expansion (\ref{expone}) (e.g., the condensate term in
Eq.~(\ref{expgold})), $A_{\rm NLO}$ consists of all 
next-to-leading contributions
(the second and third terms in Eq.~(\ref{expgold})), whereas $A \,\delta\! A$
stands for the sum of all remaining terms starting with the
next-to-next-to-leading order (NNLO). In Eq.~(\ref{expgold}), the latter is
denoted as $\delta(F^2_{\pi} M^2_{\pi}) \equiv d_{\pi}$.

With this setting, $A$  can be identified with the exact
(experimental) value of the observable $A$. Usually,
$A_{\rm LO}$ corresponds to the $O(p^2)$ contribution, $A_{\rm NLO}$ to $O(p^4)$ and
$A \,\delta\! A$ collects all higher orders starting with
$O(p^6)$~\footnote{The case
of a quantity whose expansion only starts at $O(p^4)$ or higher, requires 
particular care.}. $\delta\!A$ will be referred to as ``NNLO remainder''.
A precise definition of $\delta\!A$
involves some convention in writing the argument in the chiral logarithms
contained in the one-loop expression for $A_{\rm NLO}$. To illustrate this point
consider the typical next-to-leading expression:
\begin{equation}\label{NL}
A_{\rm NLO} = \sum_{qq'} m_q m_{q'}\left[a_{qq'}(\mu)+\sum_{PQ}
a^{PQ}_{qq'} k_{PQ}(\mu)\right]\,,
\end{equation}
or
\begin{equation}\label{NLbis}
A_{\rm NLO} = \sum_{q} m_q \left[a_q(\mu)+\sum_{PQ}
a^{PQ}_{q} k_{PQ}(\mu)\right]\,,
\end{equation}
corresponding respectively to a leading-order term
$A_{\rm LO} = O(m_{\rm quark})$ and $A_{\rm LO} = O(1)$.
Here $q,q'=(u,d,s)$ and $P,Q$ label Goldstone bosons. We have introduced
the loop factor in the general case of unequal masses~\cite{GL2}:
\begin{equation}\label{LPQ}
k_{PQ}(\mu)= \frac{1}{32\pi^2} \frac{M_P^2 \log(M_P^2/\mu^2) -
M_Q^2 \log(M_Q^2/\mu^2)}{M_P^2 - M_Q^2}\,,
\end{equation}
which in the limit of equal masses becomes:
\begin{equation}\label{LPP}
k_{PP}(\mu) = \frac{1}{32 \pi^2}
   \left[ \log \frac{M_P^2}{\mu^2} + 1 \right]\,.
\end{equation}
In Eqs.~(\ref{NL})-(\ref{NLbis}), 
the constants $a_{qq'}(\mu)$ [$a_q(\mu)$] are expressed in terms
of $O(p^4)$ LEC's $L^r_i(\mu)$ multiplied by the appropriate powers of
$B_0$ and $F_0$. These constants are defined in the chiral 
limit and are consequently independent of quark masses, 
similarly to the known numerical coefficients $a^{PQ}_{qq'}$
[$a^{PQ}_q$]. 

The only requirement from $\chi$PT is that 
$A_{\rm NLO}$ reproduces the $O(p^4)$ behaviour in the limit of small
quark masses $m_{\rm quark} \to 0$; i.e., when the Goldstone boson masses
$M_P^2$ in the loop factors (\ref{LPQ}) and (\ref{LPP})
are replaced by their respective leading-order contributions. Once this
mathematical condition is satisfied, different ways of writing the arguments
of the chiral logarithms for \emph{physical values} of
quark masses merely correspond
to different conventions in defining the NNLO remainders $\delta\!A$.
For observables of the form of Eqs.~(\ref{NL})-(\ref{NLbis}) at $O(p^4)$, 
we will use the convention which consists in writing in
Eq.~(\ref{LPQ}) the physical values of the Goldstone boson masses $M^2_P$;
alternatively, we could have used the sum of LO and NLO contributions to $M^2_P$.
This concerns, in particular, the expansion of the Goldstone boson 
masses and decay constants. In the latter case one has  
$P=Q$ and the convention simply amounts
to writing the $O(p^4)$ tadpoles, in the notation of Ref.~\cite{GL2}, as
\begin{equation}
\mu_P = \frac{1}{32 \pi^2} \frac{\left.M^2_P \right|_{\rm LO}}{F^2_0}
 \log\frac{\left.M^2_P\right|_{\rm phys.}}{\mu^2}\,.
\end{equation}
The same rule can be applied to the unitarity corrections
arising in the bare expansion of subtraction
constants that define form factors and low-energy $\pi\pi$~\cite{KMSF} 
and $\pi K$~\cite{Kpi,bkm,TBP} amplitudes. Such a prescription 
(detailed in Sec.~\ref{seclowpipi}) will suffice for the quantities 
considered in this article.

             Not much is known about the size of NNLO remainders despite
the fact that complete $SU(3) \times SU(3)$ two-loop calculations
do exist for some observables~\cite{ABT,twoloop,bijscal} and the general structure of
the generating functional is known to this order~\cite{twoloopgen}. 
Following this
line, the bare expansion can be pushed further and the NNLO remainder
$\delta\! A$ can be represented as
\begin{equation}\label{twoloops}
A\,\delta\! A = \Delta_{2L}^A (\mu) + \Delta_{1L}^A (\mu) + 
 \Delta_{\rm tree}^A (\mu) +\ldots\,,
\end{equation}
where the ellipsis stands for $O(p^8)$ and higher contributions. 
The splitting of the $O(p^6)$ part~\cite{twoloopgen} into the genuine 
two-loop contribution $\Delta_{2L}$
(containing only $O(p^2)$ vertices), the one-loop contribution $\Delta_{1L}$
(with the insertion of a single $O(p^4)$ vertex) and the tree 
$O(p^6)$ contribution $\Delta_{\rm tree}$ depends on
the renormalisation scale and scheme.

Several ingredients are actually needed to estimate  $\delta A$ 
from the representation
(\ref{twoloops}). The first two terms (loop contributions) depend 
respectively on $O(p^2)$ parameters $mB_0 , m_sB_0 , F_0$  and on 
$O(p^4)$ LEC's $L_i^r (\mu)$. Furthermore, the tree-level counterterms
$\Delta_{\rm tree}^A (\mu)$
are built up from the 90 LEC's $C_i^r (\mu)$ that define the $O(p^6)$
effective Lagrangian. Even if some of them can presumably be determined from
the momentum dependence of form factors, decay distributions and scattering
amplitudes (e.g., quadratic slopes), the remaining unknown
$O(p^6)$ constants, which merely describe the higher-order dependence
on quark masses, are probably much more numerous than the observables that
one can hope to measure experimentally. At this stage some models
(resonance saturation, large $N_c$, NJL \ldots) and/or lattice determinations
are required~\cite{modelsOp6}, but the large number of terms contributing to a
given $\Delta_{\rm tree}^A$ makes the resulting uncertainty in $\delta\!A$
delicate
to estimate. Finally, it is worth stressing that only the sum of the
three components shown in Eq.~(\ref{twoloops}) is meaningful. An estimate of
the size of NNLO remainders is therefore not possible without a
precise knowledge of the $O(p^2)$ and $O(p^4)$ constants $mB_0,F_0$
and the $L_i$'s. 

In this paper, we do not address the problem of
determining NNLO remainders on the basis of Eq.~(\ref{twoloops}). 
We are going to show 
that interesting nonperturbative conclusions can be reached, even if we
do not decompose NNLO remainders and investigate the behaviour of 
the theory as a function
of their size. We are primarily interested in the constraints imposed
by experimental data on the fundamental QCD $SU(3) \times SU(3)$ chiral      
order parameters (and quark mass ratio)
\begin{eqnarray}\label{XZr}
X(3)&=&\frac{2m\Sigma(3)}{M^2_{\pi}F^2_{\pi}}, \quad
Z(3)=\frac{F^2(3)}{F^2_{\pi}}, \quad r=\frac{m_s}{m}\,,\\
Y(3)&=&\frac{2mB_0}{M_\pi^2}=\frac{X(3)}{Z(3)}\,,
\end{eqnarray}
under various theoretical assumptions on NNLO remainders
(i.e., on higher $\chi$PT orders). A suitable approach to this problem        
is provided by Bayesian statistical inference~\cite{bayes}. 
(See App.~\ref{appbayes}
for a brief review adapted to the case of $\chi$PT.) 
The output of this analysis 
is presented as marginal probability distribution functions for 
the fundamental parameters
(\ref{XZr}) and it depends not only on the experimental input but also
on the state of our knowledge of higher $\chi$PT orders. In this approach the
latter dependence is clearly stated and can be put under control: the
analysis can be gradually refined if new information on the relevant
NNLO remainders becomes available either through Eq.~(\ref{twoloops})
or in another way.

                        We start with a very simple theoretical assumption
about higher orders: the bare chiral expansion of ``good observables''
as defined at the beginning of this section, is \emph{globally convergent}.
By these words, we mean that the NNLO remainder $\delta\! A$  in the
identity (\ref{LNLR})  is small 
compared to 1:
\begin{equation}\label{conv}
\delta\!A \ll 1\,,
\end{equation}
for the physical values of the quark masses and for the actual size
of $O(p^2)$ and $O(p^4)$ parameters.
On general grounds, one expects $\delta A = O(m_{\rm quark}^2)$.
In the worst case, its size should be $\delta A =O(m_s^2)\sim(30\%)^2= 0.1$, 
but in many situations $\delta A$ turns out to be $O(m_s m)$ or even $O(m^2)$
and is therefore more suppressed~\footnote{\label{foot:thumb}We take as order of magnitudes
10~\% for $O(m)$ contributions and 30~\% for $O(m_s)$ terms. This can be 
related to the typical sizes of violation for $SU(2)\times SU(2)$ and 
$SU(3)\times SU(3)$ flavour symmetries.}. 
These cases are usually identified
as a consequence of $SU(2) \times SU(2)$ low-energy theorems. 
 (Such suppressions are not claimed from arguments based on the Zweig rule,
since we never assume the latter.) However the NNLO remainders will not
be neglected or used as small expansion parameters in the following.

                           On the other hand, no particular hierarchy
will be assumed between the leading $O(p^2)$ and next-to-leading
$O(p^4)$ components of (\ref{LNLR}). By definition, for infinitesimally
small quark masses $m_u , m_d , m_s$ one should have
\begin{equation}\label{LNL}
A_{\rm NLO}\ll A_{\rm LO}\,,  \qquad X_A \equiv \frac{A_{\rm LO}}{A} \sim 1\,.
\end{equation}
However, due to vacuum fluctuations of $\bar qq$ pairs,
the condition (\ref{LNL}) can be easily invalidated
for physical value of $m_s \sim \Lambda_{QCD}$:
as discussed in Ref.~\cite{DGS},  the three-flavour
condensate  $\Sigma(3)$ in Eq.~(\ref{expgold})
may be of a comparable size to -- or even smaller than -- the term $m_s Z^s$,
reflecting the vacuum effects of massive $\bar ss$ pairs.
At the same time, vacuum fluctuations need not affect the overall convergence
of the bare chiral expansion (\ref{expgold}), i.e., the condition
(\ref{conv}) can still hold for ``good observables'' such as
$F^2_{\pi} M^2_{\pi}$. We will call \emph{conditionally convergent}
an observable for which $\delta A \ll 1$ but
the hierarchy condition (\ref{LNL}) does not hold.

\subsection{Instabilities in chiral series} \label{secinst}

                       Standard $\chi$PT consists of two different steps.
\begin{enumerate}
\item The first step coincides with what has been described above as
the ``bare expansion'' in powers of quark masses and external momenta.
The coefficients of this expansion are unambiguously defined in terms of
parameters of the effective Lagrangian $ B_0 , F_0 , L_i \ldots $,
independently of the convergence properties of the bare expansion.

\item The second step consists in rewriting the bare expansion as an
expansion in powers of Goldstone boson masses, by eliminating order by order
the quark masses $m$ and $m_s$ 
and the three-flavour order parameters
$\Sigma(3) , F(3)$  in favour of the physical values of Goldstone boson
masses $M^2_P$ and decay constants $F^2_P$. For this aim one inverts the
expansion of Goldstone boson masses:
\begin{equation}\label{B}
2mB_0 = M^2_{\pi} \left( 1 + \sum_{P} c^B_P M^2_P + \ldots\right)\,,
\end{equation}
where $c^B_P$ contains the low-energy constants $L_i$ and the chiral logarithms.
A similar ``inverted expansion'' is worked out for $F^2_0$:
\begin{equation}\label{F}
F^2_0 = F^2_{\pi} \left( 1 + \sum_{P} c^F_P M^2_P + \ldots \right)\,,        
\end{equation}
and for the quark mass ratio
\begin{equation}\label{r}
\frac{m_s+m}{2m} = \frac{M^2_K}{M^2_{\pi}} 
  \left( 1 +\sum_{P} c^r_P M^2_P + \ldots\right).
\end{equation}
\end{enumerate}

As a result of these two steps, observables
other than $M^2_{\pi}, M^2_K, F^2_{\pi}$ (already used in Eqs.~(\ref{B}),
(\ref{F}) and (\ref{r})) 
are expressed as expansions in powers of $M^2_P$ and
$\log M^2_P$ with their coefficients depending on the  constants 
$L_i , C_i$, etc.

We now argue that large vacuum fluctuations of $\bar ss$ pairs
could represent a serious impediment to the second step, i.e., to the
perturbative reexpression of order parameters. This may happen if  the
bare expansion (\ref{LNLR}) of Goldstone boson masses and decay constants is
only conditionally convergent: the leading and next-to-leading
contributions are then of comparable size
$A_{\rm LO} \sim A_{\rm NLO}$, despite a good global convergence $\delta\!A \ll 1$.
Let us concentrate on the three-flavour GOR ratio
$X(3)$, defined in Eq.~(\ref{XZr}), which measures the condensate $\Sigma(3)$
in the physical units $F^2_{\pi} M^2_{\pi}$. In the definition of $X(3)$ 
we can  replace $F^2_\pi M^2_\pi$ by its bare 
expansion (\ref{expgold}) and  investigate the behaviour
of $X(3)$ in limits of small quark masses. 
First of all, in the $SU(2) \times SU(2)$ chiral limit one obtains
\begin{equation}\label{mzero}
\lim_{m \to 0} X(3) = \frac{\Sigma(3)}{\Sigma(2)}
\qquad (m_s \textrm{ fixed}),
\end{equation}
from the definition of the two-flavour condensate
$\Sigma(2) \equiv \lim_{m \to 0} F^2_{\pi} M^2_{\pi}/(2 m)$.  On the other
hand, if both $m$ and $m_s$ tend to zero, we obtain by definition:
\begin{equation}\label{mszero}
\lim_{m,m_s \to 0} X(3) = 1\,.
\end{equation}
Consequently, as long as the three-flavour condensate $\Sigma(3)$ is
suppressed with respect to the two-flavour condensate $\Sigma(2)$
($\Sigma(3) \sim \Sigma(2)/2$ is suggested by the sum rule analysis in
Refs.~\cite{Bachir1,Bachir2,D}), 
 the physical value of $X(3)$ cannot
be simultaneously close to its limiting values in both
$SU(2) \times SU(2)$, Eq.~(\ref{mzero}), and $SU(3) \times SU(3)$, 
Eq.~(\ref{mszero}). 

We expect the limit ``$m\to 0$, $m_s$ physical'' to 
be a good approximation of the physical situation  of very light 
$u$ and $d$ quarks, and thus the limiting value expressed by 
Eq.~(\ref{mzero}) to be quite close to the physical
$X(3)$. On the other hand, the variation of $X(3)$
between $m_s = 0$ and $m_s$ physical may be substantial, due
to important fluctuations of the lowest modes of the (Euclidean) Dirac
operator~\cite{dirac}, which correspond to a significant Zweig rule-violating correlation 
between massless non-strange and massive strange vacuum pairs~\cite{DGS}. The 
latter contribute to the two-flavour condensate by the amount $m_s Z^s$.
(We see from Eq.~(\ref{expgold}) that $\Sigma(2)=\Sigma(3)+m_s Z^s+\ldots$.)
As long as $\Sigma(3)$ is comparable to (or smaller than) 
the ``induced condensate''
$m_s Z^s$, the hierarchy condition between LO and NLO Eq.~(\ref{LNL}) 
will be violated for the bare expansion Eq.~(\ref{expgold}),
in spite of a good global convergence $d_{\pi} \ll 1$.
The inverted expansion of $X(3)$, in which $F_\pi^2M_\pi^2$ is replaced by 
its expansion Eq.~(\ref{expgold}):
\begin{equation}\label{x}
X(3) =\frac{1}{\displaystyle 1+\frac{m_s Z^s}{\Sigma(3)}+\ldots }
   = 1 + \sum_{P} c^X_P M^2_P + \ldots\,,
\end{equation}
would then be invalidated by large coefficients, even for quite modest 
(and realistic) values of $m_s$ at 1 GeV around 150 MeV~\cite{ms}. 
In such a case, it is not a good idea to
replace in higher orders of the bare expansion  $X(3)$ by 1, 
$2m B_0$ by $M^2_{\pi}$, $F^2_0$ by $F^2_\pi$, etc.

A comment is in order before we describe in detail the nonperturbative
alternative to eliminating order by order condensate parameters
and quark masses in bare $\chi$PT expansions.
The previous example of $X(3)$ can be stated as a failure of the bare
expansion of $1/F^2_{\pi} M^2_{\pi}$. Let us remark that
this is perfectly compatible with our assumption that the bare
expansion of the QCD two-point function of
axial-current divergences (i.e., $F^2_{\pi} M^2_{\pi}$) converges globally.
Indeed, consider
a generic observable $A$ with its bare expansion Eq.~(\ref{LNL}).
The latter unambiguously defines the coefficients of the bare 
expansion of $1/A$:
\begin{equation}\label{inverseA}
B = B_{\rm LO} + B_{\rm NLO} + B\,\delta\! B\,,   \qquad  B \equiv \frac{1}{A}\,,
\end{equation}
in terms of those of $A$:
\begin{eqnarray}
B_{\rm LO} &=& \frac{1}{A_{\rm LO}}\,,\qquad B_{\rm NLO} = - \frac{A_{\rm NLO}}{A^2_{\rm LO}}\,, \\
\label{deltaB}
\delta\!B &= & 
  \frac{(1 - X_A)^2}{X^2_A} - \frac{\delta \!A}{X^2_A}\,,
\qquad\qquad  X_A \equiv \frac{A_{\rm LO}}{A}\,.
\end{eqnarray}
One observes that the global convergence of $A$ does not necessarily imply
the convergence of $B=1/A$ . If the expansion of $A$ is only conditionally
convergent, i.e., if the relative leading-order contribution $X_A$
is not close to 1, then $\delta(1/A)$ need not be small -- even in the
extreme case $\delta\!A = 0$. This explains the origin of instabilities
and large coefficients in the inverted expansion such as (\ref{B}),
(\ref{F}) or (\ref{x}). At the same time, it motivates the restriction
to a \emph{linear space}  of ``good observables'' for which $\delta \!A \ll 1$.
The latter is assumed to be represented by connected QCD correlators,
and a priori excludes nonlinear functions of them such as ratios.

\section{Constraints from Goldstone boson masses and decay constants}
  \label{secGBspec}

                             The conditional convergence of $F^2_P M^2_P$
 and/or of $F^2_P$ does not by itself bar experimental determination of
three-flavour order parameters $X(3)$ and 
 $Z(3)$. It just may prevent the use of perturbation theory in relating
 them to observable quantities. In this section a systematic
 nonperturbative alternative is considered in detail. 

 The starting point is the standard bare expansion of 
$F^2_P M^2_P$ and $F^2_P$ for
 $P = \pi , K , \eta$. (See Eq.~(\ref{expgold}) for the pion case.)
 As discussed above, these particular combinations of masses and decay
 constants are expected to converge well, since they are linearly
 related to two-point functions of axial/vector currents and of their
 divergences taken at vanishing momentum transfer. As long as
 $Z(3) = F^2_0/F^2_{\pi} \sim 1$, the convergence of
 $M^2_P$ would be as good as for $F^2_P M^2_P$. If however $F^2_P$  
 is only conditionally convergent (i.e., $Z(3)$ significantly smaller than 1),
 the expansion of $M^2_P$ could become unstable,
 in  contrast with that of $F^2_P M^2_P$; the perturbative expansion of $1/F^2_P$
 would then exhibit very poor convergence. 
 We have no prejudice in this respect: the size 
 of $Z(3)$ as well as that of $X(3)$ remain an open problem
 until they are inferred from the data. 

 The expansion of $F^2_P M^2_P$ and $F^2_P$ can be written in the generic 
 form (\ref{LNLR}),
 denoting the corresponding NNLO remainders by $d_P$ and $e_P$ respectively.
 The LO of $F^2_P M^2_P$ is given by the condensate $\Sigma(3)$ and the
 NLO contribution is fully determined by the standard $O(p^4)$ LEC's
 $L_6(\mu) , L_8(\mu)$ (and $L_7$ in the case of the $\eta$ meson). 
 Similarly, at LO
 $F^2_P$ coincides with the order parameter $F^2_0 = F^2(3)$ and the NLO
 contribution is given in terms of $L_5(\mu)$ and $L_4(\mu)$. All necessary
 formulae can be found in  Ref.~\cite{ordfluc}. Here, we  
 follow the notation of the above reference  and for the 
 reader's convenience the bare expansions of $F^2_P M^2_P$
 and $F^2_P$ are reproduced in App.~\ref{appident}.

\subsection{Pions and kaons} \label{secpika}

        For $P = \pi , K$ the mass and decay-constant
identities (Ward identities) consist of four equations that
involve $X(3) , Z(3) , r= m_s/m , L_6, L_8 , L_4, L_5$ and  four NNLO
remainders $d_\pi , d_K, e_\pi ,e_K$. 
These identities -- given in App.~\ref{appident} -- are \emph{exact} 
as long as the remainders $d_P, e_P$ are
maintained in the formulae and no expansion is performed. 

As explained in Refs.~\cite{DGS,ordfluc}, we can combine
the mass and decay constant identities (recalled in App.~\ref{appident})
to obtain two relations between the order parameters $X,Y,Z$ and
the fluctuation parameters $\rho$ and $\lambda$:
\beq \label{ordandfluc}
X(3)=1-\epsilon(r)-[Y(3)]^2\rho /4-d,\qquad
Z(3)=1-\eta(r)-Y(3)\lambda /4-e.
\eeq
where the functions of the quark mass ratio are\footnote{In this paper,
we take the following values for the Goldstone boson masses and decay
constants: $M_\pi=139.6$ MeV, $M_K=493.7$ MeV, $M_\eta=547$ MeV, 
$F_\pi=92.4$ MeV and $F_K/F_\pi=1.22$.}:
\begin{equation} \label{funcr}
\epsilon(r) = 2\frac{r_2-r}{r^2-1}, \quad
r_2= 2\left(\frac{F_KM_K}{F_{\pi}M_{\pi}}\right)^2 -1\sim 36\,,\qquad
\eta(r)=\frac{2}{r-1}\left(\frac{F_K^2}{F_\pi^2}-1\right)\,,
\end{equation}
and the following linear combinations of NNLO remainders arise:
\begin{eqnarray} \label{remainmass}
d  &=& \frac{r+1}{r-1}d_{\pi} - 
   \left(\epsilon(r)+\frac{2}{r-1}\right)d_K\,,\\
e &=&\frac{r+1}{r-1}e_{\pi}-
 \left(\eta(r)+\frac{2}{r-1}\right)e_{K}\,.
\end{eqnarray}

The LEC's $L_6$ and $L_4$ enter the discussion through the combinations:
\begin{equation}\label{lambdarho}
\lambda  = 32 \frac{M_\pi^2}{F_\pi^2} (r + 2) \Delta L_4\,,\qquad
\rho  = 64 \frac{M_\pi^2}{F_\pi^2} (r + 2) \Delta L_6\,,
\end{equation}
where the scale-independent differences
$\Delta L_i=L_i^r(\mu)-L_i^{\rm crit}(\mu)$ involve the critical
values of the LEC's defined as:
\begin{eqnarray}
L_4^{\rm crit}(\mu)
          &=& \frac{1}{256\pi^2}
	  \log\frac{M_K^2}{\mu^2}-\frac{1}{128 \pi^2} \frac{r}{(r-1)(r+2)} \left\{ \log \frac{M_K^2}{M_\pi^2} \right. \nonumber\\
 &&\left.\quad +\left(1 + \frac{1}{2 r} \right) \log
          \frac{M_\eta^2}{M_\pi^2}  \right\} 	 \label{critical4} \,,\\
L_6^{\rm crit}(\mu) &=& \frac{1}{512\pi^2}\left(
       \log\frac{M_K^2}{\mu^2} 
       + \frac{2}{9}\log \frac{M_\eta^2}{\mu^2}
         \right) \nonumber \\
 && \quad    -\frac{1}{512 \pi^2}\frac{r}{(r+2)(r-1)} \left( 3 \log
          \frac{M_K^2}{M_\pi^2}  + \log \frac{M_\eta^2}{M_K^2} \right) \,.\label{critical6}
\end{eqnarray}
The critical values of $L_4$ and $L_6$ are
only mildly dependent on $r$; for $r=25$, 
\begin{equation}\label{eq:r25}
L_6^{\rm crit}(M_\rho)=-0.26\cdot 10^{-3} ,
\quad  L_4^{\rm crit}(M_\rho)=-0.51\cdot 10^{-3}\, \qquad[r=25].
\end{equation}

The remaining two equations of the $\pi - K$ system can be reexpressed
as a relation between $\epsilon(r)$ and $L_8$ on one hand and
between $\eta(r)$ and $L_5$ on the other hand:
 \begin{eqnarray}\label{epsilon}
 \epsilon(r) &=& 16 \frac{M^2_{\pi}}{F^2_{\pi}} [Y(3)]^2 \Delta L_8 - d'\,,\\
 \label{eta}
 \eta(r) &=& 8 \frac{M^2_{\pi}}{F^2_{\pi}} Y(3) \Delta L_5 - e'\,.
 \end{eqnarray}
These relations involve the combinations of the NNLO remainders
$d' =d -d_\pi$ and $e' =e -e_{\pi}$. At large values of $r$ ($\geq 15$),
$\epsilon(r)$ and $\eta(r)$ are suppressed and $d' \ll d  \sim d_{\pi}$, 
$e' \ll e \sim e_{\pi}$.

The LEC's arise in Eqs.~(\ref{epsilon}) and (\ref{eta}) through the differences:
\begin{eqnarray}\label{eleight}
 \Delta L_8 &=& L_8^r(\mu) - \frac{1}{512\pi^2}
    \left[\log\frac{M_K^2}{\mu^2}+\frac{2}{3}\log\frac{M_\eta^2}{\mu^2}\right]
 \nonumber \\
&& \quad    -\frac{1}{512 \pi^2 (r-1)} \left( 3 \log
          \frac{M_K^2}{M_\pi^2}  + \log \frac{M_\eta^2}{M_K^2} \right) \,,   
 \\
 \label{elfive}
 \Delta L_5 &=& L_5^r(\mu) - \frac{1}{256\pi^2}
    \left[\log\frac{M_K^2}{\mu^2}+2\log\frac{M_\eta^2}{\mu^2}\right]
    \nonumber \\
&&\quad   -\frac{1}{256\pi^2(r-1)}
    \left(3\log\frac{M_\eta^2}{M_K^2}+5\log\frac{M_K^2}{M_\pi^2}\right)\,.
 \end{eqnarray}
These differences combine the (renormalized and quark-mass independent) 
constants $L_8 , L_5$ and chiral logarithms so that they are
independent of the renormalisation scale $\mu$. For $r=25$,
we obtain
\begin{equation}
\Delta L_5=L_5^r(M_\rho)+0.67\cdot 10^{-3}\,,\quad
\Delta L_8=L_8^r(M_\rho)+0.20\cdot 10^{-3}\qquad [r=25]\,.
\end{equation}

 \subsection{Perturbative reexpression of order parameters} \label{secperturb}
 
                         The four  exact equations Eqs.~(\ref{ordandfluc})
 and (\ref{epsilon})-(\ref{eta})
 can be used to illustrate explicitly the instabilities which may arise
 in the perturbative expression of $X(3)$ and $Z(3)$ in
 powers of $M^2_P$.  In the perturbative treatment of 
 three-flavour $\chi$PT~\cite{GL2},  one uses the fact that
 $Y(3) = 1 + O(M^2_P)$ to set systematically $Y(3) = 1$
 whenever it appears in the NLO term. One first uses Eqs ~(\ref{epsilon}) and
 (\ref{eta}) to eliminate $F^2_K/F^2_{\pi}$ and $r=m_s/m$ in terms of
 $\Delta L_5$ and $\Delta L_8$. The result reads:
 \begin{eqnarray}\label{FP}
 \frac{F^2_K}{F^2_{\pi}} &=& 1 +  
    8 \frac{M^2_K-M^2_{\pi}}{F^2_{\pi}}\Delta L_5+ \ldots \\
 \label{rP}
 r + 1 &=& 2 \frac{M^2_K}{M^2_{\pi}} 
   \left( 1 + 8 \frac{M^2_K-M^2_{\pi}}{F^2_{\pi}}
       [\Delta L_5 - 2 \Delta L_8] +\ldots \right)\,.
  \end{eqnarray}
 In these formulae the quark mass ratio $r$ appearing in the expressions for
 $\Delta L_i$ has to be replaced by its leading order value (obtained
 from Eq.~(\ref{rP})):
 \begin{equation}\label{r0}
 r_0 = 2 \frac{M^2_K}{M^2_{\pi}} - 1\sim 24\,.
 \end{equation}
 Strictly speaking, Eqs.~(\ref{FP}) and (\ref{rP}) do not get any direct
 contribution from the vacuum fluctuation of $\bar ss$ pairs
 which violate the Zweig rule and are tracked
 by $L_6$ and $L_4$. The situation is quite different
 in the case of the identities (\ref{ordandfluc})
 for $X(3)$ and $Z(3)$, where the terms describing fluctuations are
 potentially dangerous. Expressing them perturbatively one gets:
 \begin{eqnarray}\label{XP}
 X(3) &=& 1 - 16 \frac{M^2_{\pi}}{F^2_{\pi}} \Delta L_8 -    
                   16\frac{2M^2_K+M^2_{\pi}}{F^2_{\pi}} \Delta L_6 +\ldots \\
  \label{ZP}
 Z(3)&=& 1 - 8 \frac{M^2_{\pi}}{F^2_{\pi}} \Delta L_5 -
               8 \frac{2M^2_K+M^2_{\pi}}{F^2_{\pi}} \Delta L_4 +\ldots
  \end{eqnarray}
  The large coefficients characteristic of the perturbative treatment
  of $1/(F^2_{\pi} M^2_{\pi})$ -- and to some extent also of $1/F^2_{\pi}$ --
  now become visible. Eqs.~(\ref{XP}) and (\ref{ZP}) lead numerically to:
  \begin{eqnarray}
  X(3) &=& 1 - 37 \Delta L_8 - 950 \Delta L_6 + \ldots \\
  Z(3) &=& 1 -  18 \Delta L_5  - 475 \Delta L_4 + \ldots \label{z3pert}
  \end{eqnarray}

In Eqs.~(\ref{XP}) and (\ref{ZP}), the main NLO contribution
comes from the $M_K^2$-enhanced term proportional to 
the $O(p^4)$ Zweig-rule violating LEC's $L_6$ and $L_4$. If the latter
stay close to their critical values (corresponding to Eq.~(\ref{eq:r25}) 
for $r=25$), the NLO contributions  
remain small. On the other hand, even though the values of $L_4$ and
$L_6$ are unknown yet, dispersive relations have been used to 
constrain their values: $L_6^r (M_\rho) = (0.6\pm 0.2) \cdot 10^{-3}$ 
based on sum rules for the correlator
$\langle(\bar{u}u + \bar{d}d) \bar{s}s\rangle $~\cite{Bachir1,Bachir2,D}, and
 $L_4^r(M_\rho) = (0.2\pm 0.3)\cdot 10^{-3}$ 
from $\pi K$ scattering data~\cite{Kpi}. Such estimates -- rather different
from the critical values -- suggest a significant
violation of the Zweig rule in the scalar sector, and an important role
for  the vacuum fluctuations of $s\bar{s}$ pairs in the patterns of chiral
symmetry breaking. 

As an exercise, for illustrative purposes, we will now use 
the central values of the
above sum-rule estimates to study the convergence
of Eqs.~(\ref{XP})-(\ref{ZP}). 
We actually aim in this paper at 
providing a framework to determine more accurately
the size of vacuum fluctuations directly from experimental observables.
If we take $L_5^r(M_\rho)=1.4\cdot 10^{-3}$
and $L_8^r(M_\rho)=0.9\cdot 10^{-3}$~\cite{GL2,BEG}, 
the numerical evaluation of Eqs.~(\ref{XP})-(\ref{ZP})
leads to the decomposition:
\begin{equation}
\begin{array}{rccccccccl}
{\rm Qty} &=& {\rm LO} &+& [ {\rm fluct} & +&  {\rm other}] & + & {\rm NNLO}\\
\\
\displaystyle X(3)\equiv \frac{2m\Sigma(3)}{F_\pi^2M_\pi^2} 
   &=& 1 &-& [0.82 &+& 0.04] & + &O(p^4)\\
\\
\displaystyle  Z(3)\equiv\frac{F(3)^2}{F_\pi^2} &=& 1 &-& [0.34&+& 0.04] & +& O(p^4)\\
\\
\displaystyle (r+1)\frac{M_\pi^2}{2M_K^2} &=& 1 &-& [0.00&+& 0.06] & +& O(p^4) \,.\\
\end{array}
\end{equation}
For each quantity, the right-hand side is the sum of the leading term (1), the 
NLO term (the sum in brackets) and higher-order terms. The NLO term is 
decomposed into its two contributions: the first one comes from the 
fluctuation term (proportional to $\Delta L_4$ or $\Delta L_6$)
and the second one collects all other NLO contributions. Fluctuation parameters
have a dramatic effect on the convergence -- they are the only terms
enhanced by a factor of $M_K^2$ in Eqs.~(\ref{XP}) and (\ref{ZP}).

It should be stressed that the instability of the perturbative
expansion of $X(3)$ and $Z(3)$ does not originate from higher order
terms in the expansion (\ref{expgold}) of $F_\pi^2 M_\pi^2$.  The latter
actually factorize, and can be cleanly separated from the effect of
vacuum fluctuations. This can most easily be established in the
$SU(2) \times SU(2)$ limit (\ref{mzero}). Using a bar to indicate that a
quantity is evaluated for $m=0$ and fixed $m_s \neq 0$, one has
\begin{equation}
\Sigma(2) = \lim_{m\rightarrow 0} {F_\pi^2 M_\pi^2 \over 2 m} =
\Sigma(3) + 32 B_0^2 m_s \overline{\Delta L}_6 + \Sigma(2)
\bar{d}_\pi \ , 
\end{equation}
where
\begin{equation}
\overline{\Delta L}_6 = L_6^r (\mu) - {1 \over 512 \pi^2} \bigg(\log
{\bar{M}_K^2 \over \mu^2} + {2 \over 9} \log {\bar{M}_\eta^2
\over
\mu^2} \bigg) \ .
\end{equation}
Consequently, 
\begin{equation}\label{X3bar}
\bar{X}(3) = {\Sigma(3) \over \Sigma(2)} = {1 \over \displaystyle 1 + 32 {m_s
B_0 \over F_0^2} \overline{\Delta L}_6} (1 -\bar{d}_\pi) \ .
\end{equation}
We expect the effect of nonzero $m$ to be tiny; in particular,
$\overline{\Delta L}_6 \simeq \Delta L_6$, $d_\pi \simeq
\bar{d}_\pi =
\bar{d} = O(m_s^2)$ and $\bar{X}(3) \simeq X(3)$.  Similarly,
using the expansion of decay constants displayed in Appendix A, one
gets
\begin{equation}\label{Z3bar}
\bar{Z}(3) = {F(3)^2 \over F(2)^2} = {1 \over \displaystyle 1 + 16 {m_s B_0
\over F_0^2} \overline{\Delta L}_4} (1 -\bar{e}_\pi) \ ,
\end{equation}
where $\bar{e}_\pi = \bar{e} = O(m_s^2)$ and:
\begin{equation}
\overline{\Delta L}_4 = L_4^r (\mu) - {1 \over 256 \pi^2} \log
{\bar{M}_K^2 \over \mu^2} \ .
\end{equation}
Eqs.~(\ref{X3bar}) and (\ref{Z3bar}) are exact identities, in which
the whole effect of higher orders is gathered into the $O(m_s^2)$ NNLO
remainders $\bar{d}_\pi$ and $\bar{e}_\pi$.  Even if the
latter are small ($\sim$ 10\%), the expansion of $\bar{X}(3)$ and
$\bar{Z}(3)$ in powers of $m_s$ may break down, provided the
magnitudes of $\Delta L_6$ and $\Delta L_4$ are as mentioned above.
The $SU(2) \times SU(2)$ limit of Eqs.~(\ref{XP}) and (\ref{ZP})
just exhibits the first term of such an expansion. Let us stress
that we chose the $SU(2) \times SU(2)$ limit for simplicity here,
but that the factorisation of higher order corrections is a general
result (holding even for $m\neq 0$)~\cite{ordfluc}.

One more remark is in order, concerning the special case in which
both $\overline{\Delta L}_4$ and $\overline{\Delta L}_6$ are large,
but satisfy:
\begin{equation}
\overline{\Delta L}_4 = 2 \overline{\Delta L}_6 \ .
\end{equation}
With this particular relation between the low-energy constants, one
has:
\begin{equation}
\bar{Y}(3) = {\bar{X}(3) \over \bar{Z}(3)} = \lim_{m \rightarrow 0} {2
m B_0 \over M_\pi^2} = {1 - \bar{d}_\pi \over 1 - \bar{e}_\pi} \approx
1  \ .
\end{equation}
This would describe a situation in which the (large) vacuum
fluctuations suppress both the condensate $\Sigma(3)$ and the decay
constant $F_0^2$, i.e., partially restore the chiral symmetry, and yet
the ratio $\Sigma(3)/F_0^2 = B_0$ remains nonzero.

\subsection{Nonperturbative elimination of $O(p^4)$ LEC's}

We have just seen that the perturbative treatment of chiral series
fails if vacuum fluctuations of $\bar{q}q$ pairs are large, resulting
in instabilities in the chiral expansions. In this case, the
nonlinearities in Eq.~(\ref{ordandfluc}), relating order and fluctuation
parameters, are crucial, and we must not linearize these relations (hence
the inadequacy of a perturbative treatment).

We should therefore treat Eq.~(\ref{ordandfluc}) without performing
any approximation. Following Ref.~\cite{ordfluc}, we can exploit 
Eqs.~(\ref{ordandfluc}) to
express the chiral order parameters $X(3)$ and $Z(3)$ as functions
of the fluctuation parameters $\rho$ and $\lambda$. The ratio of order 
parameters $Y(3)$ is~\footnote{The quadratic equation for $Y(3)$ admits
two solutions, but only one of them corresponds to the physical case 
(see Ref.~\cite{ordfluc} for more detail).}:
\begin{equation} \label{y3nonpert}
Y(3)=\frac{2[1-\epsilon(r)-d]}
   {1-\eta(r)-e+\sqrt{[1-\eta(r)-e]^2+[\rho-\lambda][1-\epsilon(r)-d]}}\,.\\
\end{equation}
The nonlinear character of Eqs.~(\ref{ordandfluc}) results in
the (nonperturbative) square root. We see that the behaviour of $Y(3)$
is controlled by the fluctuation parameter $\rho-\lambda$,
i.e., $2L_6-L_4$ as can be seen from Eqs.~(\ref{lambdarho}).

The perturbative treatment sketched
in the previous section corresponds to linearising Eq.~(\ref{y3nonpert}),
assuming that the fluctuation parameter $\rho-\lambda\ll 1$. This 
is invalid if large fluctuations occur: $\rho$ and/or $\lambda$ are then
numerically
of order 1, although they count as $O(p^2)$ in the chiral limit. 
Eq.~(\ref{y3nonpert}) leads to the suppression of $Y(3)$, which
would contribute to a stabilisation of Eq.~(\ref{ordandfluc}) by reducing
the contribution proportional to the fluctuation parameters $\rho$ and 
$\lambda$. As discussed extensively in Ref.~\cite{ordfluc},
different behaviours of the fluctuation parameters can 
result in a rather varied range of patterns of chiral symmetry breaking.

We would like to extract information about $N_f=3$ chiral
symmetry breaking from physical observables, even 
in the event that the perturbative expansion breaks down.
We could proceed in the same way as in Ref.~\cite{ordfluc}
and express as many quantities as possible in terms of $L_4$ and $L_6$,
in order to stress the role played by vacuum fluctuations. 
In the present paper, we find it more
convenient to take as independent quantities 
the (more fundamental) chiral order parameters
$X(3)$ and $Z(3)$. We should emphasize that this corresponds to a different
choice from that adopted in
the perturbative treatment of chiral series:
in standard $\chi$PT, $X(3)$, $Z(3)$, $r$, $F_K/F_\pi$
are iteratively expressed in terms of $L_4, L_5, L_6, L_8$. In contrast, we start by the
same four identities and express nonperturbatively $L_4 , L_5, L_6, L_8$
in terms of $X(3)$, $Z(3)$, $r$, $F_K/F_{\pi}$; this  is a sensible
treatment provided that both LO and NLO terms are considered.

Keeping in mind that LO and NLO contributions can have a similar
size, we treat as exact identities the expansions of good observables 
in powers of quark masses,
and exploit the mass and decay constant identities to reexpress 
$O(p^2)$ and $O(p^4)$ LEC's in terms of $r$, $X(3)$, $Z(3)$,
observables quantities and NNLO remainders. This leads to:
\begin{eqnarray} \label{F0}
F_0^2&=&F_\pi^2 Z(3)\,,\\
2mB_0&=&M_\pi^2 Y(3)\,,\\
2m_s B_0&=& rM_\pi^2 Y(3)\,,
\end{eqnarray}
and to:
\begin{eqnarray} \label{L6}
[Y(3)]^2 L_6^r(\mu) &=&
 \frac{1}{16(r+2)}
   \frac{F_\pi^2}{M_\pi^2}[1-X(3)-\epsilon(r)-d] \nonumber \\
&& \quad +\frac{[Y(3)]^2}{512\pi^2}
    \left(\log\frac{M_K^2}{\mu^2}+\frac{2}{9}\log\frac{M_\eta^2}{\mu^2}\right) \nonumber \\ 
&&\quad    -\frac{[Y(3)]^2 r}{(r-1)(r+2)}\frac{1}{512 \pi^2}  \left( 3
 \log  \frac{M_K^2}{M_\pi^2}  + \log \frac{M_\eta^2}{M_K^2} \right)
\,,  \\  
{[Y(3)]}^2 L_8^r(\mu) &=& \label{L8}
   \frac{F_\pi^2}{16 M_\pi^2}[\epsilon(r)+d']   +\frac{[Y(3)]^2}{512\pi^2}
    \left(\log\frac{M_K^2}{\mu^2}+\frac{2}{3}\log\frac{M_\eta^2}{\mu^2}\right)
 \nonumber \\
&& \quad     + \frac{[Y(3)]^2}{512
 \pi^2 (r-1)}  \left( 3
 \log  \frac{M_K^2}{M_\pi^2}  + \log \frac{M_\eta^2}{M_K^2} \right)
\,, 
\\
\label{L4}
Y(3) L_4^r(\mu)&=& \frac{1}{8(r+2)} 
     \frac{F_\pi^2}{M_\pi^2}[1-Z(3)-\eta(r)-e]
     +\frac{Y(3)}{256\pi^2} \log\frac{M_K^2}{\mu^2} \nonumber \\
&&\quad
     -\frac{Y(3)}{128 \pi^2 (r+2)}\left(
    \frac{2 r + 1}{2 r - 2}\log\frac{M_\eta^2}{M_K^2}+2 \frac{4 r+1}{4
     r-4}
     \log\frac{M_K^2}{M_\pi^2}\right) 
\,, 
\\ \label{L5}
Y(3) L_5^r(\mu)&=& 
     \frac{F_\pi^2}{8 M_\pi^2}[\eta(r)+e']
     +\frac{Y(3)}{256 \pi^2} 
    \left(\log\frac{M_K^2}{\mu^2}+2\log\frac{M_\eta^2}{\mu^2}\right) \nonumber \\
&&\quad
     +\frac{Y(3)}{256 \pi^2 ( r-1)} \left(
  3 \log \frac{M_\eta^2}{M_K^2}+5\log\frac{M_K^2}{M_\pi^2}\right)
\,.
\end{eqnarray}
These equations are derived from Eqs.~(\ref{ordandfluc}), 
(\ref{elfive}), (\ref{eleight}),
and they have a much simpler expression in terms of  
$\Delta L_i$, introduced in
Eqs.~(\ref{critical4})-(\ref{critical6}) and
(\ref{eleight})-(\ref{elfive}), rather than $L_i$ ($i=4,5,6,8$):   
\begin{eqnarray}
Y^2(3)\Delta L_6 &=& \frac{1}{16(r+2)}\frac{F_\pi^2}{M_\pi^2}
  [1-\epsilon(r)-X(3)-d]\,,
\\ 
Y^2(3)\Delta L_8 &=& \frac{1}{16}\frac{F_\pi^2}{M_\pi^2}
  [\epsilon(r)+d']\,,
\\ 
Y(3)\Delta L_4 &=& \frac{1}{8(r+2)}\frac{F_\pi^2}{M_\pi^2}
  [1-\eta(r)-Z(3)-e]\,,
\\ 
Y(3)\Delta L_5 &=& \frac{1}{8}\frac{F_\pi^2}{M_\pi^2}
  [\eta(r)+e']\,.
\end{eqnarray}
The above identities are useful as long as the NNLO remainders are small.
The presence of powers of $Y(3)$, i.e., $B_0$,
follows from the normalisation of the scalar and pseudoscalar sources in
Ref.~\cite{GL2}: these powers arise only for $O(p^4)$ LEC's related
to chiral symmetry breaking (two powers for $L_6,L_7,L_8$,
one for $L_4$ and $L_5$), and are absent for LEC's associated with
purely derivative terms.

Plugging these identities into $\chi$PT expansions corresponds therefore
to a resummation of vacuum fluctuations, as opposed to the usual (iterative
and perturbative) treatment of the same chiral series. We can then reexpress
observables in terms of the three parameters of interest $X(3)$, $Z(3)$, $r$
and NNLO remainders. Before describing how to exploit experimental information
to constrain these parameters, we should first comment on the case of the $\eta$-meson.

\subsection{The $\eta$-mass and the Gell-Mann--Okubo formula} \label{secgmo}

               It remains for us to discuss the mass and decay constant of $\eta$
 as constrained by Ward identities for two-point functions of the
 eighth component of the axial current and of its divergence. This results
 into two additional relations (given in App.~\ref{appident})
 that involve one new NLO constant $L_7$  and two extra NNLO remainders 
 $d_{\eta}$ and $e_{\eta}$. These two identities will be used to
 reexpress the LEC $L_7$ in terms of order parameters and quark mass ratio,
 and to eliminate the decay-constant $F_{\eta}$,
 which is not directly accessible experimentally. This new discussion is
 closely related to the old question~\cite{GMO} 
 whether the remarkable accuracy  
 of  the Gell-Mann--Okubo (GO) formula for Goldstone 
 bosons finds a natural explanation 
 within $\chi$PT and what it says about the size 
 of the three-flavour condensate. 
 
 The combination:
 \begin{equation}\label{GMO}
 D_{GO} = 3 F^2_{\eta} M^2_{\eta}-4 F^2_K M^2_K + F^2_{\pi} M^2_{\pi}
 \end{equation}
 does not receive any $O(p^2)$ contribution from the genuine condensate
 $\Sigma(3)$.  The $\eta$-mass identity~(\ref{etamass}) leads to the following simple
 formula for $D_{GO}$, expressed in units $F^2_{\pi} M^2_{\pi}$:
 \begin{equation}\label{Lseven}
 \Delta_{GO}\equiv \frac{D_{GO}}{F^2_{\pi} M^2_{\pi}} =
 16 \frac{M^2_{\pi}}{F^2_{\pi}} (r-1)^2 [Y(3)]^2 
     (2 L_7 + \Delta L_8) + d_{GO}\,,
 \end{equation}
 where $\Delta L_8$ was defined in Eq.~(\ref{eleight}). 
 Similarly, the identity for
 $F_{\eta}$ (\ref{etadecay}) can be put into the form:
 \begin{equation}\label{Feta}
 \frac{F^2_{\eta}}{F^2_{\pi}} = 1 + \frac{4}{3} \left(\frac{F^2_K}{F^2_{\pi}} -1\right) +
 \frac{1}{48 \pi^2} \frac{M^2_{\pi}}{F^2_{\pi}} Y(3) 
 \left[(2r+1) \log \frac{M^2_{\eta}}{M^2_K} 
          - \log \frac{M^2_K}{M^2_{\pi}}\right] + e_{GO}\,.
 \end{equation}
 Eqs.~(\ref{Lseven}) and (\ref{Feta}) are exact as long as the NNLO
 remainders:
 \begin{equation}
 d_{GO}=  3 \frac{F^2_{\eta} M^2_{\eta}}{F_\pi^2M_\pi^2} d_\eta
      -4 \frac{F^2_K M^2_K}{F_\pi^2M_\pi^2} d_K + d_\pi \,,
 \qquad  e_{GO} = \frac{F_\eta^2}{F_\pi^2}e_\eta + 
       \frac{4}{3} \frac{F^2_K}{F^2_{\pi}}e_K - \frac{e_\pi}{3}
 \end{equation}
 are included. If one follows Sec.~\ref{secperturb} and 
 treats the exact formulae
 (\ref{Lseven}) and (\ref{Feta}) perturbatively, one reproduces the
 $O(p^4)$ expressions given in Ref.~\cite{GL2}, as expected.

                          Remarkably, the identities (\ref{Lseven})
and (\ref{Feta}) are simpler and more transparent than their perturbative
version, and we find them useful to make a few numerical
estimates which may be relevant for a discussion of the GO formula.
For this purpose we shall use the value  $r=r_0\sim 24$  
of the quark mass ratio and 
neglect for a moment the NNLO remainders $d_{GO}$ and $e_{GO}$, 
as well as error bars
related to the experimental inputs
on masses and decay constants.
For this exercise, we also disregard isospin breaking and 
electromagnetic corrections. First, the dependence
of $F_{\eta}$ on $Y(3)=2mB_0/M^2_{\pi}$ is negligibly small:
\begin{equation}
\frac{F_{\eta}^2}{F_{\pi}^2} = 1.651 + 0.036 \cdot Y(3)\,.
\end{equation}
In the estimate of $\Delta_{GO} = D_{GO}/F^2_{\pi} M^2_{\pi}$,
we use $F^2_{\eta}/F^2_{\pi}=1.687$ and find:
\begin{equation}\label{DeltaGO}
\Delta_{GO} = 77.70 - 74.46 + 1 = 4.24\,.
\end{equation}
We have split the result into the three contributions corresponding 
respectively to $\eta$, $K$ and $\pi$, in order to
emphasize the accuracy of the formula. If we drop the decay constants
in $\Delta_{GO}$, we obtain:
\begin{equation}
\bar \Delta_{GO} = \frac{3M^2_{\eta}- 4M^2_K + M^2_{\pi}}{M^2_{\pi}} =
                 46.06 - 50.03 + 1 = - 2.97\,.
\end{equation}
Hence, apart from a change of sign, this more familiar definition of the GO
discrepancy is of a comparable magnitude as $\Delta_{GO}$. For the reasons
already stressed, the interpretation in terms of QCD correlation functions
is more straightforward when $F^2_P M^2_P$ is used.

        If the origin of the GO formula were to be naturally explained
by the dominance of the  $O(p^2)$ condensate term
in the expansion of $F^2_P M^2_P$, the order of magnitude of the
estimate (\ref{DeltaGO}) should
be reproduced by Eq.~(\ref{Lseven}) for a typical order of magnitude of the
$O(p^4)$ LEC's $L_8$ and $L_7$ without any fine tuning of their values.
Using Eq.~(\ref{epsilon}) and neglecting the NNLO remainder $d'$, one gets:
\begin{equation} \label{L8r}
16 \frac{M^2_{\pi}}{F^2_{\pi}} (r-1)^2 [Y(3)]^2 \Delta L_8 =      
(r-1)^2 [\epsilon (r) + d'] = 22.5\qquad [r=r_0]\,.
\end{equation}
Hence, the typical $O(p^4)$ contribution $\Delta L_8$ to $\Delta_{GO}$
happens to be nearly one order of magnitude bigger than the estimate
(\ref{DeltaGO}): the latter can only be reproduced by tuning very finely
the LEC $L_7$:
\begin{equation}
[Y(3)]^2 (\Delta L_8 + 2 L_7 )  \simeq 1.3 \times 10^{-4}\,,
\end{equation}
to be compared with the above estimate
$[Y(3)]^2 \Delta L_8 \simeq 1.2 \times 10^{-3}$. 
All this of course does not reveal
any contradiction, but it invalidates the customary ``explanation'' of the
GO formula and the standard argument against a
possible suppression of the three-flavour  
condensate $\Sigma(3)$. Therefore, the fact that the GO
formula is satisfied so well remains unexplained independently of
the size of $\Sigma(3)$ and of the vacuum fluctuations. The last point
can be explicitly verified: the genuine condensate contribution $\Sigma(3)$
as well as the induced condensate $m_s Z^s$, which represents an
$O(p^4)$ contribution to $F_P^2 M_P^2$, both drop out of the GO
combination (\ref{GMO}).

We now return to our framework: we do not assume a particular 
hierarchy between LO and NLO contributions to chiral series, and 
we do not neglect any longer the  NNLO remainders 
(in the case of the $\eta$-meson, $d_{GO}$ and $e_{GO}$ might be 
sizeable and should
be kept all the way through). It is possible to use the previous formulae
to reexpress $L_7$ in a similar way to Eqs.~(\ref{L6})-(\ref{L5}):
\begin{eqnarray} \label{elseven}
[Y(3)]^2L_7 &=& \frac{1}{32(r-1)^2}\frac{F_\pi^2}{M_\pi^2}
 \Bigg[ \frac{3F_\eta^2M_\eta^2-4F_K^2M_K^2+F_\pi^2M_\pi^2}{F_\pi^2M_\pi^2}\\
&&\quad  -(r-1)^2[\epsilon(r)+d']
 -\frac{3F_\eta^2M_\eta^2d_\eta-4F_K^2M_K^2d_K+F_\pi^2M_\pi^2d_\pi}{F_\pi^2M_\pi^2}
 \Bigg]\,.\nonumber
\end{eqnarray}
This expression
should be used to reexpress nonperturbatively $L_7$ in terms
of chiral order parameters ($F_\eta^2$ is given by Eq.~(\ref{Feta})).
We can already notice that in Eq.~(\ref{elseven}), the first contribution, 
which corresponds to $\Delta_{GO}$, is 5 to 10 times suppressed with respect
to the second term $(r-1)^2[\epsilon(r)+d']$.

\section{Three-flavour analysis of $\pi\pi$ scattering}\label{secpipianalys}

The quantities $B_0 m$, $r$, $F_0$, $L_4$,...,$L_8$ appear in the bare
chiral series up to NLO.  The procedure outlined above allows us to
express these eight quantities in terms of the masses and decay
constants of Goldstone bosons. Apart from the (presumably small) six NNLO
remainders $d_P$ and $e_P$, this leaves three unknown parameters. We choose these three parameters to be the order parameters
$X(3)$ and $Z(3)$, and the quark mass ratio $r$. The remaining terms
of the $O(p^4)$ chiral Lagrangian, involving the LEC's $L_1$, $L_2$,
$L_3$, $L_9$, and $L_{10}$, do not affect the symmetry breaking sector
of the underlying theory in which we are mainly interested here.

The counting of the degrees of freedom is completely analogous to the
one described in Ref.~\cite{ABT}, where a global two-loop fit has been
performed to the masses, decay constants and $K_{e4}$ form factors. In
this reference, the LEC's $L_1$, $L_2$ and $L_3$ were also included
and constrained by the three experimental results corresponding to the
two $K_{e4}$ form factors at threshold and to their slope.
The three parameters left undetermined in the fits of Ref.~\cite{ABT}
are the ratio of quark masses $r$ and the $O(p^4)$ LEC's $L_4$ and $L_6$.
The unknown $O(p^6)$ LEC's, estimated in the above reference through
a resonance saturation assumption, introduce some theoretical
uncertainty.  In our approach, this uncertainty is included in the NNLO
remainders. We keep the latter throughout our calculation; however, at
some point of the numerical analysis we will have to make an educated
guess as to their sizes. The main differences between our
approach and that of Ref.~\cite{ABT} 
lie in our use of a nonperturbative resummation of
instabilities, compared to the canonical two-loop perturbative
elimination of $O(p^4)$ LEC's; in  our choice of $X(3),Z(3),r$ as
the three undetermined parameters, rather than $L_4,L_6,r$; and in
our treatment of higher-order remainders: in Ref.~\cite{ABT} 
they are computed up to $O(p^6)$, and the additional LEC's 
arising at this order are estimated through resonance saturation.

In order to constrain our three independent parameters, more
information is needed. In the present section we will examine the
impact of our knowledge of $\pi\pi$ scattering observables.  We have
previously analysed the $\pi\pi$ data of Ref.~\cite{E865} in terms of
two-flavour order parameters, allowing a rather precise determination
of them~\cite{DFGS}: $X(2)=0.81\pm 0.07$ and $Z(2)=0.89 \pm 0.03$.
However, the
$\pi\pi$ scattering parameters are more sensitive to the two-flavour
order parameters than to the three-flavour ones~\cite{DS,DGS,G}.
Expanding $X(2)$ in $SU(3)\times SU(3)$ $\chi$PT one can
obtain~\cite{DGS}:
\begin{equation}
X(2) (1 - \bar d_\pi ) = \frac{r}{r+2} \left[ 1 -
\epsilon(r) - d - Y(3)^2 f_1 \right] + \frac{2}{r+2}X(3)\,,  \label{eq:x2bis}
\end{equation}
where $f_1$ is a (small) combination of chiral 
logarithms, whose precise definition is recalled in Eq.~(\ref{eq:f1})
below; in Ref.~\cite{DGS} the estimate $f_1\sim 0.05$ was obtained.

If $\epsilon(r)$ is not close to 1, i.e., if $r$ larger than 15, 
the term in square brackets in Eq.~(\ref{eq:x2bis}) is dominant. Then
$X(2)$ has only a very weak ($r$-suppressed) sensitivity to $X(3)$. On
the other hand, its value is strongly correlated with $r$. We expect
therefore $\pi\pi$ scattering to provide us with valuable information
about the quark mass ratio $r$, but not about the $N_f=3$ order
parameters $X(3)$ and $Z(3)$. This section is devoted to designing
a framework testing this expectation in a quantitative way.

\subsection{Low-energy $\pi\pi$ amplitude} \label{seclowpipi}

Considerable progress has been achieved recently in the understanding
of $\pi\pi$ scattering. The solutions of the Roy equations~\cite{ACGL}
allow one to express the amplitude, in the whole energy domain below
800~MeV, in terms of only two parameters (e.g., the scalar scattering
lengths, or the parameters $\alpha_{\pi\pi}$ and $\beta_{\pi\pi}$
defined below), with very small uncertainty. It is therefore possible
to determine experimentally these two parameters in a model-independent
way.  Furthermore, at low energy the $\pi\pi$ amplitude is strongly
constrained by chiral symmetry, crossing and unitarity. It can be 
expressed, up to and including terms of order $(p/\Lambda_H)^6$, as:
\begin{equation} \label{amplpipidisp}
A_{\pi\pi}(s|t,u) = P(s|t,u)  + \bar J(s|t,u) + O[(p/\Lambda_H)^8]
\end{equation}
where $P(s|t,u)$ is a polynomial conveniently written (following the
conventions and notation of Ref.~\cite{KMSF}) in the form: 
\begin{eqnarray}
P(s|t,u) &=& \frac{\alpha_{\pi\pi}}{F_{\pi}^2}
\frac{M_\pi^2}{3} + \frac{\beta_{\pi\pi}}{F_\pi^2} \left( s - \frac{4
M_\pi^2}{3} \right) \nonumber \\
&& + \frac{\lambda_1}{F_\pi^4} \left( s - 2 M_\pi^2 \right)^2 +
\frac{\lambda_2}{F_\pi^4} \left[ \left( t - 2 M_\pi^2 \right)^2 +
\left( u - 2 M_\pi^2 \right)^2 \right] \nonumber \\
&& + \frac{\lambda_3}{F_\pi^6} \left( s - 2 M_\pi^2 \right)^3 +
\frac{\lambda_4}{F_\pi^6} \left[ \left( t - 2 M_\pi^2 \right)^3 +
\left( u - 2 M_\pi^2 \right)^3 \right]\,,
\label{polpipidisp}
\end{eqnarray}
in terms of six subthreshold parameters
$\alpha_{\pi\pi} ,\beta_{\pi\pi},\lambda_1 \ldots \lambda_4$. $\bar J(s|t,u)$
collects the unitarity cuts arising from elastic $\pi\pi$ intermediate states.
At low energy, the contributions of $K\bar{K}$ and 
$\eta\eta$ intermediate states are not neglected but 
expanded and absorbed into the polynomial $P(s|t,u)$.
The general form of $\bar J(s|t,u)$ is dictated by successive iterations of
the unitarity condition and it is entirely determined by the first four
subthreshold parameters up to $O(p^6)$:
\begin{eqnarray} \label{jdisp}
\bar J(s|t,u) & = &  U(s) \bar J_{\pi\pi}(s)\nonumber\\
                & & + [(s-u)V(t) + W(t)] \bar J_{\pi\pi}(t) \nonumber\\
		& & +  [(s-t)V(u) + W(u)] \bar J_{\pi\pi}(u) + \ldots\,, 
\end{eqnarray}
where
\begin{equation} 
\bar J_{\pi\pi}(s) = \frac{s}{16 \pi^2}
  \int ^\infty_{4M^2_\pi} \frac{dx}{x(x-s)}\sqrt{\frac{x-4M^2_\pi}{x}}\,,
\end{equation}
 $U ,V, W$ are polynomials given in terms of the parameters
 $\alpha_{\pi\pi}, \beta_{\pi\pi}, \lambda_1 , \lambda_2$ and the ellipsis
 stands for (known) $O(p^6)$ contributions. (For a more explicit form, see
 Eqs.~(3.18) and (3.47) of Ref.~\cite{KMSF}.)

    As a first step, the general representation (\ref{amplpipidisp})
 of the low-energy $\pi\pi$ amplitude can be matched with experimental
 phase shifts~\cite{E865} and with the solution of Roy 
 equations~\cite{ACGL}
 in order to determine the subthreshold parameters
$\alpha_{\pi\pi},\beta_{\pi\pi},\ldots $.
 This has been done in Ref.~\cite{DFGS}, independently of any $\chi$PT 
 expansion or predictions, leading to the following values:
 \begin{equation} \label{eq:alphabetaexp}
 \alpha_{\pi\pi} = 1.381 \pm 0.242\,, \qquad 
 \beta_{\pi\pi} = 1.081 \pm 0.023\,,
 \end{equation}
 with the correlation coefficient between the two parameters
 $\rho_{\alpha\beta} = - 0.14$. 

  At the  second stage, $\chi$PT can be used to expand the
 subthreshold parameters $\alpha_{\pi\pi}, \beta_{\pi\pi},
 \lambda_1, \ldots \lambda_4$ in powers of quark masses
 $m_u = m_d = m$ and/or $m_s$, thereby constraining the possible values of
 chiral order parameters. Notice that the expansion of the subthreshold
 parameters is expected to converge better than that of 
 scattering lengths, the latter being more sensitive to small 
 variations of the pion mass. Similarly, as already discussed in 
 Secs.~\ref{secconvinst} and \ref{secGBspec}, one should bear in mind 
 the possibility of a strong dependence of $F^2_\pi$ on $m_s$: it seems
 therefore preferable to consider the expansion of $F^4_\pi A_{\pi\pi}$, i.e., 
 the on-shell four-point function of the axial-vector current
 (rather than the scattering amplitude itself) in powers of quark 
 masses and external pion momenta. In Ref.~\cite{DFGS} the corresponding
 expansion of subthreshold parameters $\alpha_{\pi\pi}$ and 
 $\beta_{\pi\pi}$ in powers of $m$
 (with $m_s$ fixed at its physical value) was converted into a determination
 of the two-flavour order parameters $X(2)$ and $Z(2)$. 
 Now we consider the ``bare expansion'' of 
 $F^2_{\pi}M^2_{\pi}\alpha_{\pi\pi}$ and of the slope parameter 
 $F^2_{\pi}\beta_{\pi\pi}$ in powers of $m$ and $m_s$
 in order to investigate directly how tightly the available
 $\pi\pi$ experimental information constrains the three-flavour        
 condensate $X(3)$ and decay constant $F_0$ (or $Z(3)$), and the 
 quark mass ratio $r=m_s/m$.
                
 In order to establish the  ``bare'' $SU(3) \times SU(3)$
 expansion  of $F^2_\pi M^2_\pi \alpha_{\pi\pi}$ and of $F^2_\pi\beta_{\pi\pi}$
 we proceed as follows. We start by 
 rewriting LO and NLO $\chi$PT contributions in a form similar to
 Eq.~(\ref{amplpipidisp}):
\begin{equation}
F^4_\pi A_{\pi\pi}(s|t,u) = F^4_\pi P^r(s|t,u) + F^4_\pi J^r(s|t,u) + \ldots \,.
\end{equation}
$P^r(s|t,u)$ collects all LO and NLO tree and tadpole contributions
and is of the form (\ref{polpipidisp}) with the two cubic terms omitted. 
The second term collects the one-loop contribution of order 
$O(p^4)$. Both are renormalized and separately depend 
on the renormalisation scale $\mu$.
The loop part reads:
\begin{eqnarray} \label{jchiral}
F^4_\pi J^r(s|t,u) & = & s^2 \left[ \frac{1}{2} J^r_{\pi\pi}(s) 
                                   +\frac{1}{8} J^r_{KK}(s) \right] 
 \nonumber\\
                  && + M^4_\pi \left[- \frac{1}{2} J^r_{\pi\pi}(s) 
                    + \frac{1}{18} J^r_{\eta\eta}(s)\right] \nonumber\\
		  && + \frac{1}{4} [(t-2M^2_\pi)^2 J^r_{\pi\pi}(t) 
		  +(u-2M^2_\pi)^2 J^r_{\pi\pi}(u)] \nonumber\\
		  && + (s-u) t \left[ M^r_{\pi\pi}(t) 
                       + \frac{1}{2} M^r_{KK}(t) \right] 
  \nonumber\\
		  && + (s-t) u \left[ M^r_{\pi\pi}(u) + 
                         \frac{1}{2} M^r_{KK}(u) \right]\,.
\end{eqnarray}
Here $J^r_{PP}$ and $M^r_{PP}$ are the standard loop functions 
for the Goldstone boson $P$ (see, e.g., ref.~\cite{KMSF}). 
They are related to the functions $\bar J_{PP}(s)$ through:
\begin{eqnarray}
J^r_{PP}(s) & = &  \bar J_{PP}(s) - 2 k_{PP}\,, \\
M^r_{PP}(s) & = & \frac{s-4M^2_P}{12s} \bar J_{PP}(s)-
  \frac{1}{6}k_{PP}+\frac{1}{288\pi^2}\,.
\end{eqnarray}
At low energy and for $P= K,\eta$, the loop functions are replaced by
their expansion at small $s$:
\begin{equation}
J^r_{PP} = - 2 k_{PP}\,,  \quad M^r_{PP} = - \frac{1}{6} k_{PP}\,.
\end{equation}
In these equations, we have $k_{PP} = [\log(M^2_P/\mu^2) + 1]/32\pi^2 $.
                         
Multiplying Eq.~(\ref{jdisp}) by $F^4_\pi$ and dropping
all terms beyond $O(p^4)$, one should recover the formula 
(\ref{jchiral}). This fact can be used to work out 
the bare expansion of the subthreshold
parameters contained in $F^4_\pi P(s|t,u)$. Comparing (\ref{jdisp}) 
and (\ref{jchiral}) leads to:
\begin{eqnarray}
F^4_\pi P(s|t,u) & = & F^4_\pi P^r(s|t,u) 
                  - s^2 \left( k_{\pi\pi} + \frac{1}{4} k_{KK}\right) 
 \nonumber\\
                 & & + M^4_{\pi} 
  \left( k_{\pi\pi} - \frac{1}{9} k_{\eta\eta} \right) 
 \nonumber\\
		 & & - \frac{1}{2}[
              (t-2M^2_\pi)^2 + (u-2M^2_\pi)^2 ] k_{\pi\pi} 
 \nonumber\\
		 & & - \frac{1}{6} [(s-u)t + (s-t)u] 
   \left(k_{\pi\pi}+\frac{1}{2}
                 k_{KK}-\frac{1}{48\pi^2}\right)\,,  \label{pchiral}
\end{eqnarray}
which holds for all powers of $s,t,u$ provided that 
one retains just the LO and NLO powers in quark masses in the expression
of corresponding coefficients. In agreement with the convention 
explained in Sec.~\ref{secNNLO}, we keep the physical Goldstone boson masses
in the arguments of chiral logarithms arising from both tadpoles and
unitarity loops. The right-hand side of Eq.~(\ref{pchiral}) is checked to be 
independent of renormalisation scale $\mu$. The resulting 
``bare expansion'' of the two subthreshold parameters which
carry information on the symmetry-breaking sector reads:
\begin{eqnarray}
F_\pi^2 M_\pi^2 \alpha_{\pi\pi} &=& 2m\Sigma(3) + 16 m^2 A +
2 m (8 m + m_s) Z^s \nonumber \\
&& - 8 m^2B_0 ( \xi + 2 \tilde \xi) \nonumber \\
&& + \frac{1}{8 \pi^2} m^2B_0^2 \left( 4 \log
\frac{M_K^2}{M_\pi^2} - \frac{7}{3} \right) + F_\pi^2 M_\pi^2
d_{\alpha,\pi\pi}\,, \label{eq:fpi2mpi2a}\\
F_\pi^2 \beta_{\pi\pi} &=& F^2_0 + 4 m \xi + 2( 4 m + m_s)
\tilde \xi \nonumber\\
&& + \frac{1}{4 \pi^2} mB_0 \left[ \log
\frac{M_{\eta}^2}{M_K^2} + 2 \log \frac{M_K^2}{M_\pi^2} - \frac{5}{4}
\right] + F_{\pi}^2 e_{\beta,\pi\pi}\,. \label{eq:fpi2b}
\end{eqnarray}
 The subthreshold parameters are thus expressed 
 in terms of  $A,Z^s,\xi,\tilde\xi$, 
 which are defined in App.~\ref{appident} as
  scale-invariant combinations of the $O(p^4)$
 LEC's and chiral logarithms. In order to account
 for NNLO and higher chiral orders, we have added in 
 Eqs.~(\ref{eq:fpi2mpi2a})-(\ref{eq:fpi2b}) the direct
 NNLO remainders $d_{\alpha,\pi\pi}$ and $e_{\beta,\pi\pi}$.
 (Their exact role in our analysis will be discussed shortly.) 
 We would like to stress that the only quantities we really 
 subject to a chiral expansion
 are the subthreshold parameters multiplied by appropriate powers of $F^2_\pi$
 and $M^2_\pi$. The scattering amplitude as a function
 of $s,t,u$ is given by Eq.~(\ref{amplpipidisp}), 
 which holds up to and including $O(p^6)$ accuracy
 with all singularities and threshold factors correctly described using
 the physical Goldstone boson masses. 

 We have just
 illustrated how the bare expansion of subthreshold parameters 
 is obtained in practice.
 The last step now consists in replacing in the bare
 expansion (\ref{eq:fpi2mpi2a})-(\ref{eq:fpi2b})
 the parameters of the Lagrangian $mB_0 , m_sB_0$
 $L_4(\mu) , L_5(\mu) , L_6(\mu) ,L_8(\mu)$ by the three basic QCD parameters
 $X(3), Z(3) , r= m_s/m$ using the identities
 (\ref{F0})-(\ref{L5}), which yields:
\begin{eqnarray}
\alpha_{\pi\pi} &=& 1 + 3 \frac{ r \epsilon(r)}{r+2} + 2 \frac{ 1 -
  X(3)}{r+2} + 4 \frac{1 - Y(3)}{r+2} -2 \frac{Y(3) r \eta(r)}{r+2}
  \nonumber \\ 
&& -\frac{M_\pi^2}{32 \pi^2 F_\pi^2} Y(3)^2 \left\{ \frac{7}{3} +
\frac{r}{(r-1) ( r+2)} \left[ (r+2) \log \frac{M_\eta^2}{M_K^2} - (
r-2) \log \frac{M_K^2}{M_\pi^2} \right]\right\} \nonumber \\
&& - \frac{6}{r+2} d + \frac{4 Y(3)}{r+2} e - 2 Y(3) e' + [(d_\alpha -
d) + 4 d']\,, \label{eq:alpha}\\
\beta_{\pi\pi} &=& 1 +  \frac{ r \eta(r)}{r+2} + 2 \frac{ 1 -
  Z(3)}{r+2}  \nonumber \\
&&+ \frac{M_\pi^2}{32 \pi^2 F_\pi^2} Y(3) \left\{ -5 +
\frac{r}{(r-1) ( r+2)} \left[ (2 r+1) \log \frac{M_\eta^2}{M_K^2} + (
4 r+1) \log \frac{M_K^2}{M_\pi^2} \right]  \right\} \nonumber \\
&& - \frac{2}{r+2} e +[(e_\beta -e) + 2 e'] \label{eq:beta} \,. 
\end{eqnarray}
These equations relate the two observable quantities $\alpha_{\pi\pi}$
and $\beta_{\pi\pi}$ to the three independent parameters $X(3), Z(3),
r$ (recall that $Y(3) = X(3)/Z(3)$), and contain the dependence on the
{\em direct} remainders $d_{\alpha,\pi\pi}$, $e_{\beta,\pi\pi}$ as
well as on the {\em indirect} ones, stemming from the mass and decay
constant identities.  The order of magnitude of such remainders can be
estimated and will be discussed later.  

\subsection{The Bayesian approach}
The determination of the three-flavour order parameters from the
experimental knowledge of $\alpha_{\pi\pi}$ and $\beta_{\pi\pi}$ is a
subtle issue, due to the nonlinear character of
Eqs.~(\ref{eq:alpha})-(\ref{eq:beta}) and to the fact that our three
independent parameters are subject to constraints (e.g., they have to
be positive).  Moreover, the NNLO remainders are in fact unknown; we can
only estimate their order of magnitude. One is led to consider them as
sources of error, but the propagation of errors is not so
straightforward, precisely because of the above-mentioned
non-linearities.  For these reasons the standard method of maximum
likelihood is inadequate in this case. A convenient approach is
provided by Bayesian analysis~\cite{bayes}, as described in App.~\ref{appbayes}.

We introduce the correlated probability density function
$P_{\mathrm{exp}} (\alpha_{\pi\pi},\beta_{\pi\pi})$,
\begin{eqnarray} \label{eq:expdistr}
P_{\mathrm{exp}} (\alpha_{\pi\pi},\beta_{\pi\pi})= \frac{\sqrt{\det
    C}}{2 \pi} \exp \left(-\frac{1}{2} V^T C V \right), \nonumber \\
\quad
    V=\left(\begin{array}{c} \alpha_{\pi\pi} -\alpha_{\mathrm{exp}} \\
    \beta_{\pi\pi}  -   \beta_{\mathrm{exp}} \end{array} 
\right) , \quad C=\left(\begin{array}{cc} c_{11} & c_{12} \\ c_{12} &
c_{22}  \end{array}
\right) , 
\end{eqnarray}
with $1/c_{11} = \delta_{\alpha}^2 ( 1 - \rho_{\alpha\beta}^2)$,
$1/c_{22} = \delta_{\beta}^2 ( 1 - \rho_{\alpha\beta}^2)$, $1
/c_{12}=-\delta_{\alpha} \delta_{\beta}
(1-\rho_{\alpha\beta}^2)/\rho_{\alpha\beta}$, and where
$\alpha_{\mathrm{exp}}$, $\beta_{\mathrm{exp}}$, $\delta_{\alpha}$,
$\delta_{\beta}$ and $\rho_{\alpha\beta}$ are the experimental
results, uncertainties and correlations given in
Eq.~(\ref{eq:alphabetaexp}).  This distribution summarizes the
experimental result: the two numbers, $\alpha_{\mathrm{exp}}$ and
$\beta_{\mathrm{exp}}$, given with an accuracy specified by the matrix
$C$.  Suppose now that we know, independently, the true values of
$(\alpha_{\pi\pi},\beta_{\pi\pi})$; Eq.~(\ref{eq:expdistr}) can then
be interpreted as the probability of obtaining, when performing an
experiment, the values $(\alpha_{\mathrm{exp}},\beta_{\mathrm{exp}})$,
i.e., the actual observed values, given our independent knowledge of
$(\alpha_{\pi\pi},\beta_{\pi\pi})$.  Eq.~(\ref{eq:expdistr}) is completely symmetric
under exchange of $(\alpha_{\pi\pi},\beta_{\pi\pi})$ with
$(\alpha_{\mathrm{exp}},\beta_{\mathrm{exp}})$.

This last interpretation is suitable for our problem, in which theory
relates the subthreshold parameters $\alpha_{\pi\pi}$ and
$\beta_{\pi\pi}$ to the three parameters $X(3),Z(3),r$ and to the
remainders: 
\begin{eqnarray}
\alpha_{\pi\pi} &=& {\cal A} [
r,Y(3),Z(3),\delta_1,\delta_2,\delta_3,\delta_4 ]\,, \\
\beta_{\pi\pi} &=& {\cal B} [r, Y(3),Z(3),\delta_2,\delta_5 ].
\end{eqnarray}
The relevant remainders,  denoted by $\delta_i$, $i=1,...,5$, are
defined in Table~\ref{tab:remainders}; $\delta_6$ and $\delta_7$ are 
additional remainders discussed in the following section. 
\begin{table}
\begin{center}
\begin{tabular}{|c|c|c|}
\hline
Remainder & Definition & $\sigma_i$ \\
\hline
$\delta_1$ & $d$ & 0.1 \\
$\delta_2$ & $e$ & 0.1 \\
$\delta_3$ & $e'$ & 0.01 \\
$\delta_4$ & $(d_{\alpha,\pi\pi} - d) + 4 d'$ & 0.03 \\
$\delta_5$ & $(e_{\beta,\pi\pi} - e ) + 2 e'$ & 0.03 \\
$\delta_6$ & $[(1+2/r) (\bar d_\pi - d) + 2 d/r]/(1-d) $ & 0.03 \\
$\delta_7$ & $[(1+2/r) (\bar e_\pi - e) + 2 e/r]/(1-e) $ & 0.03 \\
\hline
\end{tabular}
\end{center}
\caption{Definition of NNLO remainders. $\sigma_i$ denotes the
expected order of magnitude, and barred remainders
are taken in the $N_f=2$ chiral limit. \label{tab:remainders}}
\end{table}
Therefore we can say that the probability of obtaining the data
effectively observed, for a given choice of
$r,Y(3),Z(3),\delta_i$, is
\begin{equation}
P(\mbox{data} | r,Y,Z,\delta_i) = P_{\mathrm{exp}} [{\cal A}
(r,Y,Z,\delta_i), {\cal B} ( r,Y,Z,\delta_i)].
\end{equation}
This quantity is the likelihood of the observed experimental data.
Indeed, it is not what we are interested in: data have certainly been
observed, whereas the theoretical parameters are unknown to 
us.  Instead, the probability that we need is $P(r,Y,Z,\delta_i |
\mbox{data})$. This object can be calculated as a result of a
statistical inference using Bayes' theorem, at the price of
introducing some ``subjective'' a priori knowledge of the theoretical
parameters, $\pi(r,Y,Z,\delta_i)$. The result of the experiment is
viewed as an update of our previous knowledge of the theoretical
parameters, represented by the ``prior'' $\pi(r,Y,Z,\delta_i)$,
\begin{equation} \label{eq:bayes}
P(r,Y,Z,\delta_i | \mbox{data}) = \frac{P(\mbox{data}|r,Y,Z,\delta_i) 
\cdot \pi(r,Y,Z,\delta_i)}{\int dr\ dY\ dZ\ d\delta_i \,
P(\mbox{data}|r,Y,Z,\delta_i) \cdot \pi(r,Y,Z,\delta_i)}.
\end{equation}
The stronger the significance of data, the weaker will be the
dependence of the final result on our choice of the prior.

\subsection{Choice of the prior} \label{sec:priorchoice}
The prior should reflect our beliefs about the theoretical parameters
before our learning of experimental results. If we have no reason to
prefer one value to any other, then a flat prior should be chosen.  We
certainly have to implement the requirement of positivity for $X(3)$
and $Z(3)$, so the support of the function $\pi$ will have a lower
boundary at $X(3)=0$, $Z(3)=0$.  A similar requirement can be applied
to the two-flavour order parameters $X(2)$ and $Z(2)$. 
Using their expansions in the $SU(3)\times SU(3)$ $\chi$PT one 
can obtain~\cite{DGS}
\begin{eqnarray}
X(2) (1 - \bar d_\pi ) &=& X(3) + \frac{r}{r+2} \left[ 1 - X(3) -
\epsilon(r) - d - Y(3)^2 f_1 \right]  \label{eq:x2}\\
Z(2) (1 - \bar e_\pi ) &=& Z(3) + \frac{r}{r+2} \left[ 1 - Z(3) -
\eta(r) - e - Y(3) g_1 \right] \label{eq:z2}
\end{eqnarray}
where, $f_1$ and $g_1$ are (small) combinations of chiral logarithms,
\begin{eqnarray}
f_1 &=& \frac{M_\pi^2}{32 \pi^2 F_\pi^2} \left[ \frac{r}{r-1} \left( 3
\log \frac{M_K^2}{M_\pi^2} + \log \frac{M_\eta^2}{M_K^2} \right) -
(r+2)  \left( \log \frac{M_K^2}{\bar M_K^2} + \frac{2}{9} \log
\frac{M_\eta^2}{\bar M_\eta^2} \right) \right] \nonumber \\
&& \label{eq:f1}\\
g_1 &=& \frac{M_\pi^2}{32 \pi^2 F_\pi^2} \left[ \frac{4 r +1 }{r-1}
\log \frac{M_K^2}{M_\pi^2} + \frac{2 r + 1}{r-1} \log
\frac{M_\eta^2}{M_K^2} - ( r+2) \log \frac{M_K^2}{\bar M_K^2} \right],
\end{eqnarray}
estimated in Ref.~\cite{DGS} as $f_1\sim 0.05$ and $g_1 \sim 0.07$,
and barred quantities refer to the $SU(2)\times SU(2)$ chiral limit. 
The positivity of $X(2)$ and $Z(2)$ implies then a lower bound
for the indirect remainders $\delta_1$ and $\delta_2$,
\begin{eqnarray} \label{vacuum1}
X(2) \geq 0 &\leftrightarrow& \delta_1 \leq \delta_1^{\mathrm{max}} =
1 - \epsilon(r) - Y(3)^2 f_1\,, \\
Z(2) \geq 0 &\leftrightarrow& \delta_2 \leq \delta_2^{\mathrm{max}} =
1 - \eta(r) - Y(3) g_1\,. \label{vacuum2}
\end{eqnarray}
Actually one can also establish some upper bounds for the order
parameters $X(3)$ and $Z(3)$ using the so-called paramagnetic
inequalities Eqs.~(\ref{parasig})-(\ref{paraf}) stated in Ref.~\cite{DGS}\footnote{
  The statement is that
  chiral order parameters dominated by the infrared end of the
  spectrum of the Euclidean Dirac operator should decrease as the
  number of massless flavours increases. The paramagnetic inequalities
  also apply to $X(N_f)$ and $Z(N_f)$, for $N_f=2,3$.}, 
which translate into lower bounds for the remainders $\delta_6$ and
$\delta_7$ defined in Table~\ref{tab:remainders},
\begin{eqnarray}
X(3) \leq X(2) &\leftrightarrow& \delta_6 \geq \delta_6^{\mathrm{min}}
= 1 - \frac{1 - \delta_1 - \epsilon(r) - Y(3)^2 f_1}{X(3) ( 1 -
\delta_1)}\,, \\
Z(3) \leq Z(2) &\leftrightarrow& \delta_7 \geq \delta_7^{\mathrm{min}}
= 1 - \frac{1 - \delta_2 - \eta(r) - Y(3) g_1}{Z(3) ( 1 -
\delta_2)}\,.
\end{eqnarray}
The ratio of order parameters $Y(3)$ is also bounded from above, as
can be seen from Eq.~(\ref{y3nonpert}),
\begin{equation}
Y(3) \leq Y^{\mathrm{max}} = 2 \frac{1 - \epsilon(r) - \delta_1}{1 -
\eta(r) - \delta_2}\,.
\end{equation}
We do not assume any further knowledge about the three-flavour
order parameters. The hypothesis of a strict convergence of chiral
series, in the sense that every order of the expansion should be
smaller than the previous one, would lead to the choice of a prior
concentrated around the values $X(3)\sim 1$ and $Z(3)\sim 1$. In our
approach this is not required; we allow the data to indicate whether
vacuum fluctuations destabilize the chiral series or not.  Therefore, apart
 from the constraints listed above, we will choose a flat prior
for the three-flavour order parameters. In fact,  this statement is
ambiguous, since we have introduced three different quantities,
$X(3)$, $Y(3)$ and $Z(3)$, related by $X(3)=Y(3)Z(3)$.  If we restrict
the problem to a flat prior in two parameters among $[X(3),Y(3), Z(3)]
\equiv (X,Y,Z)$, we can get three different prior p.d.f.'s in (Y,Z):
\begin{equation}
\begin{array}{cclcc}
(X,Z) &:& \pi(Y,Z) dY dZ = dX dZ = Z dY dZ &\rightarrow&
\pi(Y,Z)=Z\,, \\
(X,Y) &:& \pi(Y,Z) dY dZ = dX dY = Y dY dZ &\rightarrow&
\pi(Y,Z)=Y\,,\\
(Y,Z) &:& \pi(Y,Z) dY dZ = dY dZ           &\rightarrow&
\pi(Y,Z)=1\,.
\end{array}
\label{eq:3priors}
\end{equation}
A physical argument helps us to select one of these possibilities.
When $Z(3)$ vanishes, chiral symmetry restoration occurs and the
marginal probabilities of chiral order parameters must vanish. This
can be obtained if the prior p.d.f. vanishes in the limit $Z(3)\to 0$,
which is realized in the case of a flat prior in $(X,Z)$.  Henceforth,
we will focus on this case and therefore take $\pi(Y,Z)=Z$ inside the
range allowed by the positivity and paramagnetic constraints, and
$\pi(Y,Z)=0$ outside.

Inspired by Eq.~(\ref{ordandfluc}), 
we restrict the range of variation of $r$
such that $0\leq \epsilon(r) \leq 1$, which yields the range
\begin{equation}
r_1\leq r \leq r_2\,, \quad r_1 = 2 \frac{ F_K M_K}{F_\pi M_\pi}
-1\sim 8\,, \quad r_2=2   \frac{ F_K^2 M_K^2}{F_\pi^2 M_\pi^2} -1 \sim 36\,.
\end{equation}
No other previous knowledge is supposed for the quark mass ratio $r$,
so the prior is taken to be flat in $r$ over this allowed domain.

Finally, we must discuss how to treat the remainders
$\delta_1,\ldots,\delta_7$. We recall that they are defined as the NNLO
contributions in the chiral expansions of a selected class of
observables, for which a good overall convergence is expected.
Therefore, the remainders should be small.  This expectation can be
implemented in the prior by considering the remainders as Gaussian
distributions centered at zero, with widths corresponding to their
expected order of magnitude. Since the size of the corrections in the
chiral series is 30\% for the three-flavour expansions and 10\% for
the two-flavour ones, we will use the following rule of thumb, already
introduced in footnote~\ref{foot:thumb}: the
typical size of remainders is $(30\%)^2=0.1$ for genuine $O(m_s^2)$
remainders like $d,e$, $30\%\times 10\%=0.03$ for $O(m_s m)$
combination of remainders (such as $d_{\alpha} - d$, $e_\beta-e$: see
below), and 0.01 for $e'=O(e/r)$ and $d'=O(d/r)$. These assumptions
for the widths of the Gaussian distributions of the remainders are
collected in the last column of Table~\ref{tab:remainders}. 

As far as Table~\ref{tab:remainders} is concerned, we still
have to show that the two combinations $d_{\alpha,\pi\pi}-d_{\pi}$ and
$e_{\beta,\pi\pi}-e_{\pi}$, contributing to the remainders $\delta_4$
and $\delta_5$, are of order $O(m m_s)$, instead of $O(m_s^2)$. This
follows from a $SU(2)\times SU(2)$ 
low-energy theorem: consider the subthreshold
parameters $\alpha_{\pi\pi}$ and $\beta_{\pi\pi}$ in the $N_f=2$
chiral limit, $m_u,m_d\to 0$; from their $SU(2)\times SU(2)$ chiral
expansion in powers of $m$ -- Eqs.~(32)-(33) of Ref.~\cite{DFGS} -- it
is straightforward to conclude that
\begin{equation}
\lim_{m\to 0} \alpha_{\pi\pi} =1 , \quad \lim_{m\to 0} \beta_{\pi\pi}
=1.
\end{equation}
Then we combine $SU(3)\times SU(3)$ chiral expansions of the
subthreshold parameters and  the mass and decay constant identities
for the pion, by subtracting Eq.~(\ref{pimass}) from Eq.~(\ref{eq:alpha}) [and
Eq.~(\ref{pidecay}) from Eq.~(\ref{eq:beta})], which
yields the following equalities:
\begin{eqnarray} \label{eq:alpha-1}
& & F_\pi^2 M_\pi^2 (\alpha_{\pi\pi} -1) = 12 m^2 A + 12 m^2 Z^s -
M_\pi^2 Y(3) 4 m ( \xi + 2 \tilde \xi) \\
& &\quad + \frac{M_\pi^4}{32 \pi^2} Y(3)^2    \left( \log
\frac{M_K^2}{M_\pi^2} - \log \frac{M_\eta^2}{M_K^2} - \frac{7}{3}
\right) + F_\pi^2 M_\pi^2 ( d_{\alpha,\pi\pi} - d_\pi)\,, \nonumber \\
& & F_\pi^2 (\beta_{\pi\pi} - 1)  = 2 m \xi + 4 m \tilde \xi +
\frac{M_\pi^2 Y(3)}{16 \pi^2} \left( \log \frac{M_\eta^2}{M_K^2} + 2
\log \frac{M_K^2}{M_\pi^2} - \frac{5}{2} \right) \nonumber \\
& &\qquad + F_\pi^2 ( e_{\beta,\pi\pi} - e_{\pi})\,. \label{eq:beta-1}
\end{eqnarray}
On the right hand side of Eq.~(\ref{eq:alpha-1})
[Eq.~(\ref{eq:beta-1})], all the terms involving $O(p^4)$ LEC's and
chiral logarithms are of $O(m^2)$ [$O(m)$]. If we divide
Eq.~(\ref{eq:alpha-1}) by $F_\pi^2 M_\pi^2$ [Eq.~(\ref{eq:beta-1}) by
$F_\pi^2$] and take the $N_f=2$ chiral limit $m\to 0$, all the terms
vanish apart from the NNLO remainders. The latter must therefore
fulfill the following condition:
\begin{equation}
\lim_{m\to 0} ( d_{\alpha,\pi\pi} - d_{\pi})=0, \quad \lim_{m\to 0} (
e_{\beta,\pi\pi} - e_\pi) =0,
\end{equation}
so that the difference between $d_{\alpha,\pi\pi}$ and $d_{\pi}$
[$e_{\beta,\pi\pi}$ and $e_\pi$] is only of order $O(m m_s)$.

Having collected all the ingredients for our choice of the prior, we
can now perform the integration over the NNLO remainders to obtain
$P(r,Y,Z|\mbox{data})$. If we integrate the latter with respect to two
parameters, we end up with the marginal (posterior) probabilities:
\begin{eqnarray}
P(r|\mbox{data}) &=& \int dY dZ \, P(r,Y,Z |\mbox{data}) \label{eq:bayesr}\\
P(Y|\mbox{data}) &=& \int dr\, dZ \, P(r,Y,Z |\mbox{data}) \label{eq:bayesy}\\
P(Z|\mbox{data}) &=& \int dY dY \, P(r,Y,Z |\mbox{data}) \label{eq:bayesz}\\
P(X|\mbox{data}) &=& \int dr\, dY dZ \, \delta ( Y Z - X)\ P(r,Y,Z
|\mbox{data}) \nonumber \\
 &=& \int dr\, dZ \, \frac{1}{Z}\, P(r,X/Z,Z |\mbox{data}) .\label{eq:bayesx}
\end{eqnarray}
The precise expression of $P$ and the
 numerical evaluation of these integrals are detailed in App.~\ref{appinteg}.

\subsection{Discussion and results}

\begin{figure}[t]
\centerline{
\epsfysize=8cm
\epsffile{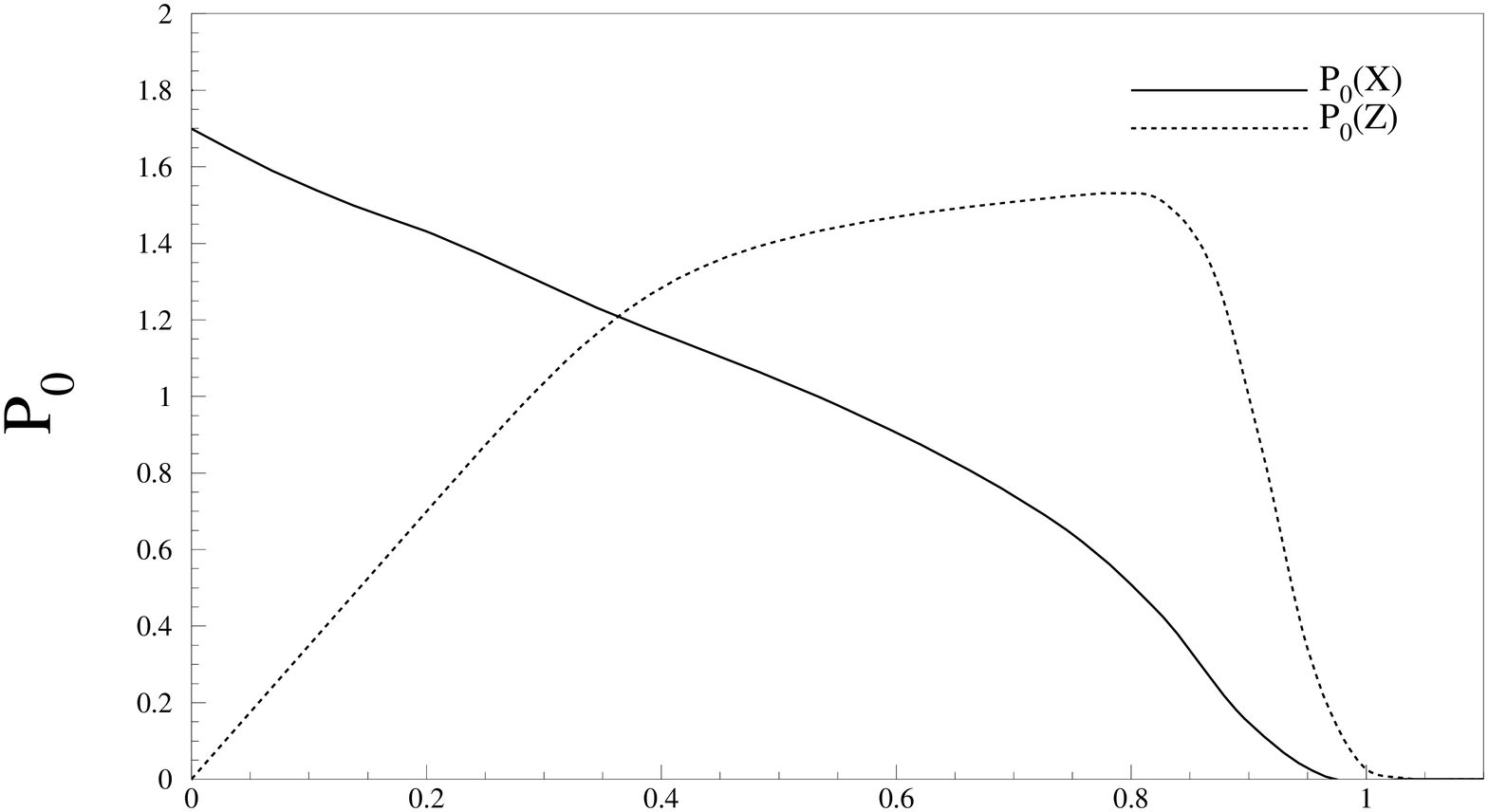}
}
\centerline{
\epsfysize=8cm
\epsffile{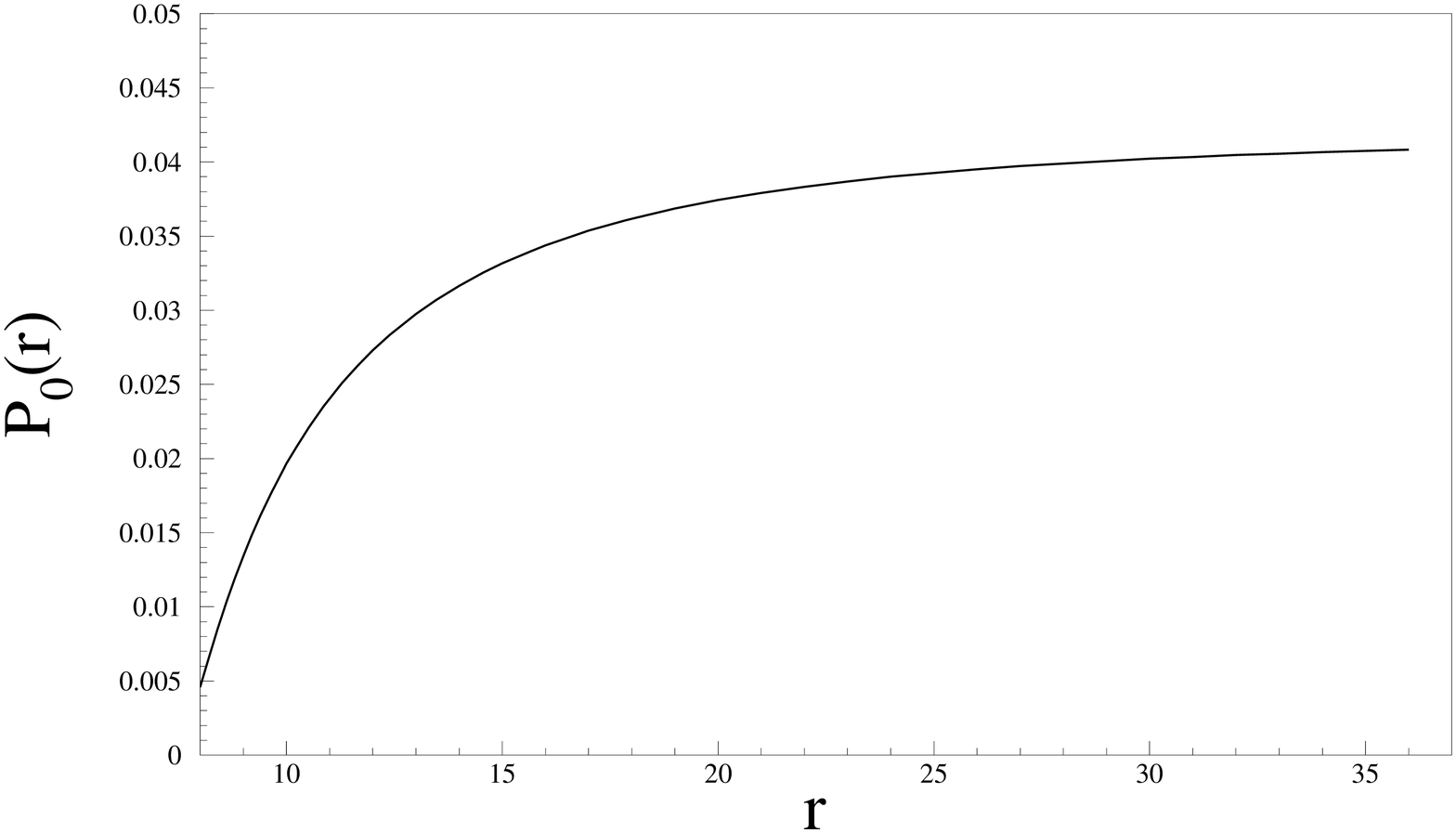}
}
\caption{The reference probability density profiles for the order
parameters $X(3)$ (full) and $Z(3)$ (dashed) [top],
and for the quark mass ratio $r$ [bottom].}
\label{fig:intprior}
\end{figure}

Even if no information from $\pi\pi$ scattering data is included, the
results of the integrations (\ref{eq:bayesr})-(\ref{eq:bayesx}) are
nontrivial, because of the interplay of the various constraints
imposed.  For instance, the prior for $r$, when integrated over all
other variables, will no longer be uniform. The probability density
profiles of Fig.~\ref{fig:intprior} are obtained by replacing
$P(r,Y,Z,\delta_i|\mbox{data})$ with $\pi(r,Y,Z,\delta_i)$ and
integrating over all but one variable. They can be interpreted as a
measure of the ``phase space'' for each parameter allowed by the
theoretical constraints. The significance of $\pi\pi$ data will be
seen in the modification induced with respect to such
``reference'' density profiles.

\begin{figure}[t]
\epsfysize=10cm
\centerline{
\epsffile{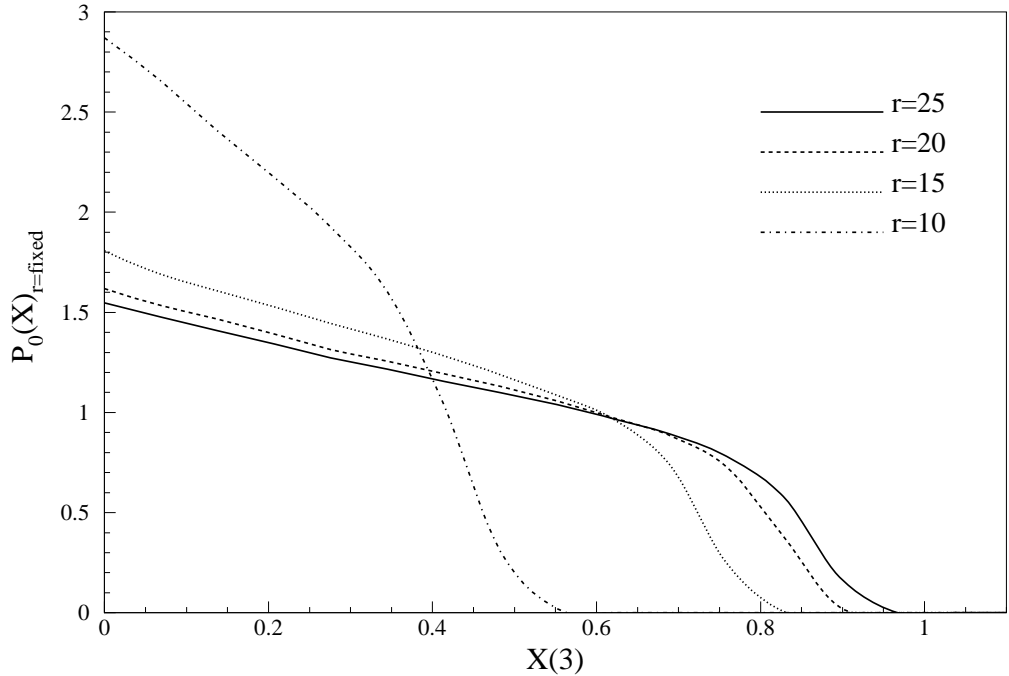}
}
\caption{The reference probability density profiles for the order
  parameter $X(3)$ when the quark mass ratio $r$ is taken fixed at
  four different values.}
\label{fig:rfixed}
\end{figure}

In the present framework, we may make more quantitative the well-known
mechanism by which a low value of $r$ implies a suppression of the
quark condensate~\cite{DGS}. In Fig.~\ref{fig:rfixed} we show the reference
probability density profiles for $X$, when the quark mass ratio $r$ is
taken fixed at different values, $r=25,20,15,10$.  (The ratio of order
parameters $Y(3)$ behaves similarly.)

\begin{figure}[t]
\epsfysize=8cm
\centerline{
\epsffile{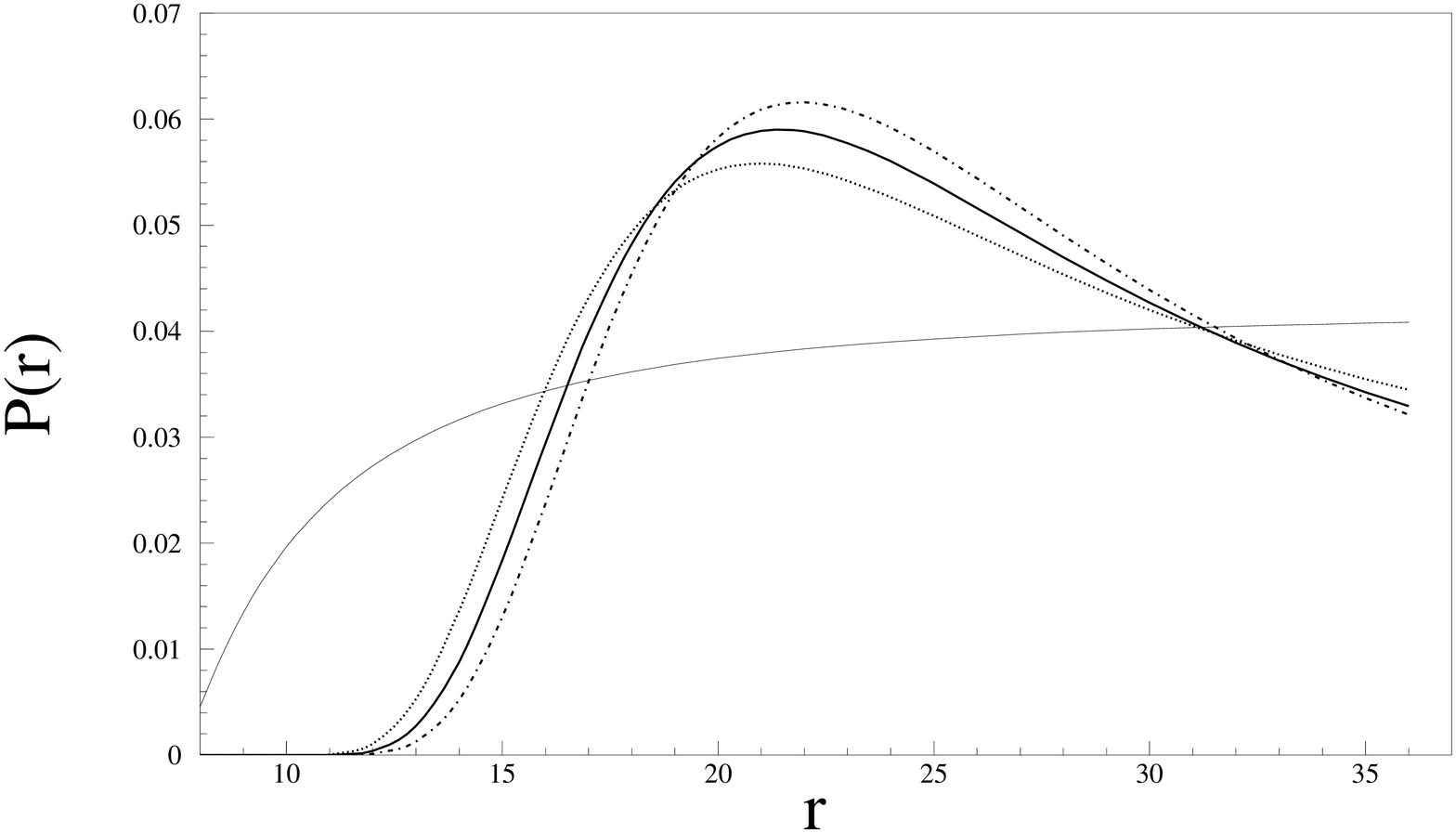}
} 
\caption{The probability density profile for $r$ inferred from the
  $\pi\pi$ data (thick lines) as compared to the reference one (thin
  dotted line). The different curves represent different choices for
  the prior. The full line corresponds to the choice of priors
  indicated in the previous section.
  The dashed-dotted line is obtained by reducing the
  widths for the prior distributions of the remainders by a factor of
  two, the dotted line by increasing the widths by a factor of two.}
\label{fig:rprofile}
\end{figure}

Notice that for small values of $r$, for which $X(3)$ is very
small,  the two-flavour order parameter $X(2)$ should also be small.
This is clear from inspection of Eqs.~(\ref{eq:x2}) and~(\ref{eq:z2}),
because $\epsilon(r)\sim 1$. Since we know from our $SU(2)\times SU(2)$ 
analysis~\cite{DFGS}
of the same $\pi\pi$ data that $X(2)$ is very close to~1,
we can expect that such data will constrain $r$ to stay away from such
small values.  This is what we observe in Fig.~\ref{fig:rprofile}, in
which we show the marginal probability density profile for $r$ with
the inclusion of the $\pi\pi$ experimental data.

These data induce a substantial change in the distribution as compared
to the reference probability density profile (restricted phase space).
However, such a broad distribution cannot be used for a real
determination of $r$: at most a lower bound for $r$ can be given in
probabilistic terms: $r\geq 14$ at 95\% confidence level.  The most
probable value of $r$ lies at $r=21-22$, very close to standard
expectations. Indeed, the motivation for the rearrangement of the
chiral expansions, namely the possible importance of vacuum
fluctuations and their potential to destabilize the chiral series, is
not essential in the case of $r$: the fluctuation parameters $L_4$ and
$L_6$ are absent from the perturbative reexpression of the chiral 
series of $r$, see Eq.~(\ref{rP}), which may thus exhibit
no instability even in the case of large fluctuations. It is therefore
not surprising that similar values of $r$ are obtained through the
perturbative expansion Eq.~(\ref{rP}) or the nonperturbative framework
presented above. Such similarity is actually a welcome check of the
approach advocated in this paper.  

We also show in Fig.~\ref{fig:rprofile} the
dependence of the inferred probability density profile on the initial
choice of the prior: the full line refers to the
choice of priors described in Sec.~\ref{sec:priorchoice}, 
while the dashed-dotted one
(dotted one) corresponds to a distribution for the remainders with half widths
(double widths), as
compared to Table~\ref{tab:remainders}. 
As expected, broader Gaussians for the NNLO remainders 
lead to a flatter p.d.f. for $r$ -- the impact of experimental
information is weakened when higher orders are allowed to be larger.
If the most probable value of
$r$ depends to some extent on the choice of the prior, the same is not
true for the 95\% lower bound, which is almost the same for all
distributions.

A comment on the so-called ``Kaplan-Manohar ambiguity'' is in order
here. In Ref.~\cite{km}, it was shown that the sum of the 
$O(p^2)$ and $O(p^4)$ chiral Lagrangian remained unchanged under a
shift in the quark masses:
\begin{equation}
m\to m+\lambda m m_s\,, \qquad m_s\to m_s+\lambda m^2\,,
\end{equation}
and a corresponding redefinition of the $O(p^4)$ LEC's $L_{6,7,8}$.
This seemingly prevents any determination of the quark mass ratio $r$.
However, the Kaplan-Manohar ambiguity also induces a shift in the $O(p^6)$
terms included in the NNLO remainders (neglected in a perturbative 
treatment of $O(p^4)$ expansions). We have assumed a good overall 
convergence of chiral series and therefore required small NNLO remainders,
which corresponds to fixing the Kaplan-Manohar ambiguity. 
We point out that this hypothesis is not specific to our framework, 
and is a fairly general assumption when dealing with chiral series.

\begin{figure}[t]
\centerline{
\epsfysize=8cm
\epsffile{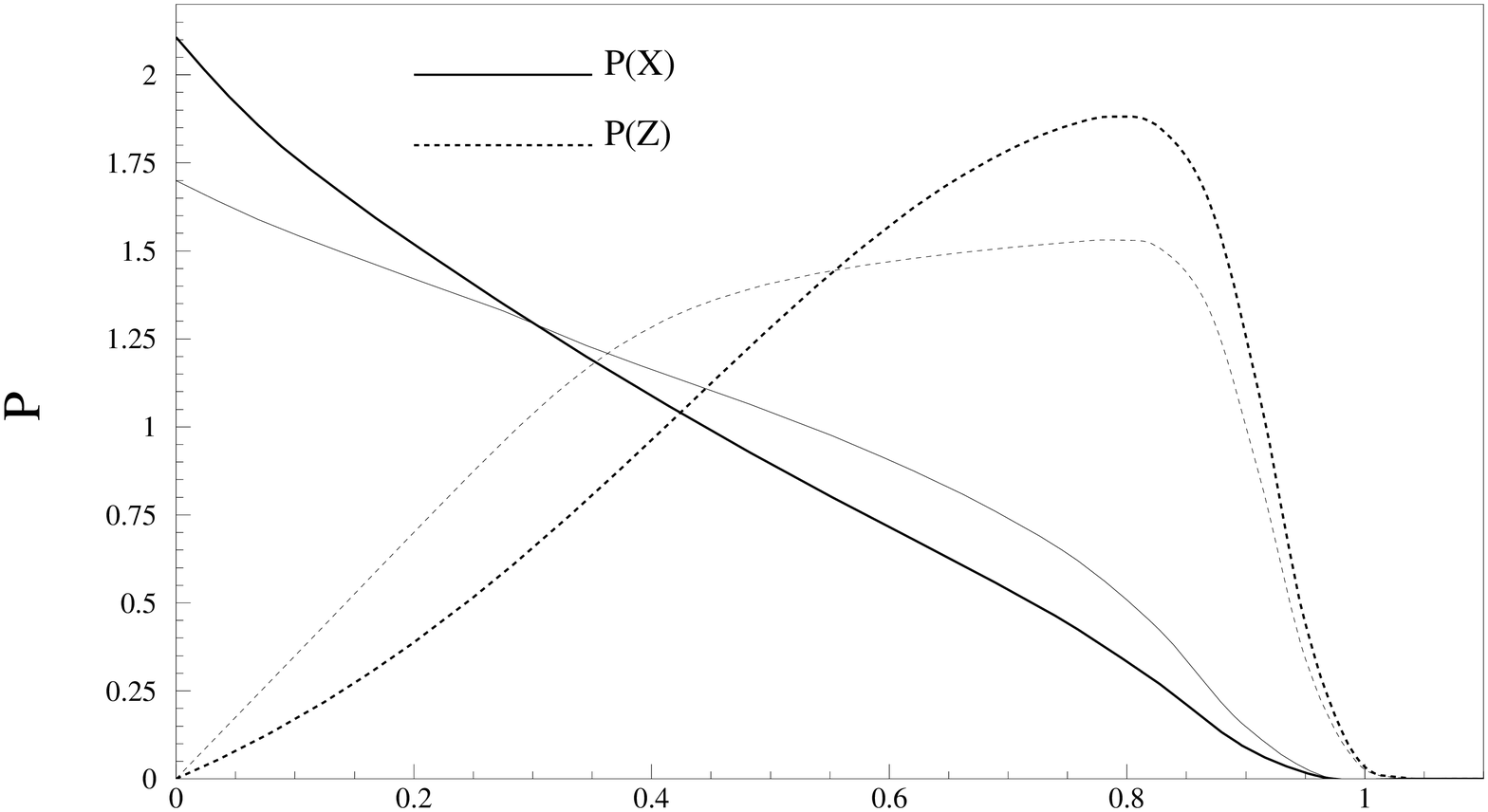}
}
\centerline{
\epsfysize=8cm
\epsffile{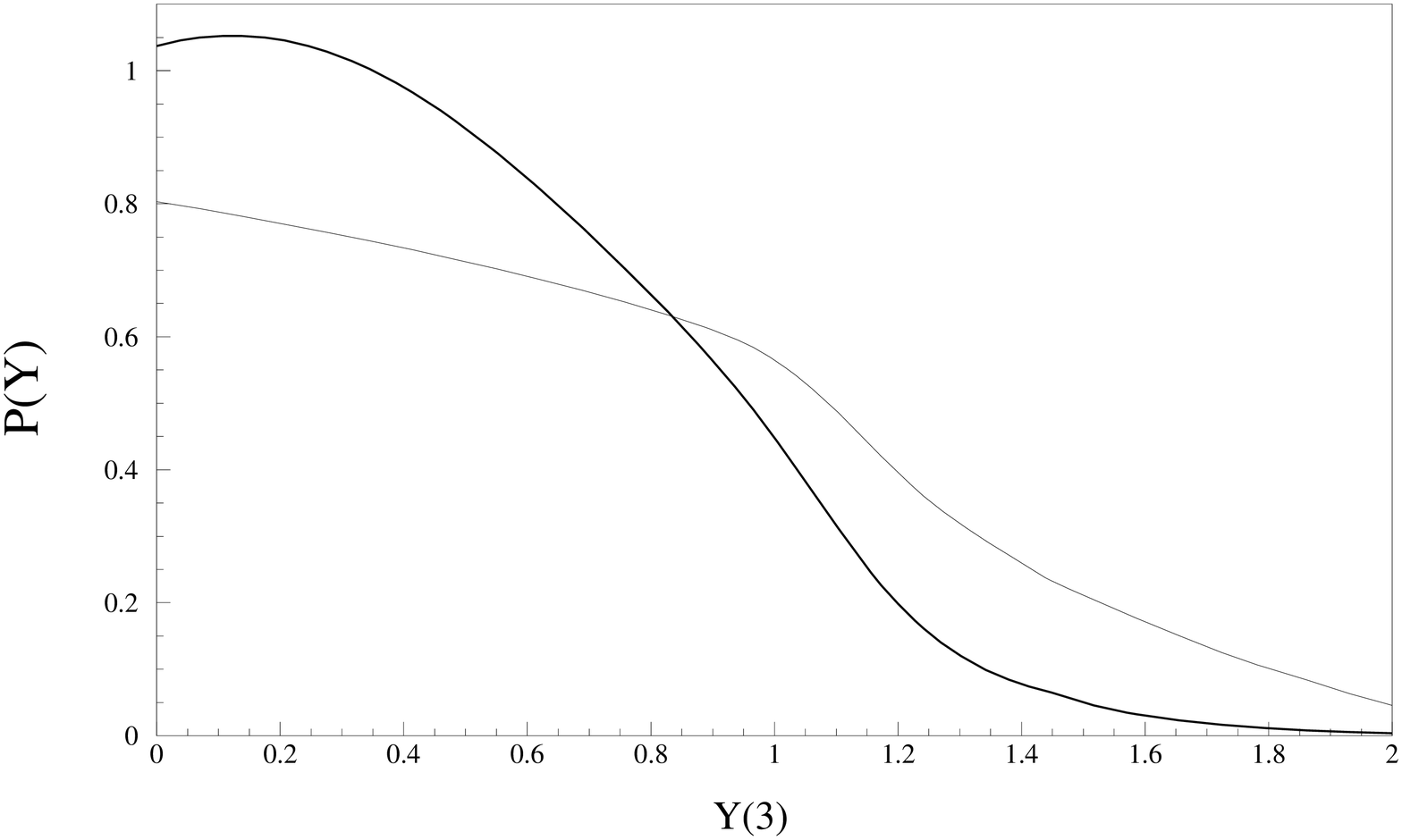}
}
\caption{The probability density profiles as inferred from the
$\pi\pi$ data (thick lines) compared to the reference one,
representing the phase space (thin lines). Full lines refer to $X(3)$,
dashed lines to $Z(3)$ [top] and full lines to the ratio of order
parameters $Y(3)$ [bottom].}
\label{fig:probxyz}
\end{figure}

On the other hand not much deviation is caused by $\pi\pi$ data with
respect to the reference profiles for $X(3)$, $Z(3)$ and $Y(3)$, as is
clear from Fig.~\ref{fig:probxyz}.
This is an indication of the fact that $\pi\pi$ data, as was expected,
are not sensitive enough to $X(3)$ and $Z(3)$. Their determination
would require the inclusion of other observables, which we discuss
now.

\section{Other sources of constraints on $X(3)$ and $Z(3)$}

We now briefly mention some possible sources of independent experimental
information about the three-flavour chiral order parameters $X(3)$ and
$Z(3)$, or equivalently about the $O(p^4)$ fluctuation parameters
$Y^2(3)L_6$ and $Y(3)L_4$. The combination of $O(p^4)$ LEC's 
$(\Delta L_8+2\Delta L_6)$, invariant under the Kaplan-Manohar
transformation, is of particular interest. It appears in many
observables and has remained undetermined so far. $Y^2(3)L_8$
is essentially given by $r$ through Eq.~(\ref{epsilon}): 
$Y^2(3)\Delta L_8=0.99\cdot 10^{-3}$ for $r=25$ (where 
NNLO remainders are neglected). From
the positivity of $X(3)$ and its relation to the fluctuation 
parameter $L_6$ Eq.~(\ref{ordandfluc}), one gets
the following upper bound for $L_6$:
\begin{equation}
Y^2(3) \Delta L_6 \leq \frac{1}{16(r+2)} 
  \frac{F_\pi^2}{M_\pi^2}[1-\epsilon(r)-d]\,,
\end{equation}
which implies that $Y^2(3) \Delta L_6 \leq 0.98\cdot 10^{-3}$ for
$r=25$ (with an expected uncertainty of 10 \% from NNLO remainders). 
Between the limit of no
fluctuations and that of maximal fluctuations, the
combination of LEC's $Y^2(3)[\Delta L_8+2\Delta L_6]$ can thus vary in the
range:
\begin{equation} \label{l6l8bd}
1.0 \cdot 10^{-3} \leq Y^2(3)[\Delta L_8+2\Delta L_6]
  \leq 3.0 \cdot 10^{-3} \qquad \qquad [r=25]
\end{equation}

This combination can be perturbatively related to the LEC
$\bar\ell_3$~\cite{GL1} of 
the $N_f=2$ chiral Lagrangian; see 
Eq.~(11.6) in Ref.~\cite{GL2}. In the limit of no fluctuations, 
$\bar\ell_3$ is a positive number of $O(1)$;  the estimate
$\bar\ell_3=2.9\pm 2.4$ was obtained in Ref.~\cite{GL1} under the
assumption  of the validity 
of the Zweig rule.
If larger vacuum fluctuations make
$Y^2(3)[\Delta L_8+2\Delta L_6]$ increase, $\bar\ell_3$ decreases 
towards negative values with a larger magnitude. 
Thus, in principle, the size of vacuum fluctuations could
be investigated through an accurate determination of $\bar\ell_3$
obtained from $\pi\pi$-scattering parameters. Unfortunately,
experimental information at such accuracy is not available
due to a large uncertainty in the scattering length $a_0^2$, 
which is not tightly constrained by available 
$K_{e4}$ decay experiments~\footnote{
The available $\pi\pi$-scattering data was analysed
in Ref.~\cite{DFGS}, yielding $\bar\ell_3=-18.5\pm 16.7$
and thus suggesting an important role played by vacuum fluctuations.
If the experimental information on the $I=2$ channel 
is replaced by a theoretical constraint
concerning the scalar radius of the pion~\cite{pipiCGL,DFGS},
$\bar\ell_3$ becomes a small positive number compatible with
the absence of fluctuations.}. Therefore, we have to consider
other sources of information in order to constrain the $N_f=3$ order
parameters.

\subsection{Goldstone boson scattering and decays}

In order to estimate the size of vacuum fluctuations of $s\bar{s}$ pairs,
processes directly involving strange mesons are required. Before
sketching how our method could be extended to the relevant processes, 
we want to comment on a few estimates of the $O(p^4)$ LEC's $L_4$ and $L_6$
that are available in the literature. These estimates show a common feature:
they rely on the standard (perturbative) treatment of vacuum
fluctuations, assuming that the latter are small, but they lead to
$L_4$ and $L_6$ significantly larger than their critical values.
These estimates should therefore be considered at most as
valuable hints of internal inconsistency of the perturbative treatment.

B\"uttiker et al.~\cite{pikroy} 
have analyzed the $\pi K \rightarrow \pi K$ and $\pi
\pi \rightarrow K \bar{K}$ amplitudes, thereby obtaining the amplitudes at
threshold and in the subthreshold region.  Comparing these results
with the $O(p^4)$ $\chi$PT expansion of Bernard et al.~\cite{bkm}, 
they determine the
LEC's $L_i\, (i=1,4)$ and the combination $L_8 + 2 L_6$:
\begin{equation}
10^3 [L_8^r + 2 L_6^r](M_\rho) = 3.66 \pm 1.52 \,.
\end{equation}
The large uncertainty quoted is reflected in the uncertainty in the
scattering lengths combination $a_0^{1/2} + 2 a_0^{3/2}$.  This could
improve considerably once experimental results from $\pi K$ atoms
become available.  
However, the analysis is based on $O(p^4)$ $\chi$PT, in which the implicit
assumption is made that $Y(3) = 1$.  From our bounds Eq.~(\ref{l6l8bd}), 
taking $Y(3)=1$, we obtain:
\begin{equation}
0.27  \leq 10^3 [L_8^r + 2  L_6^r](M_\rho)
  \leq 2.3 \qquad \qquad [r=25,Y(3)=1]\,.
\end{equation}
It is clear that the values for $L_8 + 2 L_6$
given in Ref.~\cite{pikroy} are barely compatible with the 
assumption of $Y(3) = 1$.
Furthermore, the authors of Ref.~\cite{pikroy} estimate 
$10^3 L_4^r(M_\rho) = 0.50 \pm 0.39$, which implies
$10^3 \Delta L_4 = 1.0 \pm 0.39$ and suggests 
significant violation of the Zweig rule in the scalar
sector [see Eq.~(\ref{z3pert})]. Both remarks call for
a comparison of $\pi K$ scattering amplitudes with a chiral
expansion treating nonperturbatively (possibly large)
vacuum fluctuations~\cite{TBP}.

Recently, Bijnens and Dhonte~\cite{bijscal}
have calculated the $\pi$ and $K$ scalar form factors
at two loops in $N_f=3$ $\chi$PT; they then fit their
results to the corresponding dispersive
representation based on the solution of the multi-channel
Omn\`es-Muskhelishvili equations~\cite{Bachir1,Bachir2}. 
In order to obtain ``decent fits'' for the
form factors at zero momentum transfer, they found that the LEC's
$L_4$ and $L_6$ had to satisfy the constraint
\begin{equation}
L_6^r(M_\rho) \lsim 0.6\times L_4^r(M_\rho) + 0.4\cdot 10^{-3} \,.
\end{equation}
If we take this to imply that 
$[2 L_6^r - L_4^r](M_\rho) \lsim 0.8\cdot 10^{-3}$, then we
can rewrite the constraint in terms of the fluctuation parameters
$\rho, \lambda$ introduced in Sec.~\ref{secGBspec}:
\begin{equation}
\rho - \lambda \lsim 1.6 \qquad\qquad[r=25]\,.
\end{equation}
Recalling Eq.~(\ref{y3nonpert}), the convergence of the perturbative
expansion of $Y(3)=2m B_0/M_\pi^2$ requires $\rho-\lambda\ll 1$.
Viewed in this light, the constraint of Ref.~\cite{bijscal} is not surprising:
it is simply the observation that the authors, within their
perturbative analysis, cannot find a good fit to a
perturbative $\chi$PT expansion in the presence of large vacuum
fluctuations.
In addition, they find that requiring the values of the scalar form
factors at zero not deviate too much from their lowest order values
leads to the estimate $0.3 \leq 10^3 L_4^r(M_\rho) \leq 0.6$.  Note that this
result is roughly in agreement with that of Ref.~\cite{pikroy} 
discussed above.
According to Table~2 of Ref.~\cite{bijscal}, this value of $L_4$ leads
to a suppression of $Z(3)$ down to half of $Z(2)$, as can be checked
from Eq.~(\ref{ordandfluc}). All these considerations suggest
sizeable vacuum fluctuations in the scalar sector.

We outline now how additional observables coming from Goldstone boson
scattering and decay could be incorporated naturally in the Bayesian
machinery, in order to constrain the size of vacuum fluctuations,
or equivalently $X(3)$ and $Z(3)$. The 
most obvious candidate is $\pi K$ scattering. 
The first step consists in using analyticity, unitarity, crossing
symmetry in conjunction with experimental data in order to constrain as
much as possible the low-energy $\pi K$-scattering
amplitude~\cite{pikroy}. The second stage corresponds to an analysis
similar to that in the case of $\pi\pi$ scattering: establish a
dispersive representation of the amplitude with subthreshold parameters
(i.e., subtraction constants like $\alpha_{\pi\pi}$ and 
$\beta_{\pi\pi}$), determine the value of these parameters from
experiment, then include these parameters as additional observables
for the Bayesian analysis. The expected outcome should be more 
stringent constraints on $X(3)$ and $Z(3)$~\cite{TBP}.

The decay $\eta\to 3\pi$ is a second process of interest~\cite{eta3pi}. 
The standard treatment of this decay starts with
the ratio $A_{\eta\to 3\pi}/\Delta_K$, where $A_{\eta\to 3\pi}$
is the decay amplitude and $\Delta_K=(M_{K^+}^2-M_{K^0}^2)_{\rm QCD}$.
This ratio is then perturbatively reexpressed in terms of Goldstone boson
masses and a single $O(p^4)$ LEC $L_3$. 
 As stressed previously,  a bare chiral expansion of ratios of QCD
 correlation functions that are only conditionally convergent may exhibit
 instabilities. To cope with possibly large vacuum fluctuations, a better
 starting point could be the expansion of the quantity $F^3_\pi F_\eta
 A_{\eta\to 3\pi}$, which is linearly related to a QCD correlation function.
 Such an expansion will involve more LEC's and order parameters and
 allow the extraction of the latter from the data. 
 Since the decay $\eta\to 3\pi$ is
 forbidden in the isospin limit $m_u = m_d$, we must start by 
 extending the previous
 discussion of Goldstone boson masses and decay constants to include
 isospin breaking. A dispersive representation of the amplitude
 $A_{\eta\to 3\pi}$ must then be written to define convenient observables. The
 dispersion relations are here more involved than in the case of
 $\pi\pi$- and $\pi K$-scattering: 
 they require studying $\eta\pi\to\pi\pi$ and performing
 subsequently an analytical continuation to the (physical) decay channel. 
The observables thus defined, related 
to the behaviour of the decay amplitude at the center of the 
Dalitz plot (via slope parameters), can be exploited in our Bayesian
framework in order to constrain further three-flavour chiral order
parameters and to extract the quark mass ratios $m_u/m_s$ and $m_d/m_s$.

\subsection{Two-point functions and sum rules}

In this paper, our aim was in particular to pin down $X(3)$, 
i.e., the $N_f=3$ chiral condensate
measured in physical units, which governs the behaviour of QCD  
correlation functions in the limit $m_u=m_d=m_s=0$. 
Related though different quantities
arise when the high-energy limit of the same correlation functions is 
studied through the operator product expansion (OPE). Local condensates appear
then, and those with the lowest dimension are:
\begin{equation} \label{opecond}
\Sigma_u=-\langle 0 |\bar{u}u| 0 \rangle\,,\quad
\Sigma_d=-\langle 0 |\bar{d}d| 0 \rangle\,,\quad
\Sigma_s=-\langle 0 |\bar{s}s| 0 \rangle\,,\quad
\end{equation}
where the physical vacuum of the theory is denoted $|0\rangle$ with
all the quarks carrying their physical masses: no chiral limit
is taken. 

These OPE quark condensates occur in various sum rules for
two-point correlators and could thus be determined
in this framework~\cite{gensr}. First, they arise 
(multiplied by a mass term) in 
the high-energy tail of the correlators as dimension-4 
operators. (For example, 
see~\cite{Kamborsr} for the case of pseudoscalar densities.) 
Next, in some sum rules, normal-ordered condensates of the type
(\ref{opecond}) appear through chiral Ward identities.
For instance, in the case of the divergence of the 
strangeness-changing vector current~\cite{Pichsr,Jaminsr}, 
the strange-quark mass is determined via a sum rule with no subtraction, 
but another sum rule can be written with the subtraction constant
$(m_s-m_u)(\Sigma_s-\Sigma_u)$, providing
in principle some information on the OPE quark condensates. 
Unfortunately, the high-energy tail of the (Borel transformed)
two-point function involved in this case has a QCD expansion 
which behaves quite badly and prevents any accurate
determination. Lastly, the OPE quark condensates
arise when factorisation is invoked to reexpress higher-dimensional 
four-quark operators as the square of $\bar{q}q$ vacuum expectation
values.

We stress that the OPE quark condensates $\Sigma_u,\Sigma_d,\Sigma_s$
have a different
definition (and thus value) from the chiral condensates that we have
considered here and in Ref.~\cite{DFGS}:
\begin{eqnarray}
\Sigma(2) &\equiv& -\lim_{m_u,m_d\to 0}\langle 0 |\bar{u}u| 0 \rangle
= \lim_{m_u,m_d\to 0} \Sigma_u
= \lim_{m_u,m_d\to 0} \Sigma_d
\,,\\
\Sigma(3) &\equiv& -\lim_{m_u,m_d,m_s\to
0}\langle 0 |\bar{u}u| 0 \rangle \nonumber\\ 
&=& \lim_{m_u,m_d,m_s\to 0} \Sigma_u
= \lim_{m_u,m_d,m_s\to 0} \Sigma_d
= \lim_{m_u,m_d,m_s\to 0} \Sigma_s
\,.
\end{eqnarray}
In particular, $\Sigma_u,\Sigma_d,\Sigma_s$ exhibit
an ultraviolet divergence that must be renormalized; therefore,
their definition and their value depend on the convention applied. 
It is possible to relate them to $\Sigma(3)$ using 
$N_f=3$ $\chi$PT. For instance, if we take
Eq.~(9.1) in Ref.~\cite{GL2} in the isospin limit, we get:
\begin{eqnarray}
X_{u,d}&\equiv&\frac{2m\Sigma_{u,d}}{F_\pi^2M_\pi^2}
  = 
  X(3)+[Y(3)]^2 \frac{M_\pi^2}{F_\pi^2}
     \left[16(r+2)\Delta L_6+4(2\Delta L_8+\Delta H_2)\right]
      +d_{\Sigma;u,d} \nonumber\\
 & = & 1-\frac{1}{2}\epsilon(r)+4[Y(3)]^2 \frac{M_\pi^2}{F_\pi^2}
     \Delta H_2-d+\frac{1}{2}d'+d_{\Sigma;u,d} \,, \label{sigmau}\\
X_s &\equiv&\frac{2m\Sigma_s}{F_\pi^2M_\pi^2} =
  X(3)+[Y(3)]^2 \frac{M_\pi^2}{F_\pi^2}\nonumber\\
&&\qquad\qquad\times  \left[16(r+2)\Delta L_6+4r(2\Delta L_8+\Delta H_2)
       \right]
     +d_{\Sigma;s} \nonumber\\
 & = & 1+\frac{r-2}{2}\epsilon(r)+4r[Y(3)]^2 \frac{M_\pi^2}{F_\pi^2}
     \Delta H_2 
     -d+\frac{r}{2}d'+d_{\Sigma;s} \,,
\end{eqnarray}
where NNLO remainders are denoted $d_{\Sigma;u,d}$ and $d_{\Sigma;s}$, and 
the $O(p^4)$ high-energy counterterm $H_2^r$ arises in the combination:
\begin{eqnarray}
\Delta H_2 &=& H_2^r(\mu)
  -\frac{1}{128\pi^2}
     \left(\frac{1}{2}\log\frac{M_K^2}{\mu^2}+
           \frac{1}{3}\log\frac{M_\eta^2}{\mu^2}\right) \nonumber \\
&& \quad  -\frac{1}{256 \pi^2 (r-1)}  \left( 3
 \log  \frac{M_K^2}{M_\pi^2}  + \log \frac{M_\eta^2}{M_K^2} \right)
 \,.
\end{eqnarray}
The value of such high-energy counterterms 
cannot be fixed by low-energy data only, and
their presence in the chiral expansions is 
merely a manifestation of the renormalisation-scheme dependence 
of the OPE quark condensates. 

An interesting relation, free from high-energy counterterms,
exists between the OPE condensates:
\begin{equation} \label{Xuds}
\frac{rX_{u,d}-X_s}{r-1}
  = 1-\epsilon(r)-d
  +\frac{r}{r-1}d_{\Sigma;u,d}-\frac{1}{r-1}d_{\Sigma;s} \,.
\end{equation}
Two conclusions can be drawn from this relation. First, for $r$ larger
than 15, Eq.~(\ref{Xuds}) shows that $X_{u,d}$ is close to 
$[1-\epsilon(r)-d]$, while we see from Eq.~(\ref{eq:x2bis})
that $X(2)(1-\bar{d}_\pi)$ equals $[1-\epsilon(r)-d]$ up to 
$1/r$-suppressed corrections. $\Sigma_{u,d}$ should thus be 
very close to $\Sigma(2)$, which was expected since the 
$u,d$ quarks are very light
and the physical world is near the $N_f=2$ chiral limit.

The second conclusion is that $X_s-X_{u,d}$ can hardly be obtained from 
such a relation, since $X_{u,d}$ and $1-\epsilon(r)-d$ largely cancel. 
Thus, very accurate knowledge of $r$ and $X_{u,d}$ would
be needed to determine $X_s$ this way.
More generally, the possibility of significant vacuum 
fluctuations of $s\bar{s}$ pairs makes it difficult to relate
in a quantitative way $\Sigma(3)$ and the
OPE quark condensates $\Sigma_{ud}$, $\Sigma_s$.

Such a relation between OPE and chiral quark condensates is naturally
relevant for the description of $K$-decays. 
In particular, some $K\to\pi\pi$ weak matrix 
elements are related to v.e.v.'s of four-quark operators 
in the $N_f=3$ chiral limit, thanks to sum rules for vector-axial or 
scalar-pseudoscalar correlators~\cite{sr4quark}. These sum rules 
are evaluated by 
splitting the integral in two energy domains: the low-energy region
is described by experimental data, while the high-energy behaviour 
is obtained through the operator product expansion, which involves a priori
OPE quark condensates. However, since the sum rules are evaluated in 
the chiral limit $m_u=m_d=m_s\to 0$, these condensates actually reduce to 
the $N_f=3$ chiral condensate $\Sigma(3)$. 

The dispersive estimates of $K\to\pi\pi$ matrix elements could thus
be affected at three different stages
by significant vacuum fluctuations of $s\bar{s}$
pairs leading to smaller values of $\Sigma(3)$.
Firstly, extrapolating the weak matrix elements
from the $N_f=3$ chiral limit to the physical values of the $u,d,s$-quarks
could not be carried out on the basis of the usual
treatment and values of LEC's of $N_f=3$ $\chi$PT, since the latter assume from the start
a leading contribution from $\Sigma(3)$. Moreover,
the very estimate of the sum rule could be modified
because of the change in the high-energy behaviour of the correlator in 
the chiral limit. The third question concerns dimension-6 four-quark
condensates, which appear at higher orders of OPE and are
often related to the square of a $\bar{q}q$ condensate through
factorisation, on the basis of large-$N_c$ arguments.
The presence of large $q\bar{q}$ fluctuations might render such an assumption invalid.

\subsection{Implications for lattice simulations}

In principle, the lattice should represent a
particularly favourable domain to study how QCD 
at low energy depends on the light-quark masses and how this dependence
is connected to the vacuum fluctuations of $q\bar{q}$ pairs.
Recent progress has been made in this field. Discretisations 
of the Dirac operator have been discovered with highly desirable
qualities for the simulation of light quarks. In particular, 
Ginsparg-Wilson fermions~\cite{GW} do not break chiral symmetry
explicitly. A second (cheaper) option consists 
in twisted-mass lattice QCD~\cite{twisted}, 
where a parametrized rotation of the mass
matrix allows one to restore chiral symmetry partially in observables
through an averaging procedure. Another avenue is provided
by staggered fermions~\cite{staggered}, 
which allows one to study an odd number of
flavours, at the cost of introducing unwanted flavour 
degeneracies.

Vacuum fluctuations of $q\bar{q}$ pairs are typical sea-quark 
effects. The fermionic determinant plays here an essential role,
since we are interested in chiral parameters dominated by the
infrared end of the Dirac spectrum~\cite{dirac,DGS}.
In order to study these effects on the lattice, it is therefore
mandatory to generate data for 3 dynamical flavours. For this particular
purpose, one cannot rely on quenched data (with no dynamical quark) or
on data generated with only two dynamical quarks -- even though
they can be of interest for observables relatively
insensitive to the fermion determinant, e.g., $M_\rho$.

We will now illustrate, by considering ``bare'' expansions of ``good'' 
observables such as $F_\pi^2M_\pi^2$ and $F_\pi^2$,
how lattice data could probe vacuum fluctuations of $q\bar{q}$ pairs and how
chiral extrapolations should be dealt with if the latter turn
out to be significant. We consider a slightly optimistic situation where
lattice data with reasonable accuracy can be generated for 3 light 
dynamical flavours. For simplicity, we choose to 
work in the limit of degenerate strange and
light quark masses.  The analysis could be done for independent
variations of the quark masses, but the attendant complications do not
add anything essential to our conclusions.  Moreover, it is likely
that realistic lattice calculations are more easily performed in this
simplified situation. On the lattice, we denote by $m_L$ the common mass 
of the three degenerate light flavours, and by $M_L^2$ and $F_L^2$ the 
common mass and decay constant of the eight degenerate Goldstone bosons.
We keep $m$ and $m_s$ for the physical
values of the quarks and $F_\pi^2, M_\pi^2 \ldots$ for the physical values
of Goldstone boson observables.

The observables $F_L^2$ and $F_L^2 M_L^2$ have thus the form
\begin{eqnarray}
F_L^2 M_L^2 &=& 2 m_L B_0 F_0^2 + 64 m_L^2 B_0^2 \bigg[3 L_6(\mu) +
L_8(\mu) - {1 \over 96 \pi^2} \log {M_L^2 \over \mu^2}\bigg]
+ F_L^2 M_L^2 d_L\,,\\
F_L^2 &=& F_0^2 + 16 m_L B_0 \bigg[ 3 L_4(\mu) + L_5(\mu) - {3 \over
128 \pi^2} \log {M_L^2 \over \mu^2}\bigg]
+ F_L^2 e_L \, ,
\end{eqnarray}
where the remainders $d_L, e_L$ are of order $m_L^2$.

Since $B_0, F_0$ and the LEC's $L_i$ are all defined in the $N_f=3$ chiral
limit and are thus independent of the quark masses, we may use 
Eqs.~(\ref{F0})-(\ref{L5}) to eliminate them in favour of
the real-world parameters $X(3), Y(3), Z(3), r$ and the physical values
of the masses and couplings of the Goldstone bosons, leading to:
\begin{eqnarray} \label{lattdecay}
F_L^2 M_L^2 &=& b \, F_\pi^2 M_\pi^2 X(3)
+ b^2 \, {F_\pi^2 M_\pi^2 \over r+2}
\bigg\{ 3 [1-X(3)-d] + (r-1) [\epsilon(r)+d'] \bigg\} \nonumber\\ 
&& \quad  + 2 b^2 \, M_\pi^4 Y(3)^2 \bigg\{ - \frac{1}{32 \pi^2 (
r+2)}  \bigg( 3
 \log  \frac{M_K^2}{M_\pi^2}  + \log \frac{M_\eta^2}{M_K^2} \bigg)
\nonumber \\
&& \quad  + {1 \over 16 \pi^2} \bigg(\log {M_K^2 \over M_L^2} + {1
\over 3} \log {M_\eta^2 \over M_L^2}\bigg)\bigg\} + F_L^2 M_L^2 d_L\,,
\\ 
F_L^2 &=& F_\pi^2 Z(3) \label{lattmass}
  + b \, {1 \over r+2} F_\pi^2 \bigg\{ 3[1-Z(3)-e] +
(r-1) [\eta(r)+e'] \bigg\} \nonumber \\
&& \quad - b \, {M_\pi^2 Y(3) \over 32 \pi^2 }\bigg[\frac{1}{r+2} \bigg( 3 \log
{M_\eta^2 \over M_K^2} + 7 \log {M_K^2 \over M_\pi^2}\bigg) \nonumber
\\
&& \quad  - 2  \bigg(\log
{M_\eta^2 \over M_L^2} + 2 \log {M_K^2 \over M_L^2}\bigg) \bigg] + F_L^2 e_L \, ,
\end{eqnarray}
Taking the ratio of these equations gives $M_L^2$ implicitly as a
function of $b=m_L / m$. The remainders $d, e$ enter
the resulting expressions multiplied by a factor of $b/r$; we
will ignore them as well as the $1/r$-suppressed contributions from
$d',e'$.

\begin{figure}[t]
\begin{center}
\includegraphics[angle=270,width=13cm]{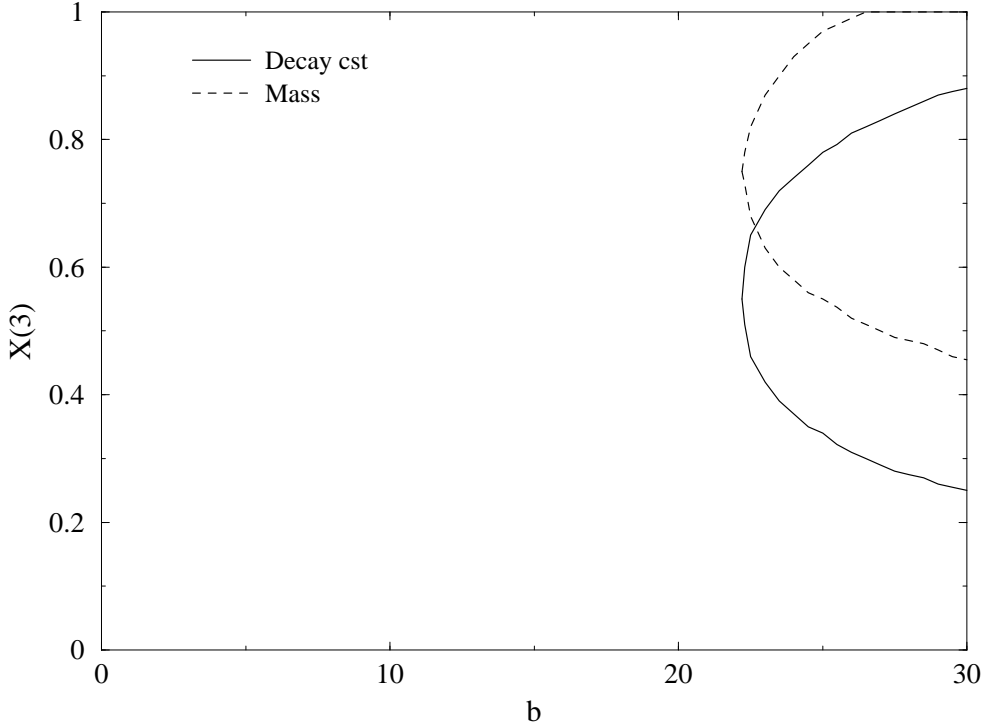}
\caption{Stability criterion for $Z(3)=0.6, r=25$. Inside the circle in full
line [dashed line], the term logarithmic in $M_L$ contributes 
more than 25\% to $F_L^2$ [$F_L^2M_L^2$]. All NNLO remainders are set to
zero.}
\label{fig:L1}
\end{center}
\end{figure}

\begin{figure}[t]
\begin{center}
\includegraphics[width=11cm]{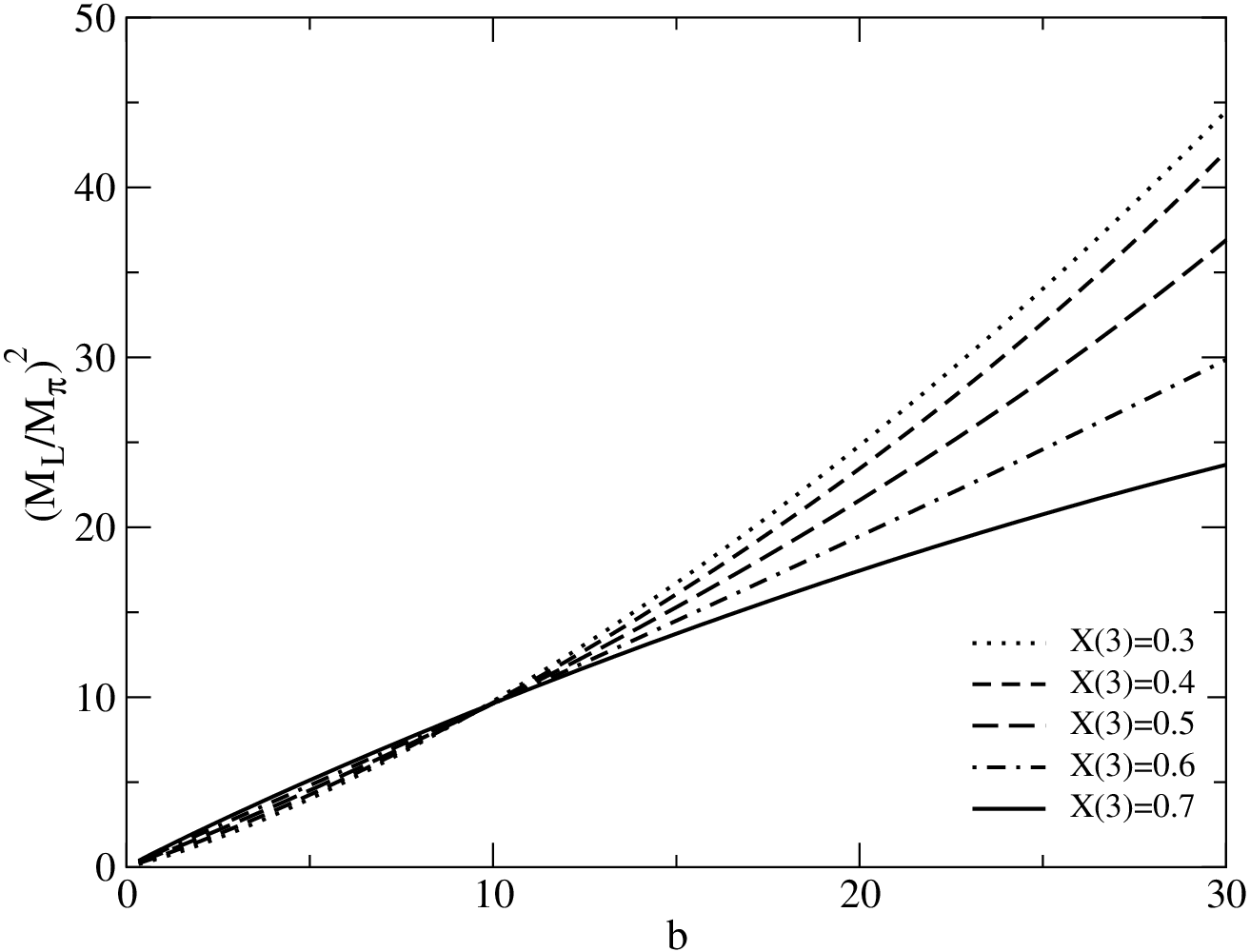}
\caption{$M_L^2/M_\pi^2$ as a function of $b=m_L/m$ for $Z(3)=0.6,
r=25$. All NNLO remainders are set to zero.
}
\label{fig:L2}
\end{center}
\end{figure}

Since the chiral expansion requires small values of $m_L$, while
present day lattice simulations prefer $m_L$ on the order of $m_s$, it
is important to ascertain if there is a range of variation for $m_L$ in
which our equations may be applied and still give valuable results.
There are two different conditions to be met.
First, as $m_L$ increases, the degenerate mass $M_L$ increases, eventually
exceeding $M_K, M_\eta$; this in itself is of concern, since 
the chiral expansion becomes unreliable as $M_L$
approaches values typical of relevant hadronic resonances.
In addition, the terms logarithmic in
$M_L^2$ become negative and eventually the procedure described above
is unstable.  Therefore, we will restrict the range of variation of
$m_L$ so that these logarithmic terms do not contribute more than 25\%
of the total.  This stability criterion constrains the allowed region in the
space of parameters $X(3), Z(3), r, m_L$; however, this region
includes values of interest.  For example, in Fig.~\ref{fig:L1} we show the
allowed region in $X(3), m_L/m$ for the illustrative choice $Z(3) = 0.6, r=25$.
For $m_L/m \le 20$, all values of $X(3)$ are possible.  As can be seen
in Fig.~\ref{fig:L2}, with such a constraint on $m_L/m$, we satisfy
also the first requirement since $M_L$ does not exceed
$1.5\cdot M_K$. 

\begin{figure}[t]
\begin{center}
\includegraphics[width=11cm]{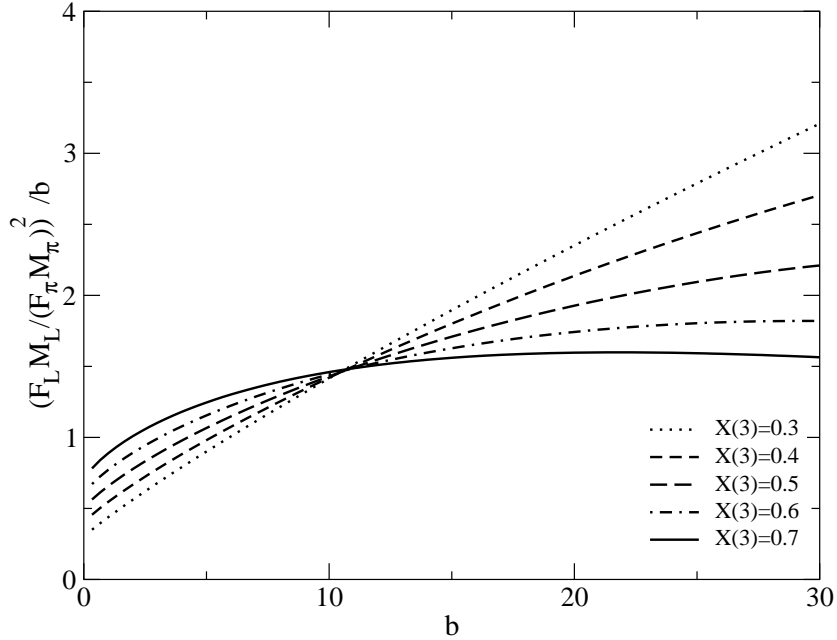}
\caption{$F_L^2 M_L^2/(F_\pi M_\pi)^2/b$ as a function of
$b$. Each line corresponds to a different value of $X(3)$, whereas $r$
is set to 25. All NNLO remainders are set to zero.
}
\label{fig:L3}
\end{center}
\end{figure}

\begin{figure}[t]
\begin{center}
\includegraphics[width=11cm]{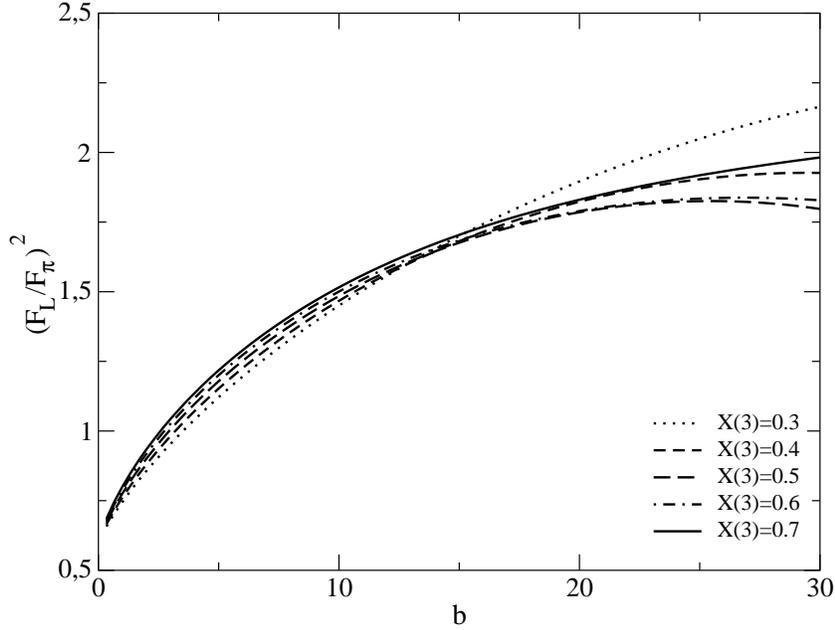}
\caption{$F_L^2/F_\pi^2$ as a function of
$b$. Each line corresponds to a different value of $X(3)$, whereas $r$
is set to 25. All NNLO remainders are set to zero.
}
\label{fig:L4}
\end{center}
\end{figure}

For values of $Z(3)$ between 0.4 and 0.8 and of $r$ between 20 and 30, 
we find that the allowed region does not qualitatively change: 
so long as we keep $m_L/m$ less than 20, then
all values of $X(3)$ are permitted according to our stability criterion.
Consider, then, the variation of $F_L^2$ and of $F_L^2 M_L^2$ as
functions of $m_L$, for fixed $Z(3)$ and $r$, as shown in
Figs.~\ref{fig:L3} and \ref{fig:L4} in the illustrative case $Z(3)=0.6, r=25$.
From Eqs.~(\ref{lattdecay})-(\ref{lattmass}), 
the general dependence on $X(3)$ is apparent: the
$X(3)$-dependence must vanish for small $b=m_L/m$ as well as for $b=m_L/m
\sim (r+2)/3 \sim 10$ (apart from a residual dependence from the logarithms).  
Therefore, the region of interest is $10 \leq
b \leq 20$.  Fortunately, even in this restricted interval, there
is considerable dependence on $X(3)$, especially for $F_L^2 M_L^2/b$. A
good knowledge of the spectrum in this range would allow us to
discriminate at least between the most extreme possibilities ($X(3)$
close to $X(2)$, $X(3)$ close to 0)\footnote{A first step 
in this direction was proposed by considering the
dependence on $m$ and $m_s$ of the Goldstone boson masses to extract
the values of some low-energy constants for partially quenched
staggered fermions~\cite{pqQCDappl}. However, this was achieved by 
relying on the chiral expansion 
of $M_P^2$, with a perturbative reexpression of the fundamental parameters
$r,X(3),Z(3)$ in terms of Goldstone boson masses.
Moreover, the $O(p^4)$ LEC's $L_4$ and $L_6$ related to the vacuum
fluctuations were eliminated through a perturbative redefinition
of the parameters involved in the chiral series.
As discussed in Secs.~\ref{secconvinst} and \ref{secGBspec}, 
this procedure need not be correct 
in the case of large vacuum fluctuations.
}. We stress that not all the
observables are equivalent for this study: for instance,
$M_L^2$ exhibits a much weaker sensitivity to vacuum fluctuations, as 
shown in Fig.~\ref{fig:L2}. This cancellation between $L_6$ and $L_4$ in
the case of the masses is quite usual~\cite{ordfluc} and does not mean
that the effect is absent for all observables, as seen from
Figs.~\ref{fig:L3} and \ref{fig:L4}.

We have not included the NNLO remainders $d_L$ and
$e_L$ here, but it is a straightforward exercise to take them into
account. We can check their size by writing them in the following
form:
\begin{eqnarray}
F_\pi^2 M_\pi^2 d_\pi &=& m_s^2 m D_1 + m_s m^2 D_2 + m^3 D_3\,, \nonumber \\
F_\pi^2  e_\pi &=&  m_s^2  E_1 + m_s m E_2 + m^2 E_3\,,
\end{eqnarray}
where $D_i$ and $E_i$ do contain $O(p^6)$ LEC's and chiral logarithms. 
If we neglect the dependence of the chiral logarithms on the quark
masses when they vary from $m,m_s$ to $m_L$, the same quantities
will appear in the
analogous expression for $F_L^2 M_L^2 d_L, F_L^2 e_L$:
\begin{eqnarray}
{d_L \over d_\pi} &=& \bigg[ {bF_\pi^2 M_\pi^2 \over F_L^2
M_L^2}\bigg] \ {b^2 \over r^2}\ \bigg[{D_1 + D_2 + D_3 \over D_1 +
D_2/r + D_3/r^2 }\bigg]\,, \nonumber \\ 
{e_L \over e_\pi} &=& \bigg[{F_\pi^2 \over F_L^2 }\bigg] \  {b^2 \over
r^2} \ \bigg[ {E_1 + E_2 + E_3 \over E_1 + E_2/r + E_3/r^2 }\bigg]\,.
\end{eqnarray}
We can check easily that the first factor is of order one in each case
(see for instance Figs.~\ref{fig:L3} and \ref{fig:L4}). Therefore, 
we conclude that the remainders $d_L, e_L$
are expected to be of order $m_L^2$, i.e., of order $0.1 \cdot b^2/r^2$
(i.e., 10~\% or less). Once the remainders are included, an accurate
determination of $Z(3)$ is quite difficult because $F_L^2$ does not
exhibit a strong sensitivity to the latter. But $X(3)$ is still
accessible because of the strong variation in the curvature of 
$F_L^2M_L^2$ as a function of $b$.

\section{Summary and conclusion}

The large-$N_c$ suppressed low-energy constants $L^r_4(\mu)$ and
$L^r_6(\mu)$ encode fluctuations of vacuum $\bar ss$ pairs.  We have
analysed the influence of these fluctuations on the convergence of
$N_f=3$ $\chi$PT.

\begin{itemize}
  
\item[i)] For the physical value of $m_s$, we assume a global (though
  possibly slow) convergence of the $SU(3) \times SU(3)$ chiral
  expansion applied to low-energy connected QCD correlation functions
  and to observables that are linearly related to the latter: typical
  examples are $F^2_P$, $F^2_P M^2_P$, $F^4_{\pi} A_{\pi\pi \to
    \pi\pi}$ , $F^2_{\pi}F^2_K A_{\pi K \to \pi K}$, etc.  For such
  quantities, the bare expansion (as defined in Sec.~\ref{secconvinst}) 
  expressed in
  terms of the order parameters $\Sigma(3)$ and $F_0$
  and in powers of quark masses $m_u,m_d, m_s$ is likely characterised
  by relatively small higher-order corrections starting at NNLO.
  
\item[ii)] Vacuum fluctuations of $\bar ss$ pairs affect in particular
  the next-to-leading-order (NLO) 
  contributions to $F^2_{\pi} M^2_{\pi}$ and $F^2_{\pi}$
  through terms $m_sL_6$ and $ m_sL_4$, which appear in $\chi$PT with
  large coefficients.  They reflect Zweig-rule violation and large
  $1/N_c$ corrections in the scalar channel. Unless $L_6$ and $L_4$
  are very precisely tuned to their critical values
  $L^{\rm crit}_6(M_{\rho})= -0.26 \cdot 10^{-3}$, $L^{\rm crit}_4
  (M_{\rho})= - 0.51 \cdot 10^{-3}$, the vacuum fluctuation NLO
  contribution to $F^2_{\pi} M^2_{\pi}$ and/or to $F^2_{\pi}$ becomes
  of comparable size or even larger than the corresponding leading-order
  (LO)
  contributions given by the three-flavour condensate $\Sigma(3)$ and
  by the pion decay constant $F^2_0$.  In this case, the expansions of
  $X(3)=2m\Sigma(3)/F^2_{\pi}M^2_{\pi}$ and $Z(3)=F^2_0/F^2_{\pi}$ in
  powers of $m_s$ break down despite the convergence of
  $F^2_{\pi}M^2_{\pi}$ and $F^2_{\pi}$. As a result, $X(3)$ and
  $Z(3)$, which measure $N_f=3$ order parameters in physical units,
  could be suppressed well below one, implying an instability of
  $N_f=3$ $\chi$PT.

\item[iii)] The instability due to vacuum fluctuations upsets the
  standard perturbative reexpression of the lowest order parameters
  $\Sigma(3)$ and $F^2_0$ as well as quark masses in favour of an
  expansion in terms of physical Goldstone boson masses, $F_\pi$ and
  $F_K$.  Instead, using the four mass and decay constant Ward
  identities, we nonperturbatively eliminate the LEC's $L_4,L_5,L_6,
  (L_7),L_8$ in terms of the order parameters $X(3)$ and $Z(3)$ , the
  quark mass ratio $r= m_s/m$ and four NNLO remainders $d_{\pi}, d_K,
  e_{\pi}, e_K$ which collect all higher-order contributions starting
  at $O(p^6)$.  This procedure amounts to an exact resummation of the
  standard perturbative reexpression of $X(3)$, $Z(3)$ and $r$ and it
  applies even if the vacuum fluctuations suppress $X(3)$ and/or
  $Z(3)$.

\item[iv)] In this way, values of the basic order parameters $X(3)$,
  $Z(3)$ and the quark mass ratio $r$ can be constrained by
  experimental data as long as NNLO remainders are under control.  In
  order to gradually incorporate theoretical conjectures about the
  order parameters $X(3)$ and $Z(3)$ (e.g., their positivity,
  paramagnetic inequalities), and on higher chiral orders (i.e., NNLO
  remainders), we propose to use Bayesian statistical inference.  In
  this approach, previous knowledge of parameters is encoded into the
  prior probability distribution function.  At this step, some degree
  of arbitrariness is introduced, but the dependence on the choice of
  the prior is guaranteed to be weak if data are significant enough.
  
\item[v)] We have applied this procedure to the three-flavour analysis
  of elastic $\pi\pi$ scattering, in order to test the ability of
  the recent high-precision low-energy data obtained by the E865
  Collaboration~\cite{E865} to constrain three-flavour chiral dynamics.  We
  have shown that, for the present state of experimental precision,
  the data cannot determine the two fundamental $N_f=3$ chiral order
  parameters $X(3)$ and $Z(3)$.  However, the low-energy $\pi\pi$ data
  is sufficient for us to put a quantitative lower bound on the quark
  mass ratio $r=m_s/m \ge 14$ at the 95\% confidence level, and to
  determine the corresponding probability distribution function.
  
\item[vi)] The Bayesian machinery is suitable for incrementally
  including further experimental information on low-energy
  observables, and seems especially well-adapted in this context, in
  view of an extension of our analysis to valence $s$-quark
  effects.  In particular, $\pi K$ scattering appears to be rather
  promising, due to recent progress obtained through the
  solution of the corresponding Roy-Steiner equations~\cite{pikroy}.  
  We plan to address this issue in future publications~\cite{TBP}.
  
\item[vii)] The possible instability of $N_f=3$ chiral expansions, and
  the prescription examined in this paper to treat it, are also
  relevant in other respects.  One example is the evaluation of $K
  \rightarrow \pi\pi$ weak matrix elements from sum rules based on
  OPE: in particular, the extrapolation of the OPE condensate from the
  chiral limit to physical values of quark masses would not be
  possible on the basis of the usual treatment of $N_f=3$ $\chi$PT
  formulas.  A similar remark applies to the study of the quark mass
  dependence of Goldstone boson masses on the lattice and to
  determinations of the LEC's by this means.  We have shown
  how such a program could be pursued even in the
  presence of large fluctuations, provided simulations with three
  dynamical flavours would be available with quark masses as small as
  the physical strange quark mass.  It remains to be seen whether such
  simulations, with proper control of the continuum and thermodynamic
  limits, will be feasible in the near future.
  
\item[viii)] In order to detect possibly large $\bar{s}s$
  fluctuations, another strategy is conceivable: one might calculate
  and analyse as many quantities as possible within the standard
  $SU(3)\times SU(3)$ $\chi$PT, up to and including two loops, using
  the standard perturbative reexpression of low-energy 
  parameters~\cite{ABT,bijscal}. The 
  instability of such $\chi$PT expansions could appear as
  an internal inconsistency of the result of corresponding fits with
  the assumptions underlying the perturbative treatment of
  standard $\chi$PT.
  
\end{itemize}

\acknowledgments

We would like to thank D.~Becirevic,
J.~Bijnens, U.-G. Mei{\ss}ner, B.~Moussallam and J.~Prades
for discussions and comments. Work partially supported by EU-RTN contract
EURIDICE (HPRN-CT2002-00311). 
LPT and IPN are Unit\'es mixtes de recherche du CNRS et de 
l'Universit\'e Paris XI (UMR 8627 and UMR 8608 respectively).

\appendix

\section{Mass and decay constant identities} \label{appident}

We recall for convenience the chiral expansion of Goldstone boson masses 
as discussed in Refs.~\cite{DGS,ordfluc}:
\begin{eqnarray}\label{pimass}
F_\pi^2 M_\pi^2 &=&2m\Sigma(3)+ 2m(m_s +2m)Z^s +4m^2 A  +4 m^2
B_{0}^2 L + F_\pi^2 M_\pi^2 d_\pi,\qquad\\
F_K^2M_K^2&=&(m_s + m)\Sigma (3) +(m_s+m)(m_s+2m) Z^s  + (m_s+m)^2 A  
   \label{kamass}\\
&&\quad + m(m_s + m)B_{0}^2 L 
	   + F_K^2 M_K^2 d_{K},\nonumber
\end{eqnarray}
The case of the $\eta$-meson is discussed in Sec.~\ref{secgmo} and below.
The connection with the standard LEC's of the $N_f \ge 3$ effective
Lagrangian is:
\begin{eqnarray}
Z^s &=& 32 B_{0}^2 \left\{L_6(\mu)  
         -\frac{1}{512\pi^2}\left[\log\frac{M_K^2}{\mu^2} 
       + \frac{2}{9}\log \frac{M_\eta^2}{\mu^2} \right]
       \right\},\label{defZ} \\
A &=& 16B_{0}^2 \left\{L_8(\mu)   
      - \frac{1}{512\pi^2}\left[ \log\frac{M_K^2}{\mu^2}
      +\frac{2}{3} \log \frac{M_\eta^2}{\mu^2} \right] \right\} . \label{defA}
\end{eqnarray}
The remainders $F_P^2M_P^2 d_P$ collect all
higher-order terms, starting at the next-to-next-to-leading order 
$O(m_q^3)$, in agreement with the definition in Ref.~\cite{ordfluc},
but different from that in Ref.~\cite{D}.
The combination of chiral logarithms $L$ is \cite{ordfluc}:
\beq \label{largelogs}
L=\frac{1}{32\pi^2}
 \left[3\log\frac{M_K^2}{M_\pi^2}+\log\frac{M_\eta^2}{M_K^2}\right].
\eeq

Similar expressions are derived for the decay constants:
\begin{eqnarray}
F_\pi^2 &=& F^2(3) + 2(m_s+2m) \tilde\xi  + 2m \xi   \label{pidecay}\\
          &&\quad +\frac{1}{16\pi^2} 2mB_{0} \left\{\log\frac{M_\eta^2}{M_K^2}
	  + 2\log\frac{M_K^2}{M_\pi^2}\right\} +F_\pi^2 e_{\pi}\qquad
	   \nonumber\\
F_K^2 &=& F^2(3) + 2 (m_s + 2m) \tilde\xi  + (m_s+m)\xi  \label{kadecay}
    + m B_{0} L +F_K^2 e_{K}
\end{eqnarray}
The scale-invariant constants $\xi $ and $\tilde\xi $ are related to
the LEC's $L_4$ and $L_5$ as follows:
\begin{eqnarray}\label{nxi}
\tilde \xi  &=& 8B_{0}\left\{L_4(\mu) 
   -\frac{1}{256\pi^2}\log\frac{M_K^2}{\mu^2}\right\}\,\\
\xi  &=& 8B_{0n}\left\{L_5(\mu)- \frac{1}{256\pi^2}
    \left[\log\frac{M_K^2}{\mu^2}
              + 2 \log\frac{M_\eta^2}{\mu^2}\right]\right\} .
\end{eqnarray}
The remainders $F_P^2 e_{P}$ collect the NNLO terms $O(m_q^2)$. 
Eqs.~(\ref{ordandfluc}), (\ref{epsilon})-(\ref{eta}) in Sec.~\ref{secperturb}
can be obtained by combining the previous identities to eliminate
the $O(p^4)$ LEC's $A,Z_S,\xi,\tilde\xi$.

The identities for the $\eta$ can 
be recast in a form reminiscent of the Gell-Mann--Okubo formula:
\begin{eqnarray} \label{etamass}
&& 3 F_\eta^2M_\eta^2 - 4 F_K^2M_K^2 + F_\pi^2M_\pi^2 \\
&&  \quad = 4(r-1)m^2 \left\{(r-1)(2 Z^p + A ) 
   - B^2_{0} L\right\} +3 F_\eta^2M_\eta^2d_\eta
   - 4 F_K^2M_K^2 d_K + F_\pi^2M_\pi^2 d_\pi\,,\nonumber \\
&& 3 F_\eta^2 - 4 F_K^2 + F_\pi^2 \label{etadecay} \\
&&   \quad = \frac{2mB_{0}}{16 \pi^2} 
  \left[(1+2r)\log\frac{M_\eta^2}{M_K^2} 
       - \log\frac{M_K^2}{M_\pi^2}
  \right] + 3 F_\eta^2e_\eta - 4 F_K^2 e_K + F_\pi^2 e_\pi\,.\nonumber
\end{eqnarray}
The $\eta$-mass identity involves the new LEC $Z^p=16B_{0}^2 L_7$. 
We have also introduced
the NNLO remainders $d_\eta$ and $e_\eta$ of order $O(m_q^2)$.

\section{$\mathbf\chi$PT and Bayesian statistical analysis} \label{appbayes}

One of the main achievements of $SU(3)\times SU(3)$ chiral
perturbation theory consists in providing a consistent framework that
includes all the constraints of chiral symmetry when one analyses
processes involving the Goldstone bosons $\pi,K,\eta$. This allows one
to express observables as expansions in powers of momenta and quark
masses up to a given order. Each order involves new low-energy
constants whose values cannot be determined from chiral constraints,
but nevertheless provide very much needed insight into the mechanism
of chiral symmetry breaking.

We would like therefore to pin down (or at least constrain) LEC's
arising at first order in $SU(3)\times SU(3)$ chiral expansions -- for
instance, the quark condensate and the pseudoscalar decay constant in
the limit $m_u,m_d,m_s\to 0$ -- by using available experimental
information.

We must therefore combine several sources of information:

\begin{itemize}
\item \emph{Experimental measurements of observables $\alpha_i$}:
$P^{\exp}_j(\alpha_i)$  

Their values have (possibly correlated) uncertainties which are
described by probability distributions. In the theoretical basis that
underlies the determination of these observables no use of $\chi$PT
should have been made. For instance, solutions of Roy equations were
required in addition to experimental phase shifts to extract $\pi\pi$
scattering parameters. As explained in Sec.~\ref{secinst}, a reasonable
choice of observables can be derived from QCD correlation functions of
currents and densities, away from kinematic singularities; e.g., one
may choose $\pi\pi$ subthreshold parameters.

\item \emph{Relations between the observables and theoretical
parameters $T_n$}: $\alpha_i={\mathcal{A}}_i(T_n)$

We have explained how mass and decay constant identities can be used
to eliminate some $O(p^4)$ LEC's in chiral expansions of
observables. We have chosen the theoretical parameters to be $X(3)$,
$Z(3)$, $r$, and remainders that comprise NNLO and higher order
corrections.

\item \emph{Theoretical constraints/assumptions about the values of
$T_n$}:  ${\mathcal{C}}_k(T_n)$

Chiral order parameters are constrained: for instance, vacuum
stability requires $X(3)\geq 0$, whereas $Z(3)$ is positive by
definition. We have also theoretical prejudices about NNLO remainders
if we require an overall convergence of chiral expansions: the
relative contribution of NNLO remainders must remain small.

\end{itemize}

It is quite easy from these elements to construct the likelihood function
\begin{equation}
\mathcal{L}(T_n)=P({\rm data}|T_n,H)
   =\prod_j  P^{\exp}_j[{\mathcal{A}}_i(T_n)]\,,
\end{equation}
which is the probability of obtaining the observed data once a
particular set of theoretical parameters is given (within the
theoretical framework $H$).  However, what we want is not
$\mathcal{L}$, but rather the reverse quantity $P(T_n|{\rm data},H)$,
i.e., the probability of having a particular set of theoretical
parameters, taking into account the data.

This problem of ``statistical inference'' has a long history.  A
possible solution is provided by Bayesian analysis~\cite{bayes}, which
relies on Bayes' theorem:
\begin{equation} \label{bayesth}
P(T_n|{\rm data},H) = \frac{\mathcal{L}(T_n) \cdot \pi(T_n|H)}
                      {\int\ [dT]\ \mathcal{L}(T_n) \cdot \pi(T_n|H)}\,,
\end{equation}
where $\pi$ is the ``prior'' distribution, i.e., the probability associated 
with 
the theoretical parameters before the experimental results have been taken in consideration:
\begin{equation}
\pi(T_n)=P(T_n|H)=\prod_k {\mathcal{C}}_k(T_n)\,.
\end{equation}
The denominator on the right-hand side of Eq.~(\ref{bayesth}) is just
a normalisation factor.  The marginal probability associated with
a given theoretical parameter is obtained by integrating the
joint probability $P(T_n|{\rm data},H)$ over all other theoretical
parameters. 

Let us mention that there is some arbitrariness in each of these
ingredients.  We could have chosen other observables, such as
combinations of scattering lengths in which one-loop chiral logarithms
cancel \cite{pipiCGL}. We could have added all $O(p^4)$ LEC's to the
set of theoretical parameters, and kept ``bare'' chiral expansions as
relations among the observables. Finally, we could have computed $O(p^6)$
contributions to NNLO remainders and used resonance saturation to
estimate the size of the remainders, following the procedure in
Ref.~\cite{ABT}.

In the present paper we have advocated a particular choice of
observables as constituting a sensible starting point for a Bayesian
analysis of data from $\pi\pi$ and $\pi K$ scattering.  Other choices
of prior p.d.f's for the theoretical parameters (especially for the
NNLO remainders) can be considered, as long as they are well motivated.
However, the posterior probabilities should not be strongly sensitive
to the choice of priors when a sufficient amount of experimental data
is included in the analysis.

\section{Integration procedure for the analysis of $\pi\pi$ scattering}
\label{appinteg}

In Sec.~\ref{secpipianalys}, we apply Bayesian methods to perform a
three-flavour analysis of $\pi\pi$ scattering, constructing a joint
probability $P(r,Y,Z,\vec{\delta}|{\rm data})$.  This gives the
probability of having a particular choice of quark mass ratio $r$,
order parameters $Y(3)$ and $Z(3)$ and NNLO remainders
$\delta_{i=1\ldots 7}$, once $\pi\pi$ scattering data is taken into
account.  By integrating over NNLO remainders, we obtain the joint
probability
\begin{eqnarray}
P(r,Y,Z|{\rm data})&=&
 \int \prod_{i=1}^7 d\delta_i\ 
  P(r,Y,Z,\vec\delta|{\rm data})\\
&\propto& \pi(Y,Z)\ \theta(r-r_1)\ \theta(r_2-r) \ 
  \theta[Z(3)]\ \theta[Y(3)]\ \nonumber\\
&&\quad \times \int^{\delta_1^{\rm max}}\!\!\! d\delta_1 
  \int^{\delta_2^{\rm max}}\!\!\! d\delta_2
  \int_{\delta_6^{\rm min}} d\delta_6 
  \int_{\delta_7^{\rm min}} d\delta_7
  \int d\delta_3\ d\delta_4\ d\delta_5\nonumber\\
&&\quad \times \prod_{i=1}^7 G(\delta_i,\sigma_i)
  \exp\left(-\frac{1}{2} {\mathcal{V}}^T C {\mathcal{V}}\right)
 \theta[Y^{\rm max}-Y(3)]\,,
\end{eqnarray}
where $\propto$ means ``equals, up to a (numerical) normalization
coefficient'', $C$ is the error matrix for the experimental data,
Eq.~(\ref{eq:expdistr}), $G(\delta_i, \sigma_i)$ is the Gaussian
function of $\delta_i$ with width $\sigma_i$, and ${\mathcal V}$ is
defined in terms of the chiral series for $\alpha_{\pi\pi}$ and
$\beta_{\pi\pi}$, Eqs.~(\ref{eq:alpha})-(\ref{eq:beta}):
\begin{equation}
{\mathcal{V}}(r,Y,Z,\vec{\delta})
  =\left(\begin{array}{c} 
{\mathcal{A}}(r,Y,Z,\vec{\delta})-\alpha_{\exp}\\
{\mathcal{B}}(r,Y,Z,\vec{\delta})-\beta_{\exp}
\end{array}\right)\,.
\end{equation}

Before any numerical evaluation, we can analytically compute some of
the integrals. $\delta_4$ and $\delta_5$ have the same width
$\sigma_4=\sigma_5$ and appear only in the experimental distribution
obtained from $\pi\pi$ scattering.  We can therefore diagonalize the
latter
\begin{equation}
R^TR=RR^T=I\,, \qquad
RCR^T=\left[\begin{array}{cc}
C_1 & 0\\ 0 & C_2
\end{array}\right]\,,
\end{equation}
to perform the integration~\footnote{A similar procedure can be
followed in the case of different widths for the Gaussian prior
p.d.f.'s for $\delta_4$ and $\delta_5$.} over $\delta_4$ and
$\delta_5$:
\begin{eqnarray}
&& \int \delta_4\, \delta_5\, G(\delta_4,\sigma_4)
   G(\delta_5,\sigma_4) \exp\left(-\frac{1}{2} {\mathcal{V}}^T C
   {\mathcal{V}}\right) \propto
   \exp\left[-\frac{1}{2}(D_1{\mathcal{W}_1}+D_2{\mathcal{W}_2})\right]\,,\\
   && D_i=\frac{C_i}{1+C_i\sigma_4^2}\,, \qquad
   {\mathcal{W}}=R\cdot{\mathcal{V}}(r,Y,Z,\delta_{1,2,3};\delta_4=\delta_5=0)\,.
\end{eqnarray}
The integrals over $\delta_6$ and $\delta_7$ are simply Gaussians
integrated over semi-infinite ranges, and are thus given in terms
of the error function ${\rm Erf}$. We obtain finally:
\begin{eqnarray}
&&P(r,Y,Z|{\rm data}) \propto \pi(Y,Z)\ \theta(r-r_1)\ \theta(r_2-r) \
  \theta[Z(3)]\ \theta[Y(3)]\nonumber\\ &&\quad \times
  \int^{\delta_1^{\rm max}}\!\!\! d\delta_1 \int^{\delta_2^{\rm
  max}}\!\!\! d\delta_2 \int d\delta_3 \prod_{i=1}^3
  G(\delta_i,\sigma_i)
  \exp\left[-\frac{1}{2}(D_1{\mathcal{W}_1}+D_2{\mathcal{W}_2})\right]
  \nonumber\\ &&\quad \times \left[1-{\rm
  Erf}\left(\frac{\delta_6^{\rm min}}{\sqrt{2}\sigma_6}\right) \right]
  \left[1-{\rm Erf}\left(\frac{\delta_7^{\rm
  min}}{\sqrt{2}\sigma_7}\right) \right] \theta[Y^{\rm max}-Y(3)]\,.
\end{eqnarray}

In order to obtain the marginal probability for either $r$, $X(3)$,
$Y(3)$ or $Z(3)$, we must perform a numerical integration over 3
remainders to obtain the joint probability $P(r,Y,Z|{\rm data})$, and
then integrate the result over two of the remaining three parameters.
We restrict the integration over the remainders $\delta_{1,2,3}$ to
the range $[-5\sigma_i,5\sigma_i]$ (the upper bound can be smaller for
$i=1,2$ due to the positivity constraints
Eqs.~(\ref{vacuum1})-(\ref{vacuum2})).  The range of integration for
the two parameters that remain to be integrated is given by the
theoretical constraints discussed in Sec.~\ref{secpipianalys}.

In order to appreciate the impact of data in the Bayesian framework,
it is quite interesting to consider the marginal probabilities
obtained when no experimental information is included. In our
particular case, this amounts to replacing the experimental
distribution $P_{\exp}(\alpha,\beta)$ by 1, or equivalently to setting
the matrix $C$ to 0. We obtain then the following p.d.f:
\begin{eqnarray}
&&P_0(r,Y,Z|{\rm data}) \propto \pi(Y,Z)\ \theta(r-r_1)\ \theta(r_2-r) \ 
  \theta[Z(3)]\ \theta[Y(3)]\nonumber\\
&&\quad \times \int^{\delta_1^{\rm max}}\!\!\! d\delta_1 
  \int^{\delta_2^{\rm max}}\!\!\! d\delta_2
  \int d\delta_3  \prod_{i=1}^3 G(\delta_i,\sigma_i)
  \theta[Y^{\rm max}-Y(3)]
  \nonumber\\
&&\quad 
 \times
 \left[1-{\rm Erf}\left(\frac{\delta_6^{\rm min}}{\sqrt{2}\sigma_6}\right)
    \right]
  \left[1-{\rm Erf}\left(\frac{\delta_7^{\rm min}}{\sqrt{2}\sigma_7}\right)
    \right]\,.
\end{eqnarray}
$P_0$ is just the normalized prior p.d.f, and corresponds to the
``phase space'' allowed by the theoretical constraints and assumptions
on the various parameters.  The resulting marginal probabilities can
be found in Sec.~\ref{secpipianalys}, where they are compared with the
ones that include experimental knowledge about $\pi\pi$ scattering.


\begin{thebibliography}{100}

\bibitem{GL1} J.~Gasser and H.~Leutwyler,
\ap{158}{1984}{142}.

\bibitem{ms}
S.~Chen, M.~Davier, E.~G\'amiz, A.~Hocker, A.~Pich and J.~Prades,
Eur.\ Phys.\ J.\ C {\bf 22} (2001) 31
[\hepph{0105253}].

E.~G\'amiz, M.~Jamin, A.~Pich, J.~Prades and F.~Schwab,
\jhep{0301}{2003}{060}
[\hepph{0212230}].

\bibitem{GL2}
J.~Gasser and H.~Leutwyler,
\npb{250}{1985}{465}.

\bibitem{scalar}
M.~Boglione and M.~R.~Pennington,
\prd{65}{2002}{114010}
[\hepph{0203149}].

S.~Spanier and N.~T\"ornqvist
in K.~Hagiwara {\it et al.}  [Particle Data Group Collaboration],
\prd{66}{2002}{010001}.

\bibitem{Bachir1} 
B.~Moussallam,
\epjc{14}{2000}{111}
[\hepph{9909292}].

\bibitem{Bachir2}
B.~Moussallam,
\jhep{0008}{2000}{005}
[\hepph{0005245}].

\bibitem{DS} S.~Descotes-Genon and J.~Stern,
\plb{488}{2000}{274}
[\hepph{0007082}].

\bibitem{D} S.~Descotes-Genon,
\jhep{0103}{2001}{002}
[\hepph{0012221}].

\bibitem{instant}
T.~Appelquist and S.~B.~Selipsky,
\plb{400}{1997}{364}
[\hepph{9702404}].

M.~Velkovsky and E.~V.~Shuryak,
\plb{437}{1998}{398}
[\hepph{9703345}].

\bibitem{dirac}
T.~Banks and A.~Casher,
\npb{169}{1980}{103}.

H.~Leutwyler and A.~Smilga,
\prd{46}{1992}{5607}.

J.~Stern,
\hepph{9801282}.

S.~Descotes-Genon and J.~Stern,
\prd{62}{2000}{054011}
[\hepph{9912234}].

\bibitem{DGS} S.~Descotes-Genon, L.~Girlanda and J.~Stern,
\jhep{0001}{2000}{041}
[\hepph{9910537}].

\bibitem{GST}
L.~Girlanda, J.~Stern and P.~Talavera,
\prl{86}{2001}{5858} 
[\hepph{0103221}].

\bibitem{ordfluc}
S.~Descotes-Genon, L.~Girlanda and J.~Stern,
\epjc{27}{2003}{115}
[\hepph{0207337}].

\bibitem{E865}
S.~Pislak {\it et al.}  [BNL-E865 Collaboration],
\prl{87}{2001}{221801}
[\hepex{0106071}];
\prd{67}{2003}{072004}
[\hepex{0301040}].

\bibitem{DFGS} 
S.~Descotes-Genon, N.~H.~Fuchs, L.~Girlanda and J.~Stern,
\epjc{24}{2002}{469}
[\hepph{0112088}].

\bibitem{BEG}
J.~Bijnens, G.~Ecker and J.~Gasser,
published in 2nd DAPHNE Physics Handbook (125-140), L.~Maiani,
  G.~Pancheri and N.~Paver Eds.
[\hepph{9411232}].

\bibitem{ABT}
G.~Amoros, J.~Bijnens and P.~Talavera,
\npb{568}{2000}{319}
[\hepph{9907264}];
\npb{585}{2000}{293}
[Erratum-ibid. {\bf B 598} (2000) 665]
[\hepph{0003258}].

\bibitem{twoloop}
J.~Bijnens, G.~Colangelo and P.~Talavera,
\jhep{9805}{1998}{014}
[\hepph{9805389}].

J.~Bijnens, P.~Dhonte and F.~Persson,
\npb{648}{2003}{317}
[\hepph{0205341}].

\bibitem{Kpi}
B.~Ananthanarayan and P.~B\"uttiker,
\epjc{19}{2001}{517}
[\hepph{0012023}].

B.~Ananthanarayan, P.~B\"uttiker and B.~Moussallam,
\epjc{22}{2001}{133}
[\hepph{0106230}].

\bibitem{bijscal}
J.~Bijnens and P.~Dhonte,
\hepph{0307044}.

\bibitem{GChPT} 
N.~H.~Fuchs, H.~Sazdjian and J.~Stern,
\plb{269}{1991}{183};
J.~Stern, H.~Sazdjian and N.~H.~Fuchs,
\prd{47}{1993}{3814}
[\hepph{9301244}].

M.~Knecht and J.~Stern,
published in 2nd DAPHNE Physics Handbook (169-190), L.~Maiani,
  G.~Pancheri and N.~Paver Eds.  [\hepph{9411253}].

\bibitem{bayes}
D.~Silvia, \emph{Data Analysis - A Bayesian Tutorial},
Clarendon Press (1996).

G.~D'Agostini,
Cern Yellow Report, CERN-99-03,
[\href{
http://preprints.cern.ch/cgi-bin/setlink?base=cernrep&categ=Yellow_Report&id=99-03
}{http://preprints.cern.ch/cgi-bin/setlink?base=cernrep{\&}categ=Yellow{\_}Report{\&}id=99-03}]

Additional references can be found on\\
\href{http://astrosun.tn.cornell.edu/staff/loredo/bayes/}
{http://astrosun.tn.cornell.edu/staff/loredo/bayes/}

For recent discussions of the Bayesian and frequentist approaches
to the determination of CKM-matrix elements, see

M.~Ciuchini {\it et al.},
\jhep{0107}{2001}{013}
[\hepph{0012308}].

A.~H\"ocker, H.~Lacker, S.~Laplace and F.~Le Diberder,
\epjc{21}{2001}{225}
[\hepph{0104062}].


\bibitem{twoloopgen}
J.~Bijnens, G.~Colangelo and G.~Ecker,
\jhep{9902}{1999}{020}
[\hepph{9902437}].

J.~Bijnens, G.~Colangelo and G.~Ecker,
\ap{280}{2000}{100}
[\hepph{9907333}].

J.~Bijnens, L.~Girlanda and P.~Talavera,
\epjc{23}{2002}{539}
[\hepph{0110400}].

\bibitem{KMSF}
M.~Knecht, B.~Moussallam, J.~Stern and N.~H.~Fuchs,
\npb{457}{1995}{513}
[\hepph{9507319}].

\bibitem{bkm}
V.~Bernard, N.~Kaiser and U.~G.~Meissner,
\npb{357}{1991}{129};
\prd{43}{1991}{2757}.

\bibitem{TBP}
S.~Descotes-Genon, N.~H.~Fuchs, L.~Girlanda and J.~Stern,
in preparation.

\bibitem{modelsOp6}
J.~Bijnens,
\prep{265}{1996}{369}
[\hepph{9502335}].

G.~Ecker, J.~Gasser, A.~Pich and E.~de Rafael,
\npb{321}{1989}{311}.

\bibitem{GMO}
H.~Leutwyler,
\npb{337}{1990}{108}.

\bibitem{G}
L.~Girlanda,
\plb{513}{2001}{103}
[\hepph{0104270}].

\bibitem{ACGL}
B.~Ananthanarayan, G.~Colangelo, J.~Gasser and H.~Leutwyler,
\pr{353}{2001}{207}
[\hepph{0005297}].

\bibitem{km}
D.~B.~Kaplan and A.~V.~Manohar,
\prl{56}{1986}{2004}.

\bibitem{pipiCGL}
G.~Colangelo, J.~Gasser and H.~Leutwyler,
\plb{488}{2000}{261}
[\hepph{0007112}].

G.~Colangelo, J.~Gasser and H.~Leutwyler,
\npb{603}{2001}{125}
[\hepph{0103088}].

G.~Colangelo, J.~Gasser and H.~Leutwyler,
\prl{86}{2001}{5008}
[\hepph{0103063}].

\bibitem{pikroy}
P.~B\"uttiker, S.~Descotes-Genon and B.~Moussallam, 
\hepph{0310283}.

\bibitem{eta3pi}
J.~Gasser and H.~Leutwyler,
\npb{250}{1985}{539}.

A.~V.~Anisovich and H.~Leutwyler,
\plb{375}{1996}{335}
[\hepph{9601237}].

M.~Walker, ``$\eta\to 3\pi$'', Master Thesis, University of Bern (1998).
Can be obtained from 
\href{http://www-itp.unibe.ch/research.shtml#diploma}{http://www-itp.unibe.ch/research.shtml{\#}diploma}.

J.~Bijnens and J.~Gasser,
{\it Phys.\ Scripta} {\bf T99} (2002) 34
[\hepph{0202242}].

\bibitem{gensr}
J.~Bijnens, J.~Prades and E.~de Rafael,
\plb{348}{1995}{226}
[\hepph{9411285}].

S.~Narison,
\hepph{0202200}.


\bibitem{Kamborsr}
J.~Kambor and K.~Maltman,
\prd{62}{2000}{093023}
[\hepph{0005156}];
\plb{517}{2001}{332}
[\hepph{0107060}].

\bibitem{Pichsr}
M.~Jamin, J.~A.~Oller and A.~Pich,
\epjc{24}{2002}{237}
[\hepph{0110194}].

\bibitem{Jaminsr}
M.~Jamin,
\plb{538}{2002}{71}
[\hepph{0201174}].

\bibitem{sr4quark}
V.~Cirigliano, E.~Golowich and K.~Maltman,
\prd{68}{2003}{054013}
[\hepph{0305118}].

J.~F.~Donoghue and E.~Golowich,
\plb{478}{2000}{172}
[\hepph{9911309}].

J.~Bijnens, E.~Gamiz and J.~Prades,
\jhep{0110}{2001}{009}
[\hepph{0108240}].

\bibitem{GW}
P.~H.~Ginsparg and K.~G.~Wilson,
\prd{25}{1982}{2649}.

D.~B.~Kaplan,
\plb{288}{1992}{342}
[\heplat{9206013}].

H.~Neuberger,
\plb{417}{1998}{141}
[\heplat{9707022}],
\plb{427}{1998}{353}
[\heplat{9801031}].

M.~L\"uscher,
\plb{428}{1998}{342}
[\heplat{9802011}].

\bibitem{twisted}
R.~Frezzotti and G.~C.~Rossi,
\heplat{0306014}.

\bibitem{staggered}
L.~Susskind,
\prd{16}{1977}{3031}.

H.~S.~Sharatchandra, H.~J.~Thun and P.~Weisz,
\npb{192}{1981}{205}.

H.~Kluberg-Stern, A.~Morel, O.~Napoly and B.~Petersson,
\npb{220}{1983}{447}.

\bibitem{pqQCDappl}
D.~R.~Nelson, G.~T.~Fleming and G.~W.~Kilcup,
\prl{90}{2003}{021601}
[\heplat{0112029}].

\end{thebibliography}
\end{document}